\documentclass[11pt,a4paper]{article}
\usepackage[utf8]{inputenc}
\usepackage{amsmath, amssymb, amsfonts}
\usepackage{graphicx}
\usepackage{hyperref}
\usepackage{geometry}
\usepackage{cite}
\usepackage{braket}
\usepackage{subcaption}
\usepackage{soul}
\usepackage{xcolor}

\usepackage{authblk}

\title{Proton-Neutron Pairing in $N=Z$ Nuclei within the Quark-Meson-Coupling Energy Density Functional}

\author[1]{T. Popa}
\author[1]{N. Sandulescu\footnote{corresponding author, email: sandulescu@theory.nipne.ro}}
\author[2]{D. Gambacurta}

\affil[1]{National Institute of Physics and Nuclear Engineering, 077125 Magurele, 
Romania}
\affil[2]{Laboratori Nazionali del Sud, 95123 Catania, Italy}

\begin{document}

\maketitle

\begin{abstract}
We investigate the impact of isovector and isoscalar proton–neutron pairing correlations on the ground-state 
properties of even–even $N=Z$ nuclei with mass numbers in the range $16 \leq A \leq 120$. 
Nuclear mean fields are 
generated using the quark–meson coupling (QMC) energy density functional, while pairing correlations are 
treated within the quartet condensation model (QCM). Ground-state energies are obtained from axially deformed, 
self-consistent QMC+QCM calculations employing a zero-range pairing interaction with a density-dependent term derived consistently within the QMC framework. We show that proton–neutron pairing provides a significant contribution
to the binding energies of $N=Z$ nuclei, leading to improved agreement with experimental data.

\end{abstract}

\section{Introduction}

The role of isovector ($T=1$) and isoscalar ($T=0$) proton–neutron ($pn$) pairing in nuclei close 
to the  $N=Z$ line has been extensively discussed over many years 
(see the reviews \cite{frauendorf,sagawa}). 
Among the key issues addressed is the influence of $pn$ pairing on the excess binding 
of $N=Z$ nuclei relative to their neighboring nuclei \cite{satula,chasman,bentley,qcm_wigner}, 
the delayed alignment observed in high-spin states of $N=Z$ nuclei compared with
$N > Z$ systems \cite{kaneko,cederwall}, the superallowed $\alpha$ decay of
$N=Z$ nuclei above $^{100}$Sn \cite{auranen,liotta,clark} and the emergence of
four-body, $\alpha$-like correlations in $N=Z$ nuclei 
\cite{flowers,yamamura,senkov,dobes,qcm_t1}. Another long-standing and
highly debated question is whether nuclei can sustain a condensate of isoscalar, deuteron-like
pairs. Experimental signatures of such a pairing phase have been sought in a variety of observables \cite{frauendorf,sagawa}, most recently in proton–neutron transfer reactions between $N=Z$ nuclei \cite{assie}; however, to date, no unambiguous experimental evidence has been found.

From a theoretical point of view, $pn$ pairing is commonly described within 
the Hartree – Fock – Bogoliubov (HFB) framework, in which all pairing channels are treated on an equal 
footing through a generalized Bogoliubov transformation that mixes proton and neutron single-particle states \cite{goodman_review}. 
Most HFB calculations predict that isovector pairing correlations dominate the ground states of light
and medium-mass $N=Z$ nuclei and, in most cases, do not coexist with isoscalar
pairing \cite{goodman2001}. 
More recent HFB studies  suggest that isoscalar pairing may become dominant in the ground states 
of some heavy $N=Z$ nuclei with mass numbers $A>100$ \cite{gezerlis_bertsch}. 
However, these predictions are strongly dependent on nuclear
deformation \cite{gezerlis_def} and have not yet been tested in fully self-consistent HFB calculations in which deformation
and pairing correlations are allowed to compete together.

Furthermore, it remains unclear to what extent the predictions of proton–neutron HFB calculations are affected by the fact that they do not exactly conserve particle number, angular momentum, and isospin. Proton–neutron HFB calculations including particle-number and angular-momentum projection have been performed recently \cite{hfb_augusto}; however, these projections were carried out after variation and did not account for deformation effects, which are essential for a realistic description of $N=Z$ nuclei.

An alternative to the HFB approach, which exactly conserves particle number, angular momentum, and isospin, is the Quartet Condensation Model (QCM) \cite{qcm_t1,qcm_t0t1,qm_qcm_t0t1}. In the QCM framework, the ground state of even–even $N=Z$ nuclei is described as a condensate of quartets composed of two neutrons and two protons, coupled to total isospin $T=0$ and, in the case of spherical symmetry, to total angular momentum $J=0$. It has been shown that the QCM provides a highly accurate description of correlation energies in $N=Z$ systems interacting through both $T=1$ and $T=0$ pairing forces, with relative errors below $1\%$.

Recently, the QCM has been applied to study the effects of isovector and isoscalar pairing correlations on the ground-state energies of nuclei close to the $N=Z$ line \cite{skyrme_qcm}. This analysis was performed using self-consistent axially deformed Skyrme+QCM calculations with a zero-range pairing interaction. The results demonstrated that proton–neutron pairing significantly improves the agreement between calculated and experimental binding energies. 
It is worth recalling that $N=Z$ nuclei are typically underbound by several MeV in Skyrme and Gogny HFB calculations that neglect proton–neutron pairing correlations.

The purpose of the present work is to extend the above analysis of $N=Z$ nuclei within the framework of the Quark–Meson Coupling (QMC) model \cite{guichon1988,guichon1996,qmc_edf,qmc_progr}. Recent mean-field calculations based on the energy density functional derived from the QMC model have shown remarkable accuracy in reproducing nuclear properties \cite{stone,qmc_martinez,qmc3,thesis_martinez}. For instance, the latest version of the QMC model \cite{qmc3}, employed in the present study, predicts nuclear binding energies with a root-mean-square deviation of 1.74, which is significantly  smaller than the value of 3.11 obtained with the Skyrme SV-min \cite{sv-min}, despite relying on only five adjustable parameters compared with the fourteen parameters of SV-min.

Unlike Relativistic Mean-Field (RMF) and Skyrme- or Gogny-type energy density functionals, which treat nucleons as point-like particles, the QMC model describes nucleons as clusters of confined quarks interacting through scalar and vector meson fields. The most important consequences of explicitly including the quark substructure are the polarization of nucleons in the nuclear medium and the density dependence of the QMC energy density functional, which is microscopically derived. Since the QMC functional is obtained from the non-relativistic limit of the underlying relativistic QMC model, it automatically incorporates spin–orbit and tensor interactions, without the need to introduce additional parameters, in contrast to Skyrme and Gogny functionals. Moreover, in the most recent version of the QMC functional, the like-particle pairing interaction is also derived consistently within the QMC framework.

In QMC calculations performed so far, pairing correlations have been restricted to neutron–neutron and proton–proton channels and treated within the BCS approximation. In the present work, we extend the QMC framework by explicitly including proton–neutron pairing correlations, which are treated using the QCM approach. In this framework, we perform self-consistent axially deformed QMC+QCM calculations and investigate the impact of proton–neutron pairing on the ground states of even–even $N=Z$ nuclei with mass numbers in the range $16 \leq A \leq 120$.

The paper is organized as follows. In Section 1, we briefly review the QMC and QCM models and introduce the proton–neutron pairing interaction employed in this study. Section 2 describes the computational scheme used to carry out the self-consistent QMC+QCM calculations. In Section 3, we present and discuss the results.

\section{Theoretical Framework}
\subsection{Quark-Meson Coupling Model}

The Quark-Meson Coupling (QMC) model and the associated energy density functional (EDF) was already presented in detail
in previous publications (e.g., see the review \cite{qmc_progr}). 
For the consistency reason, below we recall the basic assumptions.

The QMC model, developped by Guichon, Thomas and collaborators \cite{guichon1988,guichon1996}, provides a self-consistent framework for describing nuclear structure by incorporating medium modifications to nucleon properties through quark-meson interactions. In QMC the nucleons are treated as clusters of three confined quarks. The quarks from different bags interact via scalar (\(\sigma\)), vector (\(\omega\)), and isovector (\(\rho\)) mesons. These mesons mediate the effective nuclear force, while the quark substructure introduces density-dependent modifications to the nucleon properties, a feature absent in the conventional mean-field approaches.

The derivation of the QMC EDF starts from the total energy of the system \cite{qmc_progr}
\begin{equation}
    E_{QMC}=\sum_{i=1,...}\sqrt{P_i^2 + M_i^{*2}(\sigma(\Vec{R_i}))} + g^i_\omega\omega(\Vec{R_i})+g_\rho\Vec{I_i}\cdot\Vec{B}(\Vec{R_i})+E_\sigma+E_{\omega,\rho}
\end{equation}
where $M_i^{*}$, $\Vec{R_i}$ and $\Vec{P_i}$ are the effective mass, the position and the momentum of the baryon $i$. The quantities  $\sigma(\Vec{R_i})$, $\omega(\Vec{R_i})$ and $\Vec{B}(\Vec{R_i})$ are the meson fields, whereas $\Vec{I}_i$ is the isospin operator. 

A key consequence of the quark nucleon structure  is the quadratic dependence on the effective Dirac mass on the $\sigma$ field:
\begin{equation}
M_N^* = M_N - g_\sigma \sigma + \frac{d}{2} (g_\sigma \sigma)^2,
\end{equation}
where \(g_\sigma\) is the coupling constant for the scalar meson.
The quantity $d$ denotes the scalar polarizability, which accounts for medium effects on the internal structure of the nucleon. Its value depends on the bag radius that confines the quarks. For a bag radius of 1 fm, commonly adopted in nuclear-structure applications \cite{stone,qmc_martinez}, the corresponding value is 
$d$ = 0.18 fm.

The meson contributions to the total energy (1),  $E_\sigma$ and $E_{\omega,\rho}$,  have a similar form as in RMF \cite{rmf}. For example, the mean-field energy coming from the $\sigma$ meson is:
\begin{equation}
E_{\sigma} = \int d^3r \left[ \frac{1}{2} (\Vec{\nabla} \sigma)^2 + V(\sigma) \right].
\end{equation}
The potential $V(\sigma)$ is chosen of the form:
\begin{equation}
V(\sigma) = \frac{1}{2} m_\sigma^2 \sigma^2 + \frac{\lambda_3}{3} \sigma^3.
\end{equation}
As in the case of RMF, the cubic term is essential for providing an accurate description of surface properties of nuclei. 

The derivation of the QMC EDF from the total energy of the system is presented in Refs. \cite{qmc_edf,qmc_progr}. The EDF, obtained from the non-relativistic limit of the QMC equations, consists of several zero-range terms with explicit density dependence. In contrast to Skyrme EDFs, these terms are derived microscopically and exhibit a more involved density dependence, including inverse powers of
$(1+d\rho G_\sigma)$. 
 
Since the QMC EDF originates from an underlying relativistic framework, both the spin–orbit and tensor interactions are generated self-consistently within the model, in contrast to Skyrme or Gogny functionals where these terms are introduced explicitly. 
As a consequence, the QMC EDF depends on a small number of parameters, which are adjusted to experimental data. They are: the coupling constants of the mesons $g_\sigma$, $g_\omega$ and $g_\rho$, the mass of the sigma meson (the masses of the other two mesons are taken at their physical values) and the parameter $\lambda_3$.

\subsection{Quartet Condensation Model}

In this study, we employ the version of the quartet condensation model (QCM) specifically developed to treat both isovector and isoscalar pairing correlations in axially deformed N=Z nuclei \cite{qcm_t0t1}. In this framework, the pairing interaction is assumed to scatter pairs of nucleons in time-reversed single-particle states of an axially deformed mean field, which in the present case is generated by the QMC energy density functional.
The corresponding pairing Hamiltonian is 
\begin{equation}\label{QCMH}
    H=\sum_i (\epsilon_i^\nu N_i^\nu + \epsilon_i^\pi N_i^\pi) + \sum_{i,j} V_{i,j}^{T=1}\sum_\tau P_{i,\tau}^{\dagger}P_{j,\tau} + \sum_{i,j} V_{i,j}^{T=0} D_{i0}^{\dagger}D_{j0} 
\end{equation}
where $\epsilon_i^\nu$ and $\epsilon_i^\pi$ are the single-particle energies of the neutrons and protons, respectively,
$N_i^{\nu}$ and $N_i^{\pi}$ are the corresponding number operators, and  $V_{i,j}^T$ are the matrix 
elements of the pairing interaction in the isospin channel $T$.

The isovector pair creation operators \(P_{i,\tau}^{\dagger}\) (where $\tau=-1,0,1$) are defined as $P_{i,1}^{\dagger} = \nu_i^{\dagger}\nu_{\bar{i}}^{\dagger}$, $P_{i,-1}^{\dagger} = \pi_i^{\dagger}\pi_{\bar{i}}^{\dagger}$, and $P_{i,0}^{\dagger} = (\nu_i^{\dagger}\pi_{\bar{i}}^{\dagger}+\pi_i^{\dagger}\nu_{\bar{i}}^{\dagger})/\sqrt{2}$, where $\bar{i}$ denotes the time-reversed state of $i$. The isoscalar pair creation operators are defined as
$D_{i,0}^{\dagger} = (\nu_i^{\dagger}\pi_{\bar{i}}^{\dagger}-\pi_i^{\dagger}\nu_{\bar{i}}^{\dagger})/\sqrt{2}$. 

The ground state of the Hamiltonian (5) for even-even systems with an equal number of neutrons and protons is approximated by the QCM state \cite{qcm_t0t1}:
\begin{equation}\label{QCMWF}
    \ket{QCM}=[Q^{\dagger}_{T=1} + (\Delta^{\dagger}_0)^2]^{n_q}\ket{0}
\end{equation}
where $n_q=(N+Z)/4$ is the number of quartets, and $N$ and $Z$ denote the numbers of 
neutrons and the protons above the " core " state $\ket{0}$, which are included in the pairing calculations. 

In the QCM state (6) the isovector pairing correlations are taken into account
by the isovector quartet $Q^{\dagger}_{T=1}$ built by two neutrons and two 
protons coupled to the total isospin T=0. Its expression is
\begin{equation}
Q^{\dagger}_{T=1} = 2\Gamma_{1}^{\dagger}\Gamma_{-1}^{\dagger} - (\Gamma_0^{\dagger})^2
\end{equation}
where 
\begin{equation}
\Gamma^{\dagger}_{\tau}=\sum_i x_i P_{\tau,i}^{\dagger}
\end{equation}
are the collective isovector pair operators. 
The isoscalar proton-neutron correlations are taken into account through the collective
isoscalar pair
\begin{equation}
\Delta^{\dagger}_0=\sum_i y_i D_{i0}^{\dagger}
\end{equation}

The QCM state (6) has the following properties: (a) it provides  an exact solution for the ground state of the Hamiltonian (6) in the case of degenerate single-particle levels; (b) for non-degenerate single-particle states and realistic pairing interactions, for which the Hamiltonian (6) can be diagonalized exactly, it gives a very good
approximation to the ground-state, with errors in the correlation energy below 1$\%$.

The QCM state is constructed by applying a collective four-body quartet operator - composed of two neutrons and two protons - 
$n_q$ times to the vacuum state $\ket{0}$. For this reason, and by analogy with pair condensation, the QCM state it is referred to as a quartet condensate. It is important to note, however, that this state does not correspond to a Bose–Einstein condensate, since the quartet operators do not obey bosonic commutation relations.

By construction, and in contrast to the HFB approximation, the QCM state has well-defined particle number and total isospin, which is zero for even–even $N=Z$ systems. In the case of an axially symmetric mean field, the QCM state has also a well-defined projection of the total angular momentum, $J_z$=0, although it does not possess a well-defined total angular momentum.

The QCM state depends on the mixing amplitudes of the collective isovector (8) and 
isoscalar (9) pairs, $x_i$ and $y_i$, respectively. These amplitudes  are determined
 variationally by minimizing  the expectation value of the
pairing Hamiltonian in the QCM state,
\begin{equation}
\delta_{x,y}\bra{QCM}H\ket{QCM}=0
\end{equation}
subject to the normalisation condition $\bra{QCM}\ket{QCM}=1$.

In the mean field QMC+QCM calculations, the quantities of interest generated by the 
QCM state are the occupation probabilities of the proton and neutron
single-particle states involved in the pairing calculations, $ v^2_{p,i}$ and $v^2_{n,i}$,
as well as the isovector and isoscalar pairing energies. The pairing energies are defined as
\begin{subequations} \label{PEdef}
\begin{align}
E_P^{T=1}= E_I^{T=1} - \sum_i V_{i,i}^{T=1} (v^4_{p,i} + v^4_{n,i} + v^2_{p,i} v^2_{n,i}) \\
E_P^{T=0}= E_I^{T=0} - \sum_i V_{i,i}^{T=0} v^2_{p,i} v^2_{n,i}
\end{align}
\end{subequations}
where $E_I^T$  are  the expectation values of the isovector and isoscalar pairing 
interactions in the  QCM state:
\begin{align*}
E_I^{T=1}= \bra{QCM}\sum_{i,j} V_{i,j}^{T=1}\sum_\tau P_{i,\tau}^{\dagger}P_{j,\tau}\ket{QCM} \\
E_I^{T=0}= \bra{QCM} \sum_{i,j} V_{i,j}^{T=0} D_{i0}^{\dagger}D_{j0}\ket{QCM}.
\end{align*}

As can be noticed from Eqs. (11), the contributions of the self-energies (the second terms) are not included in the pairing energies, since they would merely renormalize the single-particle energies generated by the mean field of the QMC energy density functional.

\subsubsection{Isovector and isoscalar pairing interactions}
\vskip 0.3cm
{\it (a) Neutron-neutron and proton-neutron pairing} 
\vskip 0.3cm

In the present study, we adopt a like-particle pairing interaction proportional to the
pairing force derived within the QMC model. This interaction can be written as
\begin{equation}
V_P^{(T=1)}(\vec{r},\vec{r}') = - s V_0 \left( 1- \eta \frac{\rho(\Vec{r})}{1+d'G_\sigma \rho(\Vec{r})} \right) 
\delta(\Vec{r} - \Vec{r}'),
\end{equation}
where $\rho$ is the baryon density, 
$V_0=G_\sigma-G_\omega - G_\rho/4 $ 
and $\eta =d'G^2_\sigma/V_0$.
The pairing force depends on the coupling constants for the scalar, vector and isovector mesons, $G_i={g_i^2}/{m_i^2}$, as well as on $d' = d + G_\sigma \lambda_3/3$, which account for the modification of the polarisability $d$ due to the contribution from the last term of Eq. (4).

In the definition above, we have introduced a scaling factor $s$, which is unity in the original formulation. In the present QMC+QCM calculations, this factor is adjusted to achieve a better description of the odd–even mass differences in 
$N>Z$ nuclei (see section 3.3 below).

\begin{figure}[htbp]
    \centering
\includegraphics[width=0.75\textwidth, angle=0]{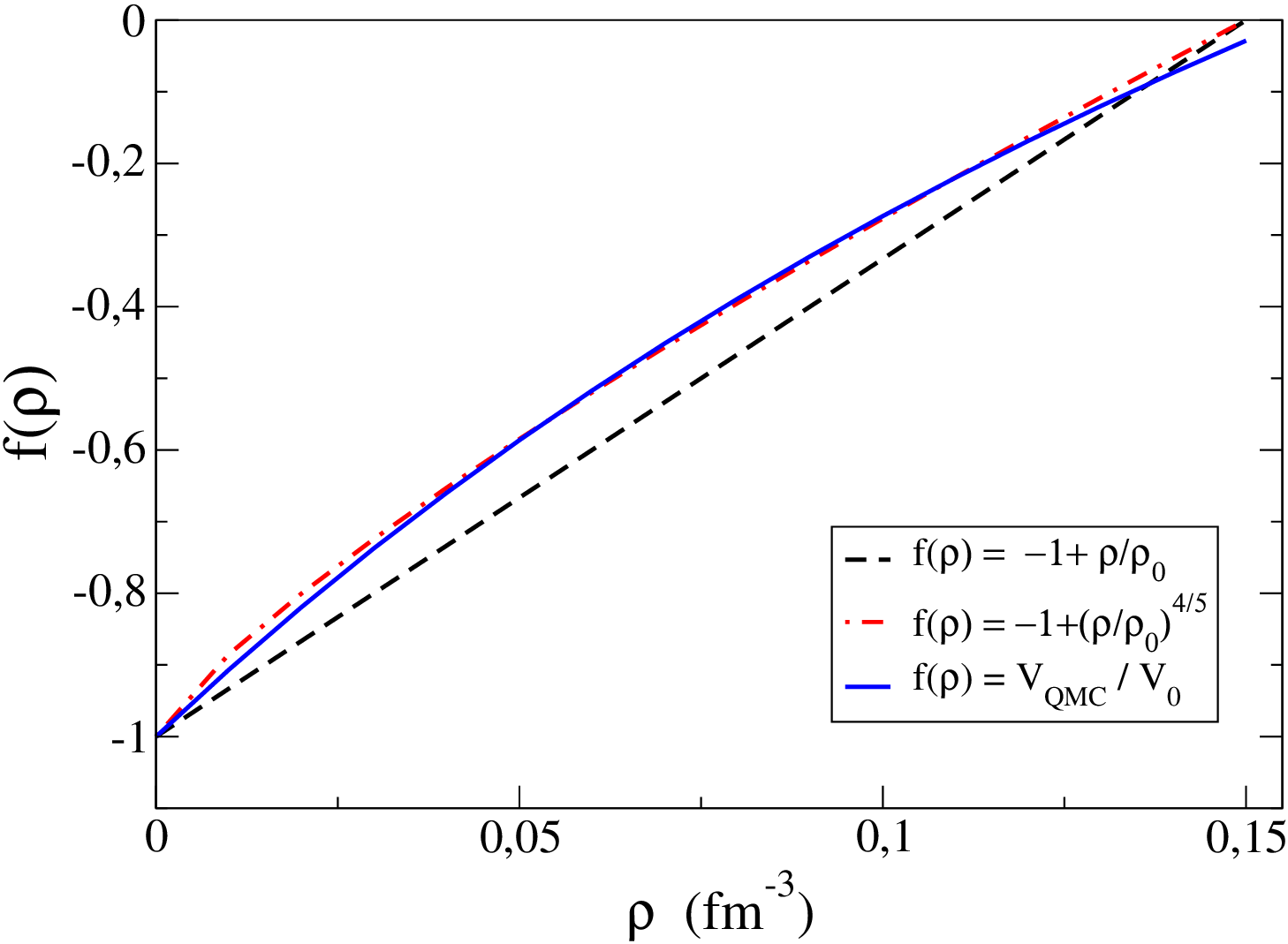}
\caption{Density dependence of the interaction (12) compared to  the force (13) for the parameters shown
in the figure.}
\end{figure}

The pairing force (12) has a non-trivial density dependence expression, 
different from the majority of mean field BCS or HFB calculations. 
In the latter cases the pairing force  is taken as
\begin{equation}
V(\vec{r},\vec{r}') = - \bar{V}_0 \left( 1- \bar{\eta} (\frac{\rho(\Vec{r})}{\rho_0})^{\alpha} \right) 
\delta(\Vec{r} - \Vec{r}')
\end{equation} 
where $\rho_0$ is the saturation density. 

Figure 1 shows the QMC pairing force normalized to its strength. It is evident that this interaction differs substantially from the volume ($\bar{\eta}$=1) and surface ($\bar{\eta}$=1, $\alpha$=1) pairing forces 
commonly employed in mean-field calculations. The same figure also demonstrates that the QMC pairing force can be reasonably approximated by the pairing interaction of Eq. (13) with parameters $\bar{\eta}$=1 and  $\alpha$=4/5. The fit was performed using $\rho_0$=0.15fm$^3$, corresponding to the nuclear-matter saturation density in the latest version of the QMC model employed in the present study.

\vskip 0.2cm
\noindent
{\it (b) Proton-neutron isovector and isoscalar pairing}
\vskip 0.2cm
 
Due to isospin invariance (with the Coulomb interaction neglected in the pairing channel), it is commonly assumed that the isovector proton–neutron pairing interaction is identical to the like-particle pairing interaction. Accordingly, for all three components of the isovector pairing force, we adopt the like-particle pairing interaction given in Eq. (12).

In principle, the isoscalar proton–neutron pairing interaction can also be derived within the QMC approach. However, no unambiguous derivation is currently available. For this reason, in treating the isoscalar proton–neutron pairing channel we follow the standard procedure commonly adopted in HFB calculations, assuming that the isoscalar pairing interaction is proportional to the isovector one. Namely
\begin{equation}\label{ISStrength}
    V_P^{(T=0)}=w V_P^{(T=1)} .
\end{equation}

\subsection{ Matrix elements of pairing interactions}

In the present QMC+QCM calculations the pairing interaction is supposed to scatter pairs in time-reversed axially-deformed single particle states $\psi_K$ and $\psi_{\Bar{K}}$ generated by the
QMC EDF mean field. They are defined by
\begin{equation*}
    \psi_K(\boldsymbol{r})=\varphi_{K\uparrow}(r_{\perp},z)e^{i \Lambda_K^{-}\phi}\ket{\uparrow} + \varphi_{K\downarrow}(r_{\perp},z)e^{i \Lambda_K^{+}\phi}\ket{\downarrow}
\end{equation*}
\begin{equation*}
    \psi_{\Bar{K}}(\boldsymbol{r})=\varphi_{K\downarrow}(r_{\perp},z)e^{-i \Lambda_K^{-}\phi}\ket{\downarrow} + \varphi_{K\uparrow}(r_{\perp},z)e^{-i \Lambda_K^{+}\phi}\ket{\uparrow}
\end{equation*}
where $K=(a,\Lambda_K^{\pm})$ and 
$\Lambda_K^{\pm}= \Omega_K \pm 1/2 $ is
the projection of the total angular momentum on $z$ axis, composed by the orbital and spin projections.

The matrix elements of the zero-range pairing interaction (Eqs. 12 and 14) in the isovector and isoscalar channels are derived in the Appendix. Here, we present the final expressions and discuss their key properties.

The matrix elements of the $T=1$ and $T=0$ interactions are the
following:
\begin{equation}
\bra{I\Bar{I}}V_P^{T=1}\ket{K\Bar{K}}= - \int r dr dz d\phi
V^{T=1}(r,z) [\varphi^2_{I\uparrow}(r_{\perp},z)+\varphi^2_{I\downarrow}(r_{\perp},z)][\varphi^2_{K\uparrow}(r_{\perp},z)+\varphi^2_{K\downarrow}(r_{\perp},z)] 
\end{equation}
\begin{multline}
\bra{I\Bar{I}}V_P^{T=0}\ket{K\Bar{K}}=
- \int r dr dz d\phi V^{T=0}(r,z) \{ [\varphi^2_{I\uparrow}(r_{\perp},z)-\varphi^2_{I\downarrow}(r_{\perp},z)][\varphi^2_{K\uparrow}(r_{\perp},z)-\varphi^2_{K\downarrow}(r_{\perp},z)] \\ 
 +
4 \varphi_{I\uparrow}(r_{\perp},z)\varphi_{I\downarrow}(r_{\perp},z)\varphi_{K\uparrow}(r_{\perp},z)\varphi_{K\downarrow}(r_{\perp},z) \}    
\end{multline}
In the expressions above, $V^{T}$ denotes the axially-symmetric part of the pairing interactions 
(Eqs. 12 and 14) which multiplies the delta function, i.e., $V_P^{T} = - V^{T} \delta$.

It can be observed that all matrix elements of the isovector pairing interaction are negative, as expected. The diagonal matrix elements of the isoscalar interaction are also negative and proportional to the isovector ones, since $V^{T=0}=wV^{T=1}$. In contrast, the off-diagonal matrix elements of the isoscalar interaction can take positive values. An example of the off-diagonal matrix elements in the two pairing channels is shown in Fig. 2, where the isoscalar matrix elements are seen to fluctuate between negative and positive values. The presence of positive off-diagonal matrix elements in the isoscalar channel is the main reason why isoscalar pairing correlations are usually suppressed relative to isovector pairing. This important physical effect is not captured in calculations that employ purely negative off-diagonal matrix elements in the isoscalar channel.

\begin{figure}[htbp]
 \centering
 \includegraphics[width=0.75\textwidth]{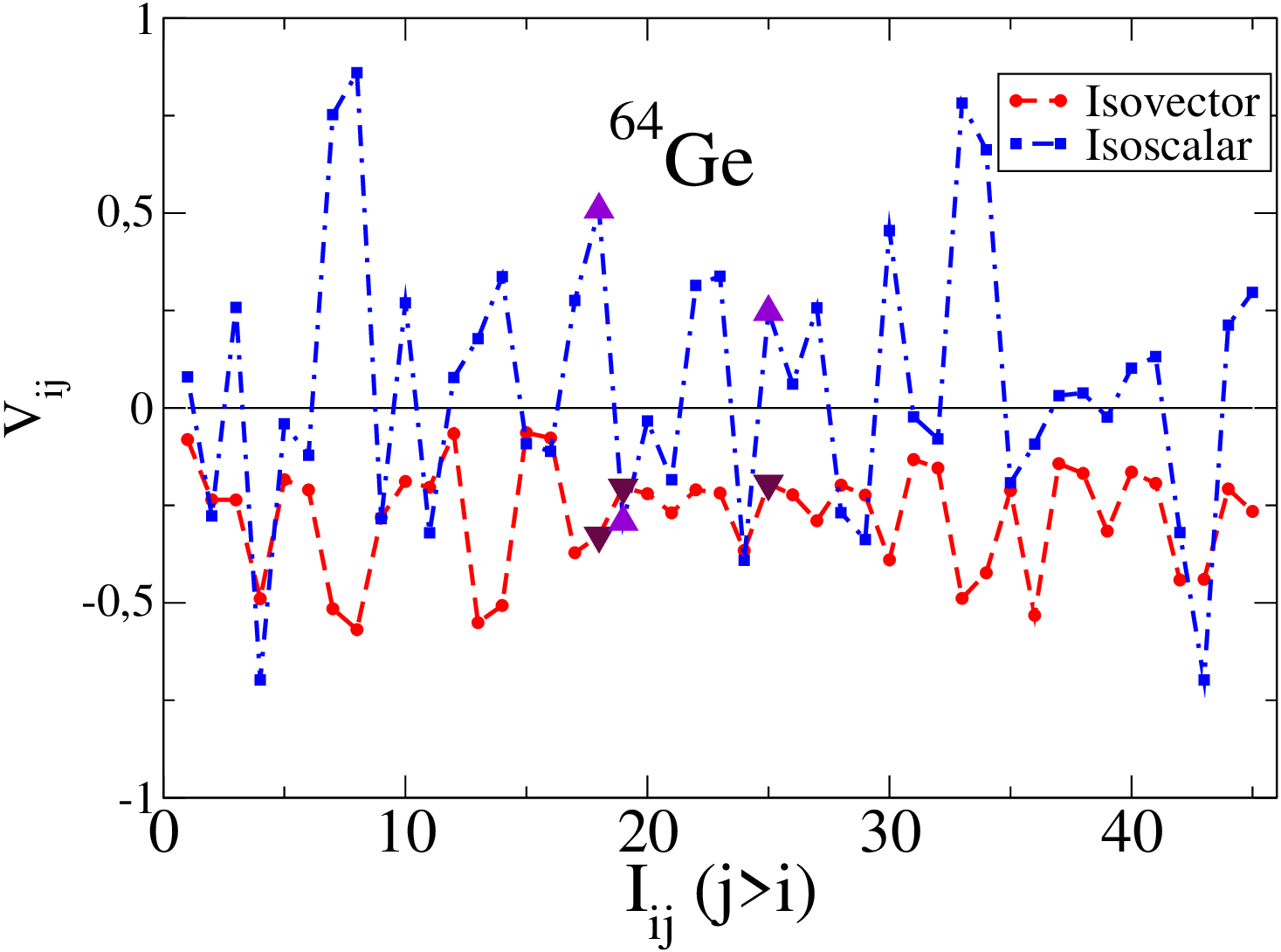}

 \caption{Off-diagonal matrix elements of the isovector (12) and isoscalar (14) 
pairing interactions in $^{64}$Ge, calculated with the scalling factors $s=1.5$ 
and $w=1.6$. Triangles indicate the matrix elements associated with states
near the Fermi level, which play a dominant role. The quantity $I_{ij}$ (with $j > i$) 
denotes the pair indices corresponding to $V_{ij}$, where $i$ and $j$ run from 1 to 10.}
   
 \label{fig:ME_pf}
\end{figure}

\section{Computational Scheme and Parametrisation}

\subsection{QMC calculations}

The mean field is calculated using the latest version of the QMC energy density functional, QMC$\pi$-III \cite{qmc3}. This version includes the spin–tensor component of the EDF, in addition to the pion-exchange Fock term, the $\sigma$-meson self-interaction, 
and the full spin–orbit interaction—containing both space and time components—that were already present in earlier QMC versions. The QMC calculations are performed using the code developed by the QMC group, which has been modified to incorporate proton–neutron pairing correlations.

The QMC mean-field depends on the meson-quark coupling constants, the mass of the $\sigma$ meson (the masses of the other mesons are taken at their physical values) and the parameter $\lambda_3$. These parameters are obtained from a fit to the binding energies and 
root-mean-square radii of doubly magic and semi-magic nuclei. The fitting procedure is described in Refs.\cite{qmc3,thesis_martinez}. 
The resulting parameter values are $G_\sigma = 9.62$, $G_\omega = 5.21$, $G_\rho =4.71$, $M_\sigma =503$ MeV, and $\lambda_3 = 0.05 $. 

The parameters discussed above were obtained within the QMC+BCS framework, which includes like-particle pairing described by the pairing interaction (12) derived within the QCM approach. In this framework, the BCS approximation yields zero pairing energy for doubly magic nuclei, as well as for the closed-shell subsystems of semi-magic nuclei. This is no longer the case when pairing correlations are treated beyond the BCS approximation, as in particle-number–projected HFB calculations or within the QCM approach, 
as discussed below (see section 5.2).

Consequently, when performing QMC+QCM calculations using the parameters obtained from the QMC+BCS fit, one expects an overbinding of the doubly magic $N=Z$ nuclei included in the fit, relative to the corresponding QMC+BCS results. A more consistent alternative would be to refit the QMC parameters using binding energies calculated within the QMC+QCM framework. Such a refit, however, is beyond the scope of the present study.

\subsection{QCM calculations}

The variational solution of the QCM equations (10) is obtained by analytically evaluating the expectation value of the pairing Hamiltonian and the norm using the Cadabra algorithm \cite{cadabra}
(for details, see the Appendix of Ref. \cite{qcm_t0t1_ngz})
Since the analytical derivation becomes increasingly cumbersome as the number of particles involved in the pairing calculations increases, we perform the calculations using the QCM wave function (6) with $n_q$=3. This choice implies that, for a given 
(N,Z) nucleus, the pairing interaction is assumed to act on the six protons and six neutrons outside the core (N-6,Z-6). These nucleons are allowed to scatter among ten single-particle states above the core. In the QCM calculations presented below, the occupation probability of the first level above the core is close to unity. Since the states within the core (N-6,Z-6) are more deeply bound, their occupation probabilities are even higher; consequently, their contribution to the pairing correlations can be neglected.

The QCM calculations are performed using an isospin-invariant pairing Hamiltonian, in which the proton single-particle states are taken to be identical to the neutron single-particle states. This approximation does not affect significantly the total pairing energies.

\subsection{Pairing Interaction Strengths}

The strength of the zero-range pairing force (13) derived within QCM is not well-defined. This is because such an interaction, when applied in the particle-particle channel, produces results that depend strongly on the model space in which the pairing is active. 
A typical example is the BCS pairing gap, which diverges as the dimension of the model space increases.

To fix the pairing strength of the interaction (13), we follow the standard procedure used in BCS-type calculations, namely, through  the odd-even mass differences (OEMD).

For the N=Z nuclei with the atomic mass $16 \leq A \leq 120$ analyzed in this paper, the strength of the isovector pairing force is adjusted separately for nuclei with valence nucleons above $^{16}$O, $^{40}$Ca and $^{100}$Sn. Since it is difficult to disentangle the contribution of proton-neutron pairing to the binding energy of N=Z nuclei, the pairing strength is fixed from the OEMD of $N>Z$ nuclei with $Z=8,20,50$. For the OEMD we use the 3-point expression
\begin{equation}
\Delta^{(3)} (N) = \frac{1}{2} [B(N+1,Z)+B(N-1,Z)-2B(N,Z)]
\end{equation}
where B(N,Z) are the experimental binding energies and N is an odd number. 

The OEMD is compared to the BCS average pairing calculated with the interaction (13). The average pairing gap is estimated by 
\begin{equation}
\bar{\Delta} = \frac{\sum_i \Delta_i u_i v_i}{\sum_i u_i v_i}
\end{equation}
where 
\begin{equation}
\Delta_i =- \frac{1}{2} \sum_j V^{(T=1)}_{ij} u_j v_j 
\end{equation}

The results obtained for various values of the scaling 
factors $s$ are shown in Fig. 3. These results are derived from BCS calculations performed with ten states above the doubly magic nuclei $^{16}$O, $^{40}$Ca and $^{100}$Sn.

As shown in Fig.~3, for all nuclei considered, the average pairing gaps predicted by the interaction
with the strength derived within the QMC approach ($s$=1), are significantly smaller than the experimental odd–even mass differences. Better agreement for the oxygen, calcium, and tin isotopes is obtained with scaling factors $s=1.3$, $1.5$, and 
$1.8$, respectively. These scaling factors are employed in the QCM calculations presented below.

\begin{figure}[htbp]
    \centering
    \begin{subfigure}{0.3\textwidth}
        \centering
        \includegraphics[width=\textwidth]{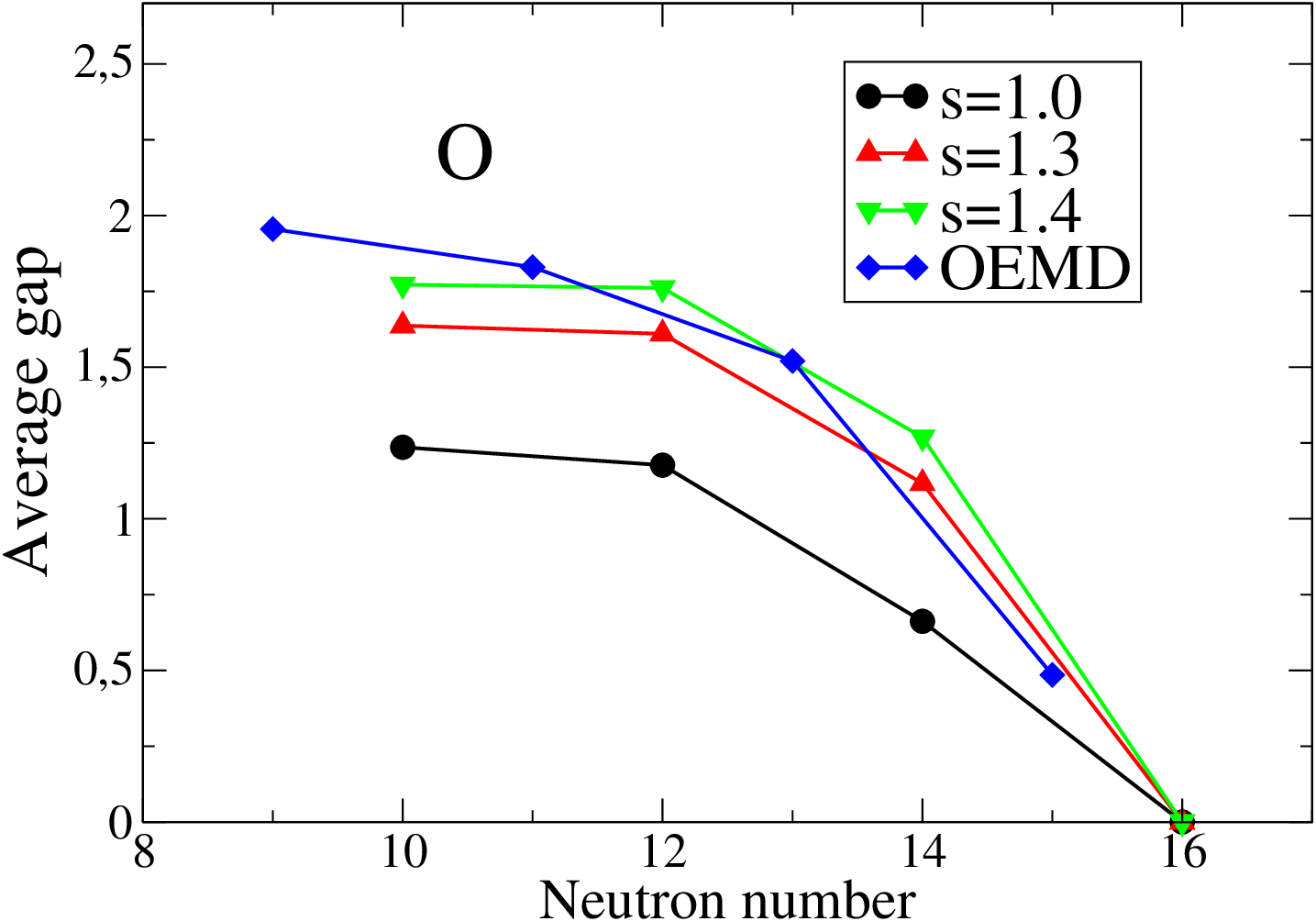}
    \end{subfigure}
    \begin{subfigure}{0.3\textwidth}
        \centering
        \includegraphics[width=\textwidth]{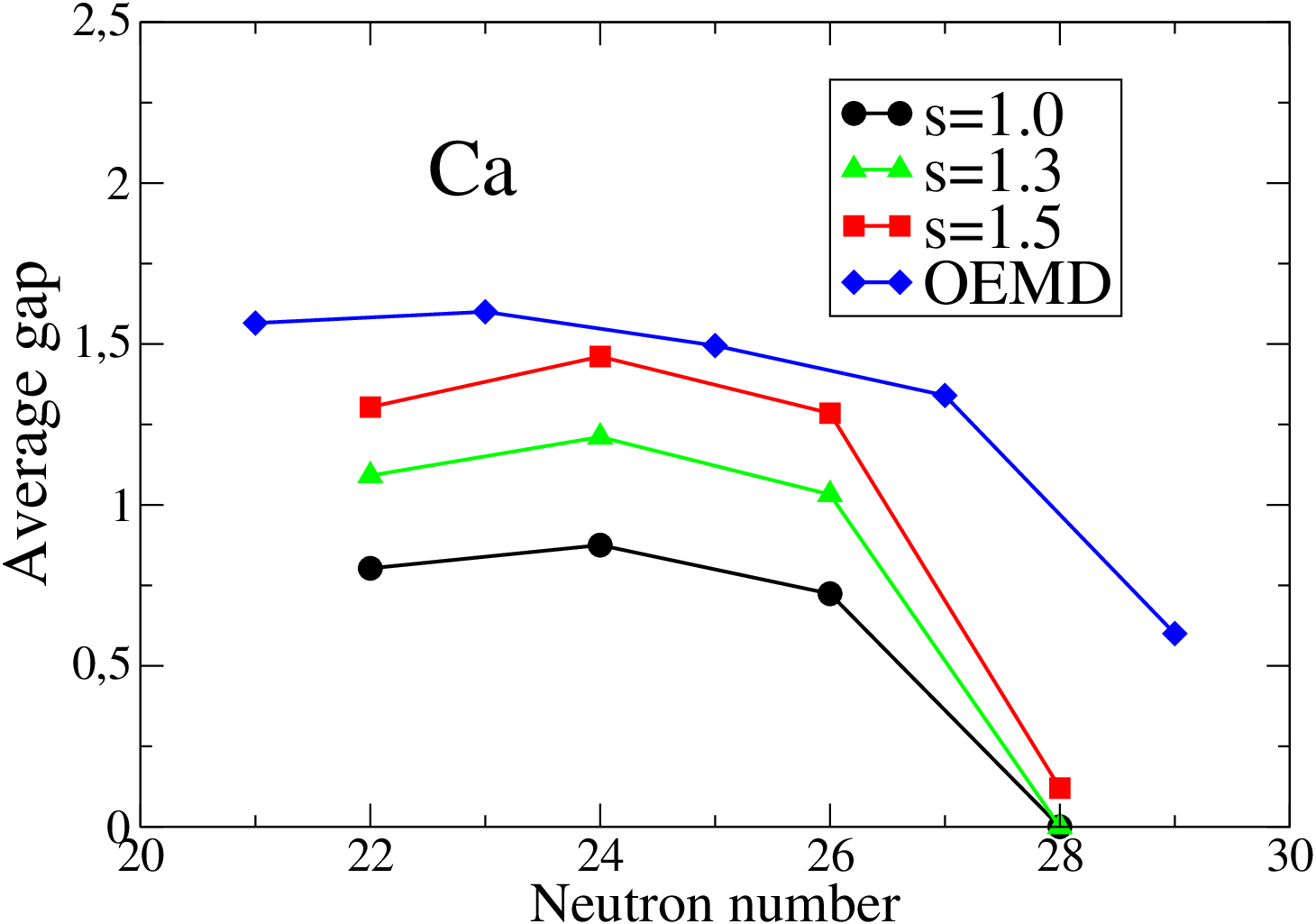}
    \end{subfigure}
    \begin{subfigure}{0.3\textwidth}
        \centering
        \includegraphics[width=\textwidth]{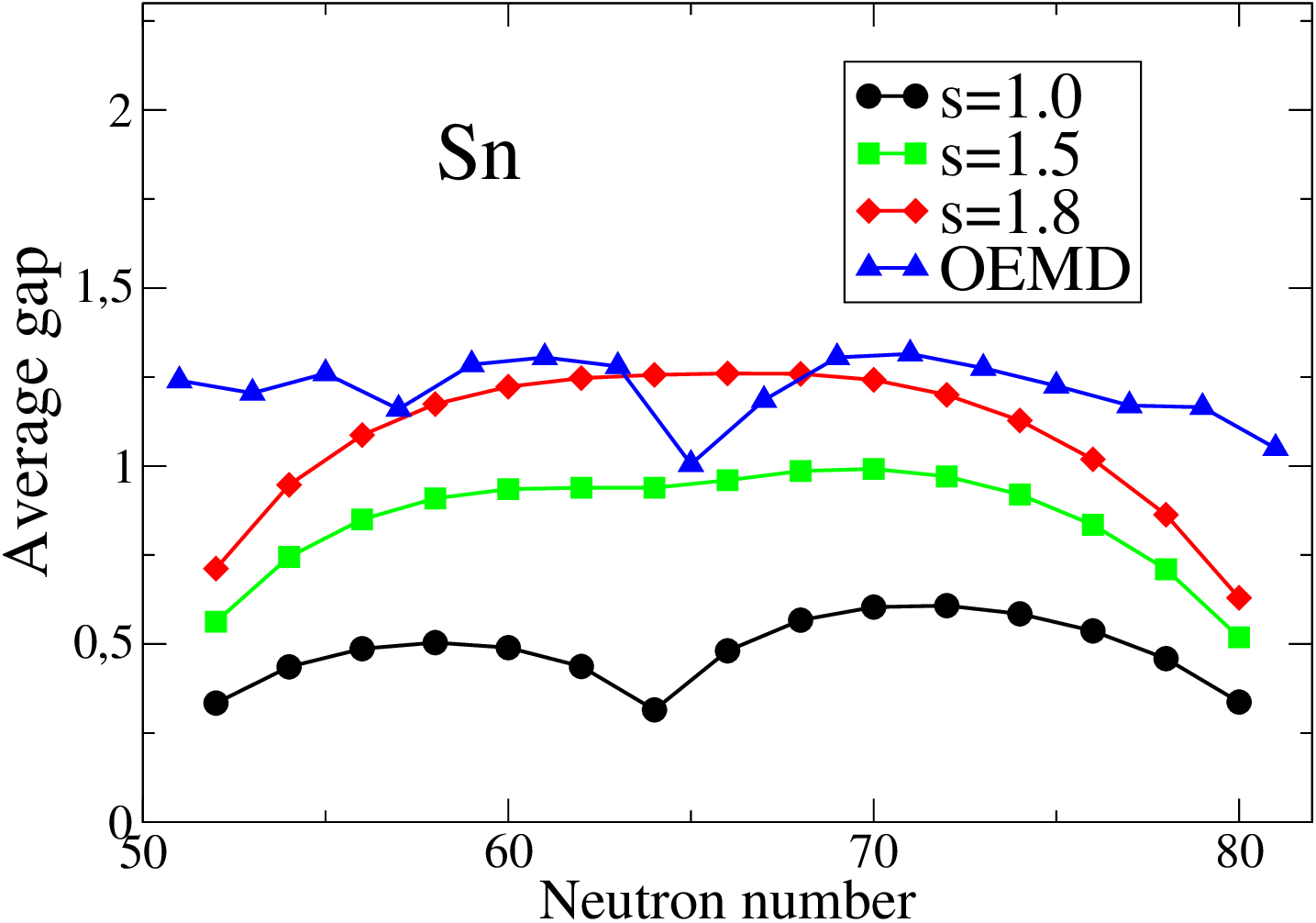}
    \end{subfigure}
    \caption{Average pairing gaps (18) for various scaling factors $s$, compared with 
     OEMD (17) for oxygen, calcium, and tin isotopes.}
    \label{fig:BE_deformation_sd}
\end{figure}

At present, there is no clear prescription for deriving the isoscalar pairing force or determining its parameters within mean-field calculations. This issue becomes even more challenging in approaches where the T=1 and T=0 pairing channels coexist, as in the QCM framework.

As mentioned in section 2.2.1, in the present calculations we assume the isoscalar pairing force to be proportional to the isovector one, as indicated in Eq. (14). The proportionality factor 
$w$ is treated as a parameter. Guidance on its value comes from HFB+QRPA calculations of
Gamow–Teller states, in which only the T=0 pairing channel contributes. In this work, we adopt 
$w$=1.6, which lies within the range of values for which HFB+QRPA calculations 
 provide a reasonable description of the Gamow–Teller strength in $^{56}$Ni \cite{sagawa}.

With the mean-field and pairing parameters fixed as described above, we perform self-consistent QMC+QCM calculations. The procedure is as follows. First, the QMC equations are solved iteratively until convergence is achieved. Using the single-particle states within the selected pairing window, we then compute the matrix elements of the pairing interaction, solve the QCM equations, and determine the occupation probabilities of the single-particle states.
In the subsequent iteration, these occupation probabilities are used to construct updated densities, which generate a new mean field and revised single-particle energies. This cycle is repeated until full self-consistency is reached. Finally, the pairing energies defined in Eq. (11) are evaluated and added to the binding energy obtained from the QMC mean field.

\section{Results and Discussion}

Below, we present results for the ground-state properties of $N=Z$ nuclei with mass
numbers between 16 and 120. For simplicity, the presentation is organized separately for the $sd$-shell
nuclei ( $16 < A < 40$), the $pfg$-shell nuclei ($40 < A < 100$ ) and the nuclei above $^{100}$Sn. 
We then discuss the doubly magic $N=Z$ nuclei $^{16}$O, $^{40}$Ca, $^{56}$Ni, and $^{100}$Sn.

For all nuclei, we present results obtained from QMC+QCM calculations using the isovector pairing interaction (12) and the combined isovector plus isoscalar pairing interaction (14), hereafter referred to as QCM1 and QCM, respectively. The QMC+QCM results are compared with QMC+BCS calculations, in which proton–neutron pairing is not included, as well as with QMC results, where pairing correlations are neglected altogether. In addition, we compare our results with 
Gogny-HFB calculations \cite{gogny}, which include only neutron–neutron and proton–proton pairing, and, when available, with experimental data.

\subsubsection{Nuclei with $16 < A < 40$}

We begin by discussing how the calculated binding energies compare with the experimental values. The binding-energy residuals, defined as the difference between the experimental and calculated binding energies, are shown in Fig.~4. It can be seen that both the QMC+BCS and Gogny-HFB calculations underestimate the binding energies of all 
N=Z nuclei in this mass region. This behavior is, in fact, common to all BCS and HFB calculations, including those employing Skyrme-type energy density functionals.

From Fig.~4, one can also observe that, within the BCS framework, pairing correlations contribute only in the case of $^{20}$Ne. The same behavior is found in the Gogny-HFB calculations. This implies that, for $A>20$, the underestimation of the binding energies in these approaches is entirely related to the mean-field properties of the energy density functional. As shown in Fig.~4, the QMC 
underestimates the binding energies by about 2 MeV, which is approximately a factor of two smaller than the corresponding deviation obtained with the Gogny functional. The QMC residuals are also significantly smaller than those obtained in Skyrme-HF calculations \cite{skyrme_qcm} with the UNEDF1 functional \cite{unedf1}.

\begin{figure}[h]
    \centering
    \begin{minipage}{0.45\textwidth}
        \centering
        \includegraphics[width=\textwidth]{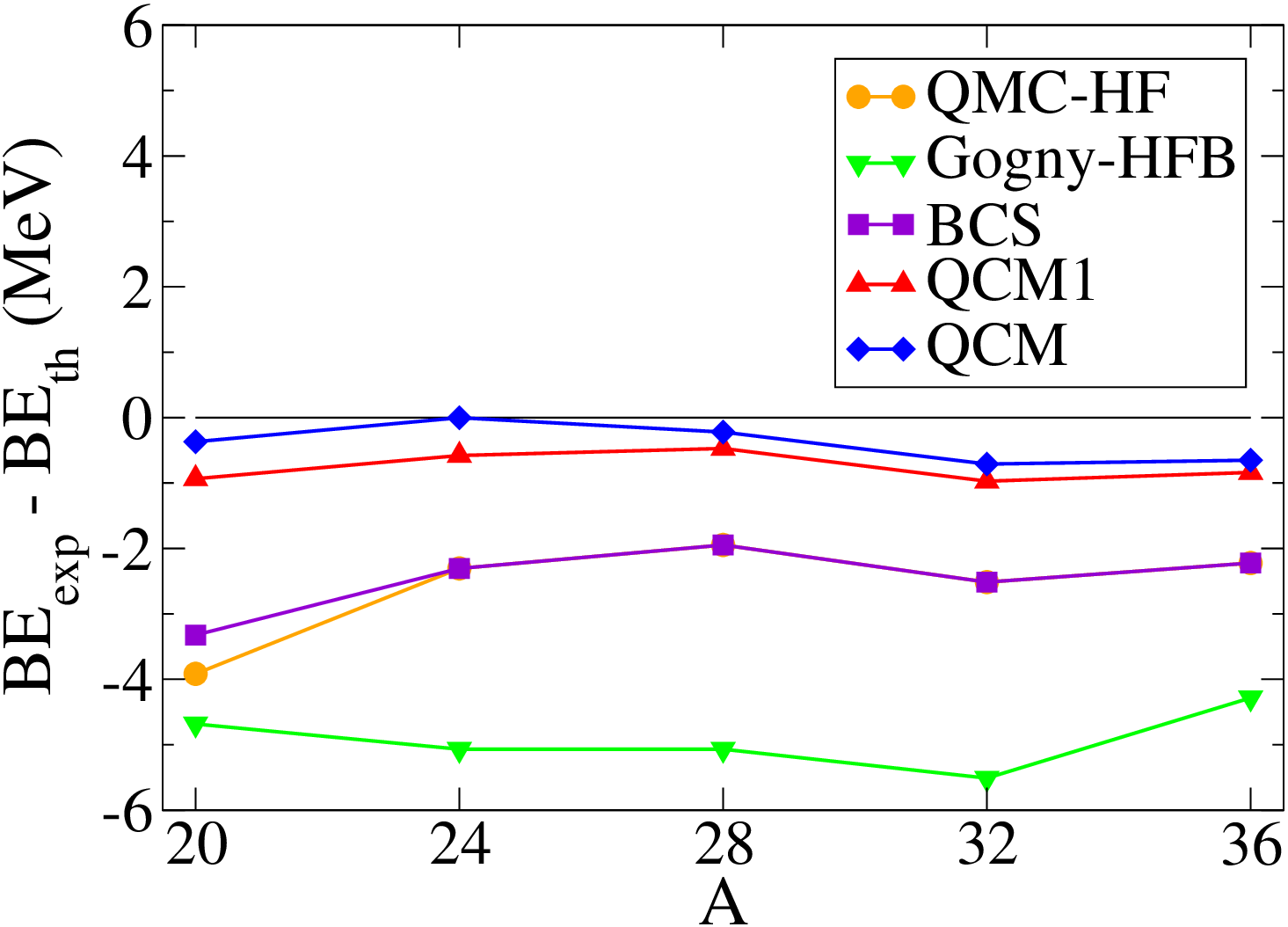}
        \caption{Binding-energy residuals of  $sd$-shell nuclei obtained within various 
          approximations indicated in the figure.}
        \label{fig:BE_sd}
    \end{minipage}
    \hfill
    \begin{minipage}{0.45\textwidth}
        \centering
        \includegraphics[width=\textwidth]{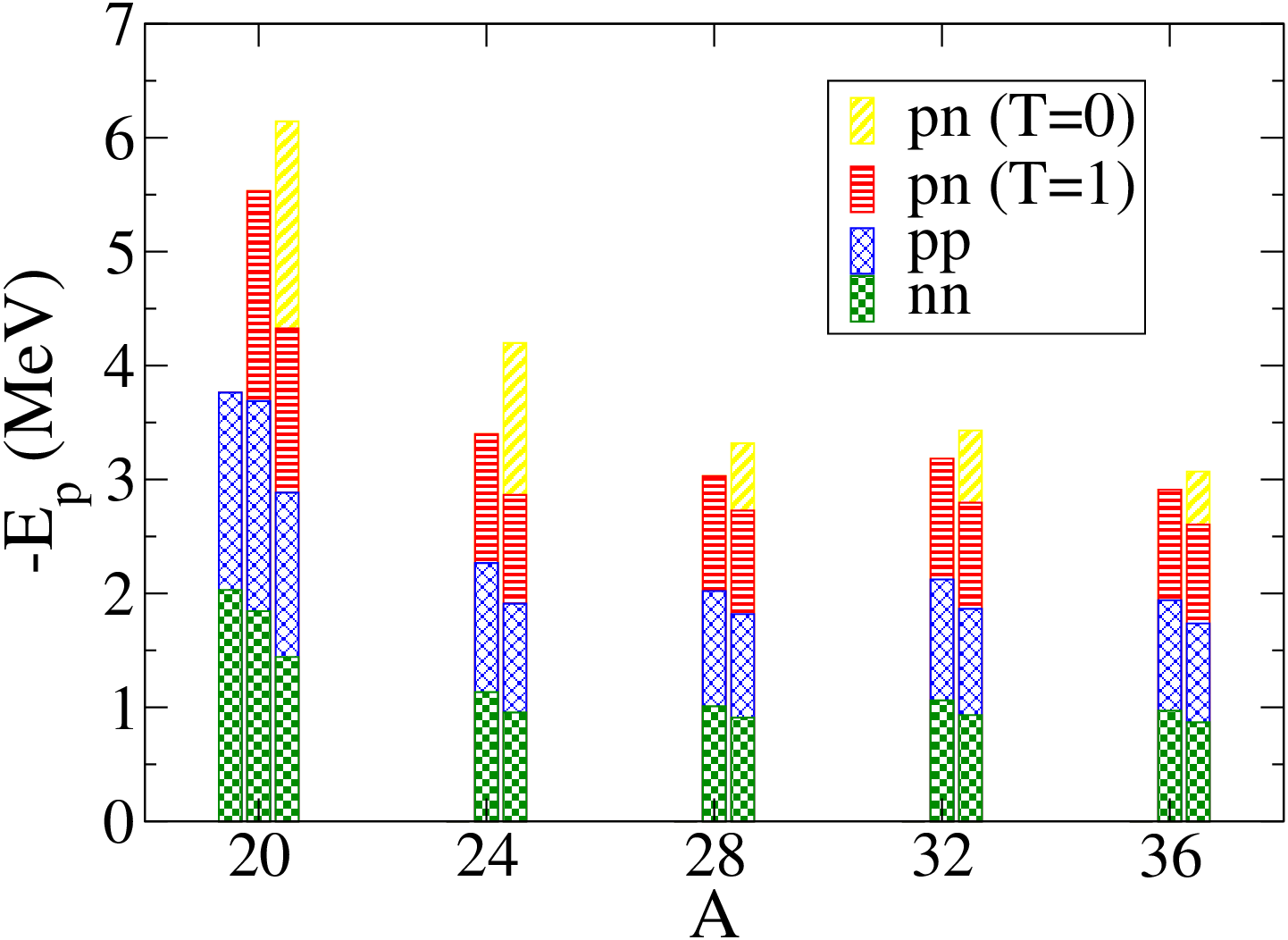}
        \caption{Pairing energies for the sd-shell nuclei. For each nucleus, the results are displayed (from right to left) for QCM, QCM1 and, when available, BCS.}
        \label{fig:PE_sd}
    \end{minipage}
\end{figure}

The most noticeable feature in Fig.~4 is the substantial improvement in the binding energies predicted by the QMC+QCM calculations compared with the QMC+BCS results. As a consequence, the energy residuals become very small—below 100 keV for the first three nuclei and about 700 keV for the last two. From Fig.~4, it can also be seen that switching on the isoscalar pairing interaction does not significantly change the binding energies. This behavior is related to the competition between the isovector and isoscalar pairing correlations, which is discussed below.

The contributions of pairing correlations to the binding energies are shown in Fig.~5. The pairing energies are calculated at the deformation corresponding to the minimum of the total binding energy. We display the neutron–neutron ($nn$), proton–proton ($pp$), and the isovector 
proton–neutron ($pn$) pairing energies, which are identical due the isospin invariance assumed in the QCM calculations, as well as the isoscalar $pn$ pairing energies. The latter includes contributions from the three spin configurations with total spin projections $S_z$=-1,0,1.  

From Fig. 5, one can see  that the isoscalar $pn$ pairing has its largest contribution in $^{20}$Ne and $^{24}$Mg, which is consistent with the binding-energy residuals shown in Fig.~4. For all nuclei, the variation of the total pairing energy induced by $pn$ pairing follows a similar trend. We discuss in more detail the case of $^{20}$Ne, for which a BCS solution also exists. The total BCS pairing energy for the $nn$ and $pp$ channels, shown in the left column of Fig.~5, is about 3.8 MeV, which is close to the Gogny-HFB pairing energy of approximately 4 MeV. When the isovector $pn$ pairing is included, the total pairing energy increases to about 5.5 MeV, with all T=1 channels contributing equally. This explains the significant reduction of the binding-energy residual observed in Fig.~4.

When the isoscalar interaction is included, the total pairing energy increases further, but only by about 600 keV, despite the fact that the isoscalar $pn$ pairing energy itself is large, approximately 1.8 MeV. This behavior is due to a reduction of the isovector pairing energy by about 1.2 MeV. Such a pattern reflects the competition between isovector and isoscalar pairing correlations, which draw coherence from the same model space. 

\begin{figure}[htbp]
    \centering
    \begin{subfigure}{0.3\textwidth}
        \centering
        \includegraphics[width=\textwidth]{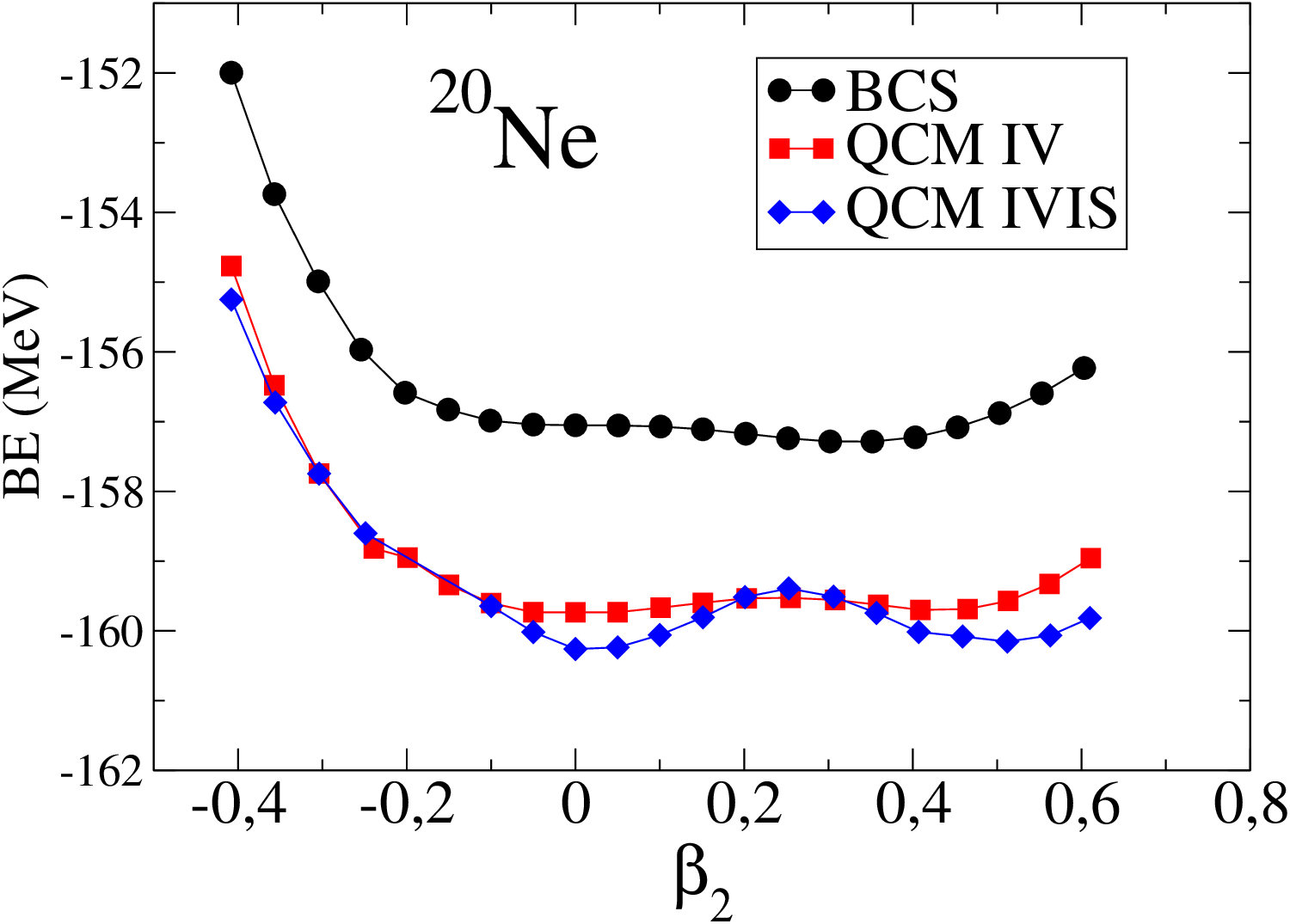}
    \end{subfigure}
    \begin{subfigure}{0.3\textwidth}
        \centering
        \includegraphics[width=\textwidth]{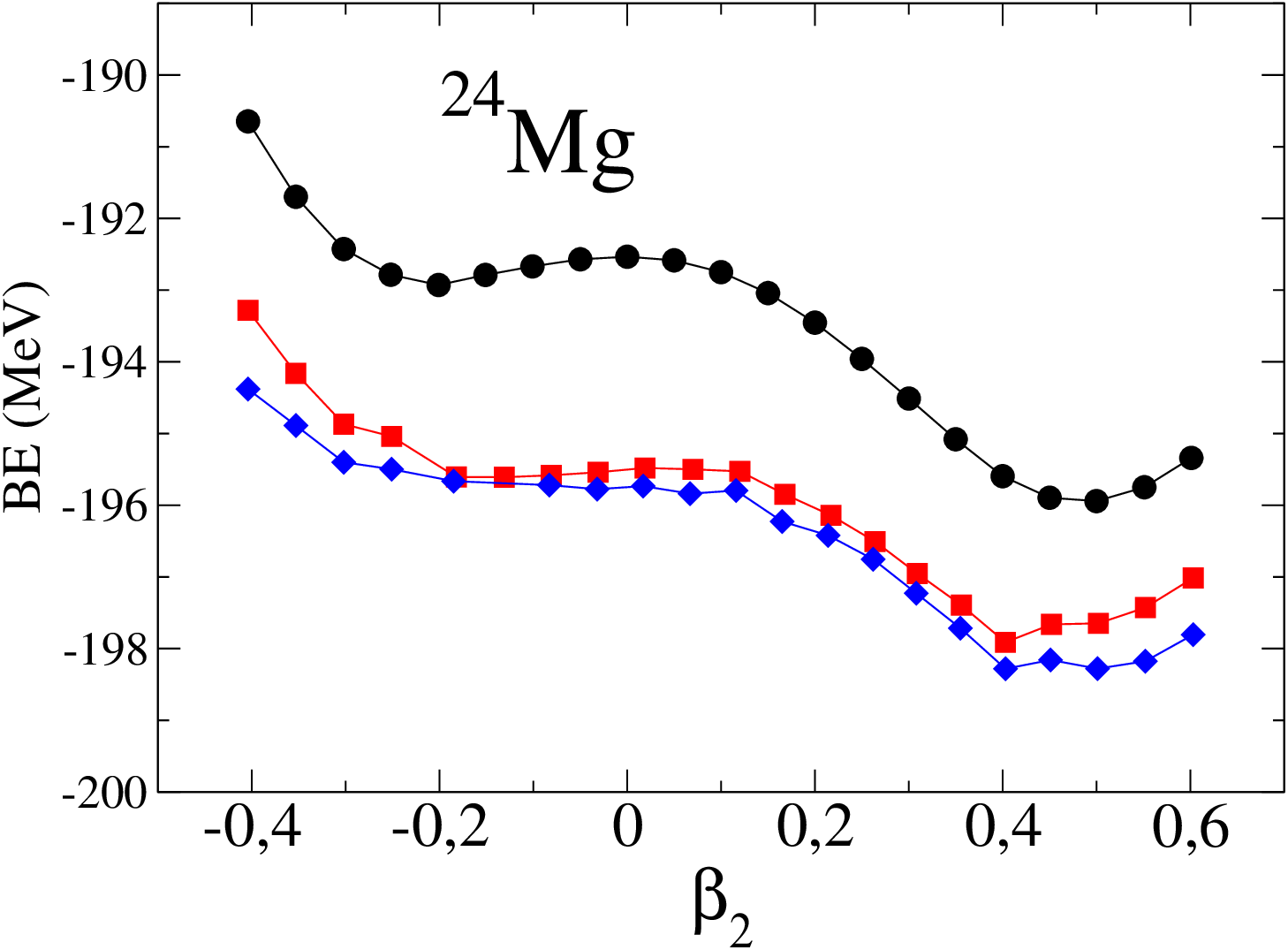}
    \end{subfigure}
    \begin{subfigure}{0.3\textwidth}
        \centering
        \includegraphics[width=\textwidth]{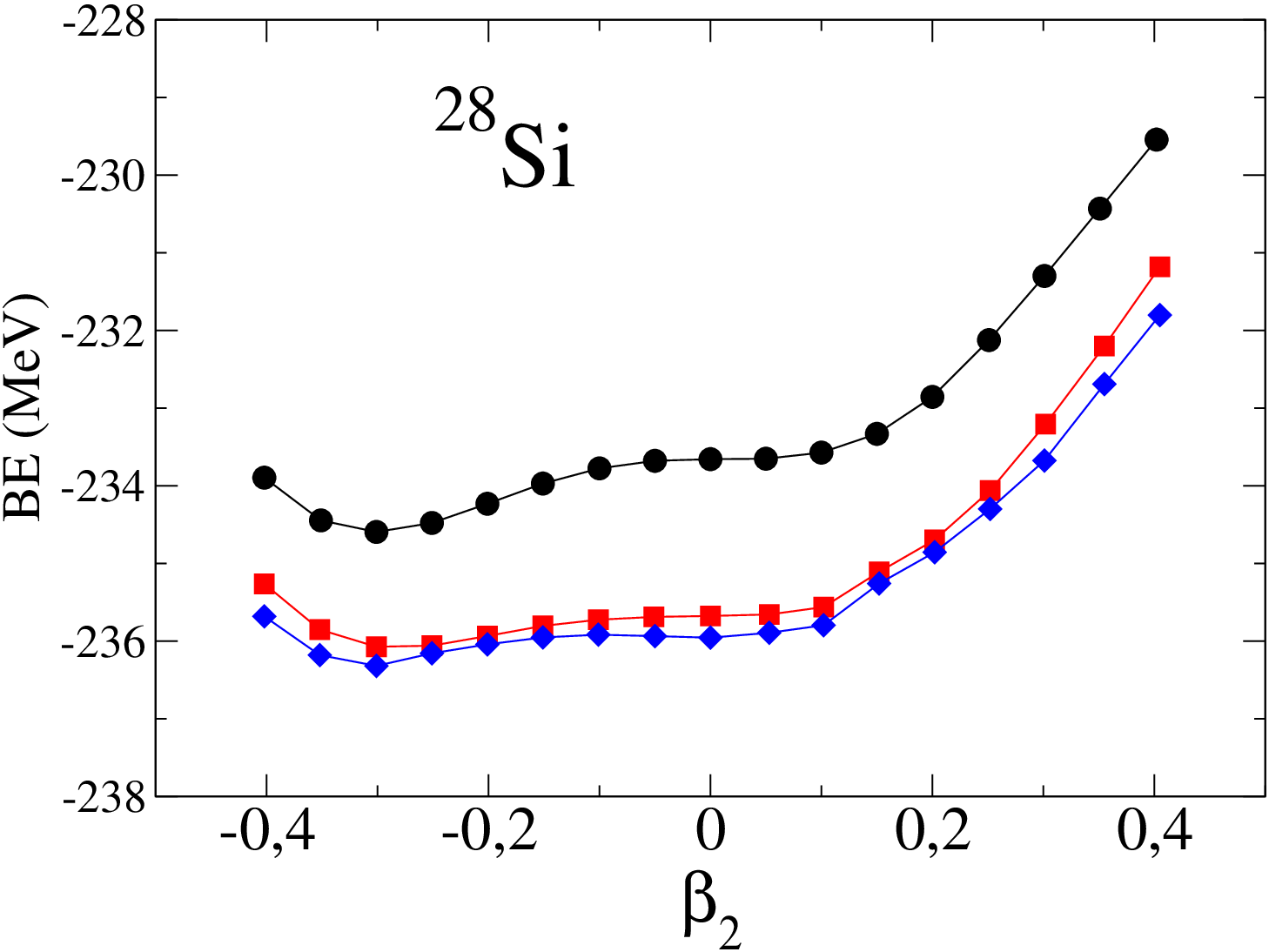}
    \end{subfigure}

    \begin{subfigure}{0.3\textwidth}
        \centering
        \includegraphics[width=\textwidth]{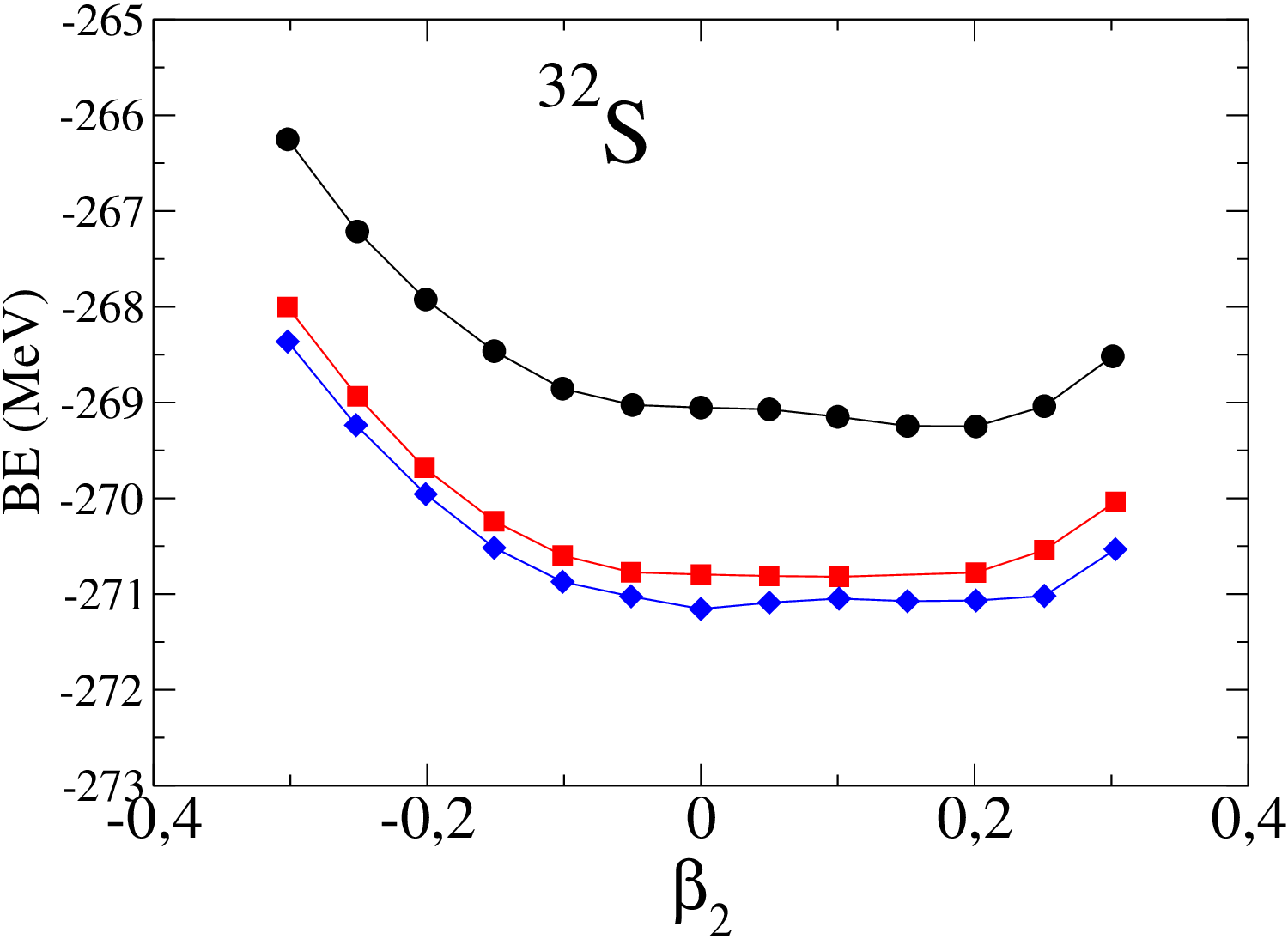}
    \end{subfigure}
    \begin{subfigure}{0.3\textwidth}
        \centering
        \includegraphics[width=\textwidth]{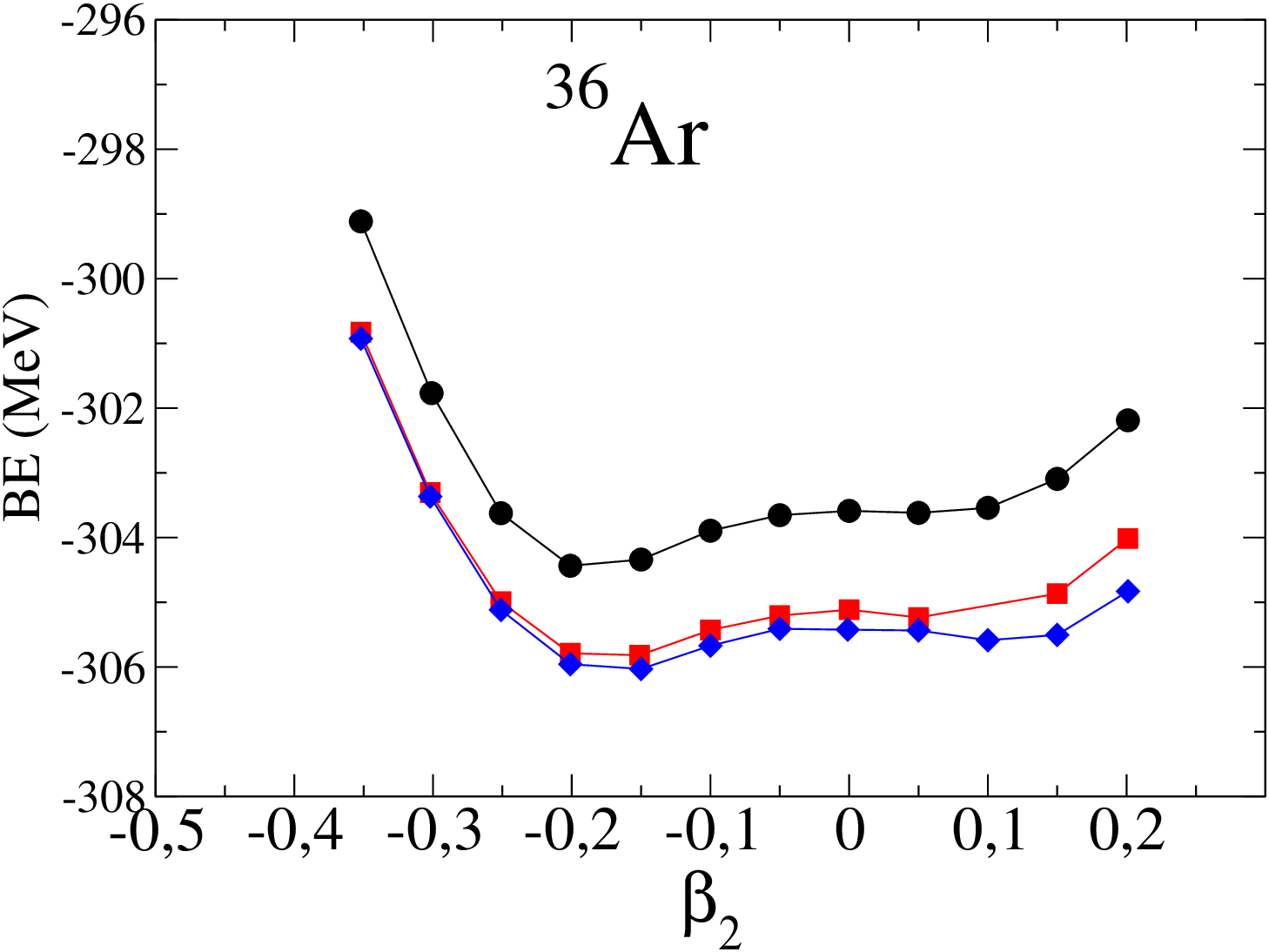}
    \end{subfigure}
    \begin{subfigure}{0.3\textwidth}
        \centering
        \includegraphics[width=\textwidth]{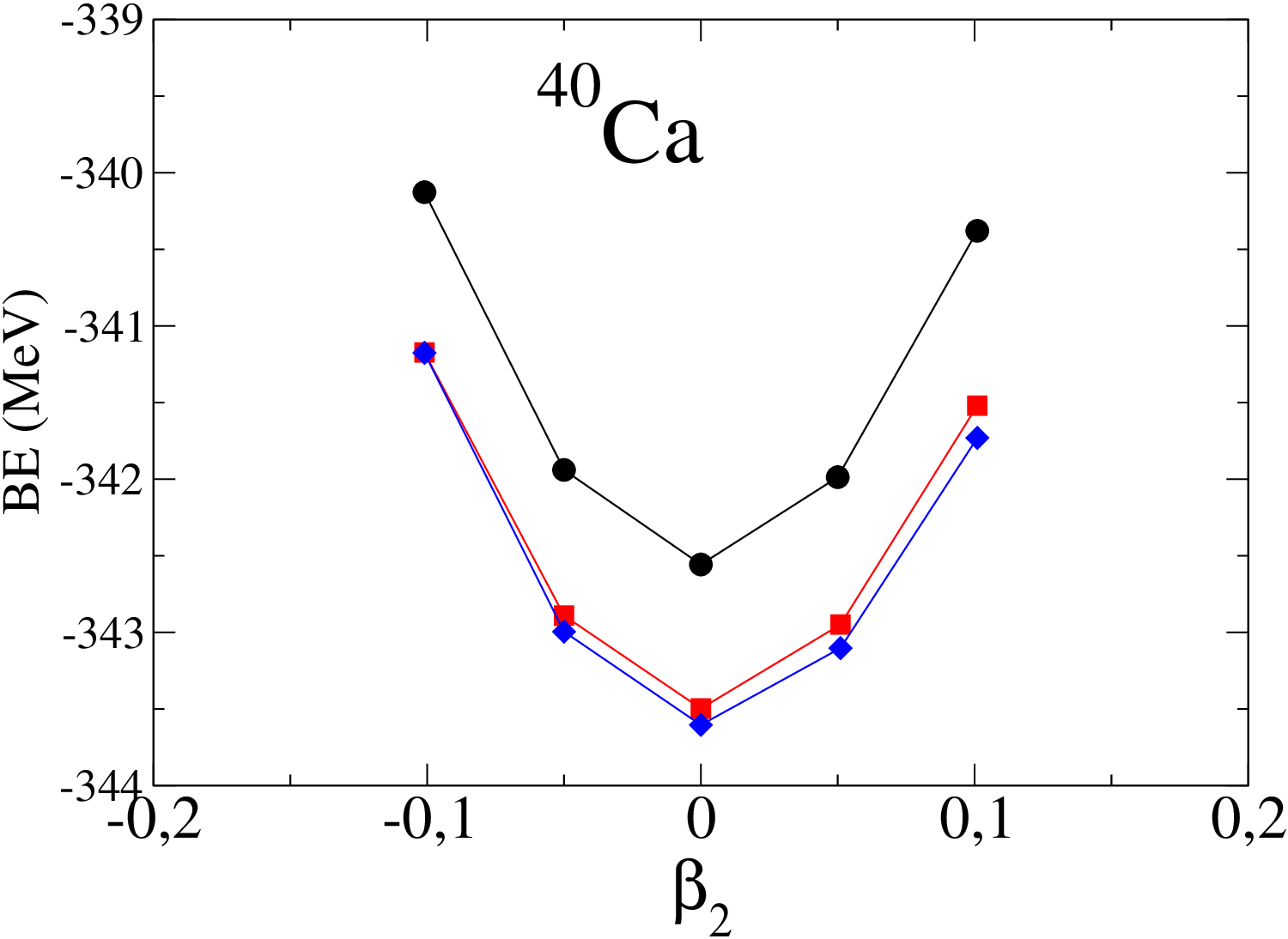}
    \end{subfigure}

\caption{Binding energies as a function of quadrupole deformation for
the $sd$-shell nuclei.}
    \label{fig:BE_deformation_sd}
\end{figure}

Figure 6 illustrates the dependence of the binding energy on quadrupole deformation. 
The following general features can be observed:
(a) the inclusion of isovector pairing uniformly lowers the binding energies for all deformations by a comparable amount relative to the BCS results;
(b) except for the cases discussed below, the isoscalar pairing has only a minor effect on the deformation dependence of the binding energy and does not significantly alter the position of the energy minima.

The most pronounced effect of the isoscalar interaction is found in 
$^{20}$Ne. When the isoscalar interaction is switched on, two nearly degenerate local minima appear: one at approximately zero deformation and another around $\beta_2 \approx$ 0.47. The latter is slightly more bound, by a few keV. This minimum, which is considered the ground state of $^{20}$Ne, is expected to become more bound after angular-momentum restoration, which is broken in the present calculations.

The isoscalar interaction also affects the binding energies of 
$^{24}$Mg at large deformations, on both the prolate and oblate sides. On the prolate side, the isoscalar interaction shifts the isovector minimum toward larger deformation. Significant effects of the isoscalar interaction at large deformation are also observed for 
$^{32}$S and $^{36}$Ar. For $^{36}$Ar the isocalar pairing generates
a local prolate minimum at $\beta_2 \approx $ 0.1. A peculiar situation is observed in $^{32}$S, where the binding energy remains rather flat for $\beta_2$ values between 0.0 and 0.25, making it difficult to draw reliable conclusions about the shape and shape coexistence of this nucleus.

Table 1 presents the nuclear deformations at the energy minima obtained from unconstrained QMC+QCM calculations, compared with FRDM \cite{frdm} and experimental values \cite{nudat}. The comparison with experiment should be regarded as indicative only, since for some nuclei, such as 
$^{32}$S, the binding-energy curves are rather flat in the vicinity of the minimum. As a consequence, for these nuclei the position of the minimum is expected to vary significantly with the parameters of the pairing interaction. 
 
\begin{table}[htbp]
    \centering

    \begin{minipage}{0.48\textwidth}
        \centering
        \renewcommand{\arraystretch}{1.1}
        \setlength{\tabcolsep}{5pt}
        \begin{tabular}{|c|c|c|c|}
            \hline
            \textbf{A} & \textbf{QCM}  & \textbf{FRDM} & \textbf{Exp} \\
            \hline
            20  & 0.474  & 0.36  & 0.720 \\
            24  & 0.478  & 0.39  & 0.606 \\
            28  & -0.308 & -0.36 & 0.412 \\
            32  & 0.193  & 0.22  & 0.314 \\
            36  & -0.18  & -0.26 & 0.353 \\
            \hline
        \end{tabular}
        \caption{Quadrupole deformations obtained from QCM calculations, 
        compared with FRDM \cite{frdm} and experimental values \cite{nudat}.}
        \label{tab:beta2_comparison_sd}
    \end{minipage}
    \hfill
    \begin{minipage}{0.48\textwidth}
        \centering
        \renewcommand{\arraystretch}{1.1}
        \setlength{\tabcolsep}{7pt}
        \begin{tabular}{|c|c|c|c|}
            \hline
            \textbf{A} & \textbf{BCS}  & \textbf{QCM} & \textbf{Exp} \\
            \hline
	      20 &  2.816 & 2.868 & 3.0055 \\
            24 &  2.967 & 2.972 & 3.0570 \\
            28 &  3.037 & 3.042 & 3.1224 \\
            32 &  3.138 & 3.145 & 3.2611 \\
            36 &  3.277 & 3.278 & 3.3905 \\
            \hline
        \end{tabular}
        \caption{Charge radii obtained from BCS and QCM calculations, compared with  
         experimental values \cite{iaea_radii}.}
        \label{tab:cradii}
    \end{minipage}

\end{table}

Table 2 presents the root-mean-square (rms) charge radii obtained from the QCM and BCS calculations, compared with the experimental values. With the exception of $^{36}$Ar, we observe a slight improvement in the agreement with the experiment when using QCM instead of BCS. This difference has two main sources: (a) the QCM and BCS solutions correspond to different equilibrium deformations; and (b) the occupation probabilities of the single-particle states within the pairing window, which contribute to the charge radii through the charge density, differ between the two approaches.

The effect of pairing on charge radii is nevertheless expected to be small, since pairing modifies only the occupation probabilities of the states within the pairing window, whose contribution to the total charge radius is relatively minor compared to that of all other states.

\section{Nuclei with $40 < A < 100$}

The binding-energy residuals and pairing energies for the 
$pfg$-shell nuclei are shown in Figs.~7 and 8. One first notices that the QMC+BCS calculations yield binding energies that are larger than those obtained with the Gogny-HFB approach by about 2–4 MeV for nuclei up to $^{80}$Zr. This occurs despite the fact that the HFB pairing energies are significantly larger than the corresponding BCS values. For example, in $^{44}$Ti and $^{64}$Ge, the Gogny-HFB pairing energies are approximately 10 MeV and 8 MeV, respectively, whereas the BCS pairing energies are about 4.7 MeV and 1.7 MeV.

\begin{figure}[h]
    \centering
    \begin{minipage}{0.45\textwidth}
        \centering
        \includegraphics[width=\textwidth]{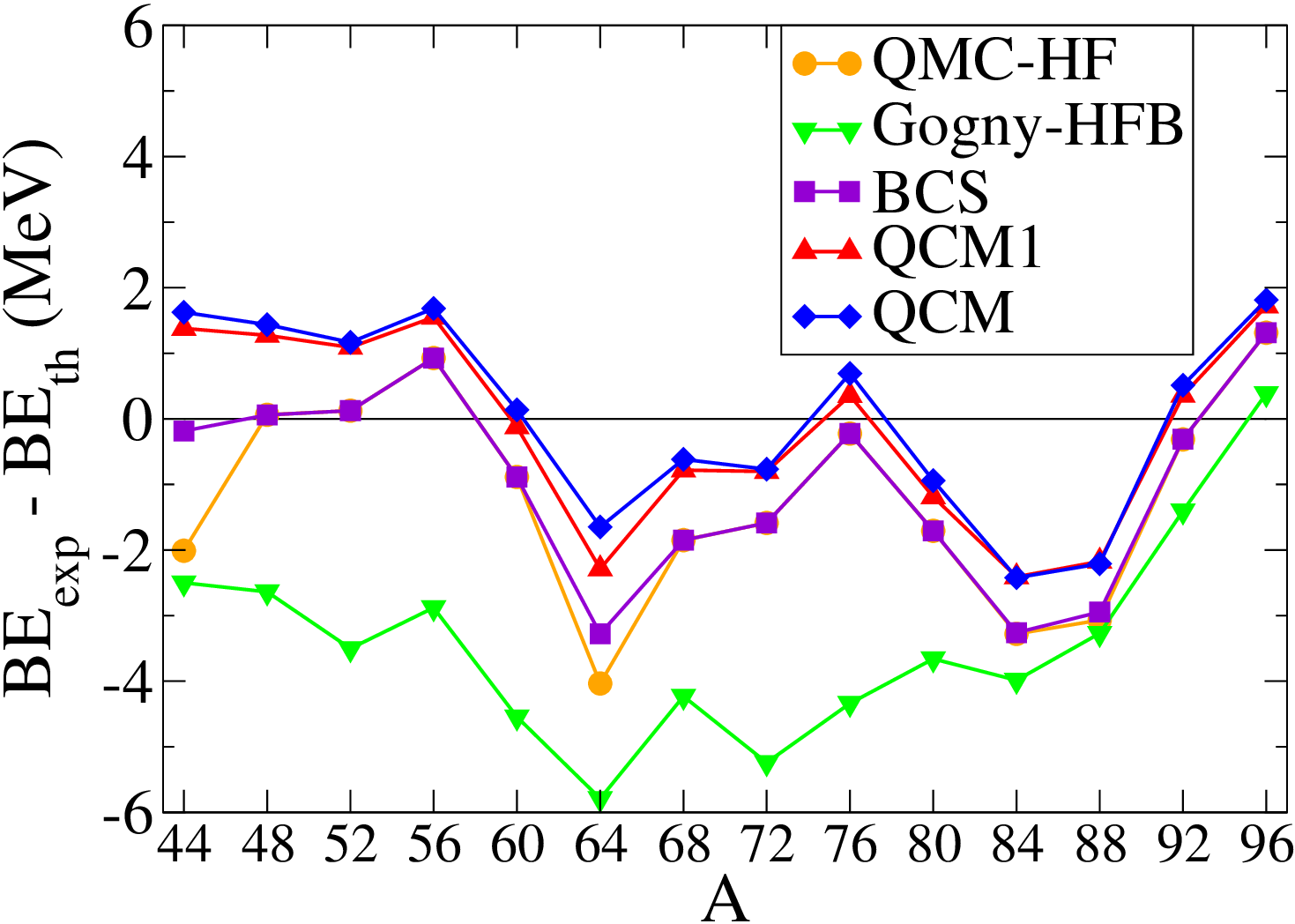}
        \caption{Binding-energy residuals for $pfg$-shell nuclei in
various approximations indicated in the figure.}
        \label{fig:BE_pf}
    \end{minipage}
    \hfill
    \begin{minipage}{0.45\textwidth}
        \centering
        \includegraphics[width=\textwidth]{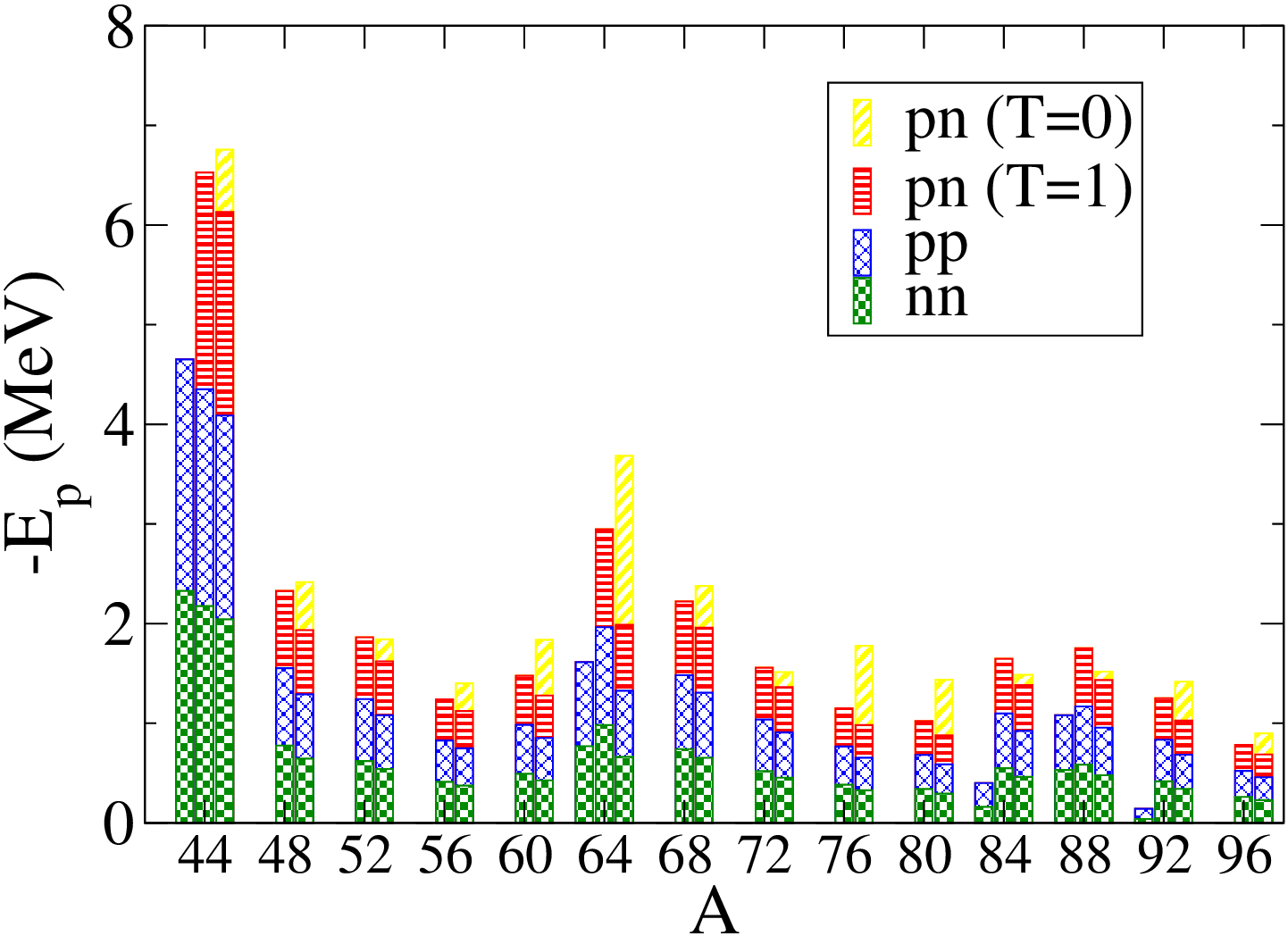}
        \caption{Pairing energies for $pfg$-shell nuclei. For each nucleus, the results are shown
(from right to left) for QCM, QCM1 and, when available, BCS.}
        \label{fig:PE_pf}
    \end{minipage}
\end{figure}

Moreover, for nuclei with $A>$60, the Gogny-HFB pairing energies remain very large, ranging from 6 to 15 MeV, while the BCS pairing energies are zero or negligible. This indicates that the differences in binding energies between the QMC+BCS and Gogny-HFB results arise primarily from the mean-field contributions, with the QMC mean field providing substantially more binding than the Gogny-HFB mean field.

From  Fig.~7 it can be seen that the binding-energy residuals for   
 $^{44}$Ti, $^{48}$Cr and $^{52}$Fe are close to  zero within the
QMC+BCS approximation. Since BCS pairing does not contribute  for $^{48}$Cr and $^{52}$Fe, this indicates that the binding energies of these nuclei are already well reproduced at the QMC level. 
In contrast, the doubly magic nucleus $^{56}$Ni is overbound by about 1 MeV in QMC. Because this nucleus is included in the fitting procedure of the QMC parameters, it is likely that the binding energies of the neigboring $f$-shell nuclei are also overestimated at the QMC level 
by a comparable amount. 

\begin{figure}[htbp]
    \centering
    \begin{subfigure}{0.32\textwidth}
        \centering
        \includegraphics[width=\textwidth]{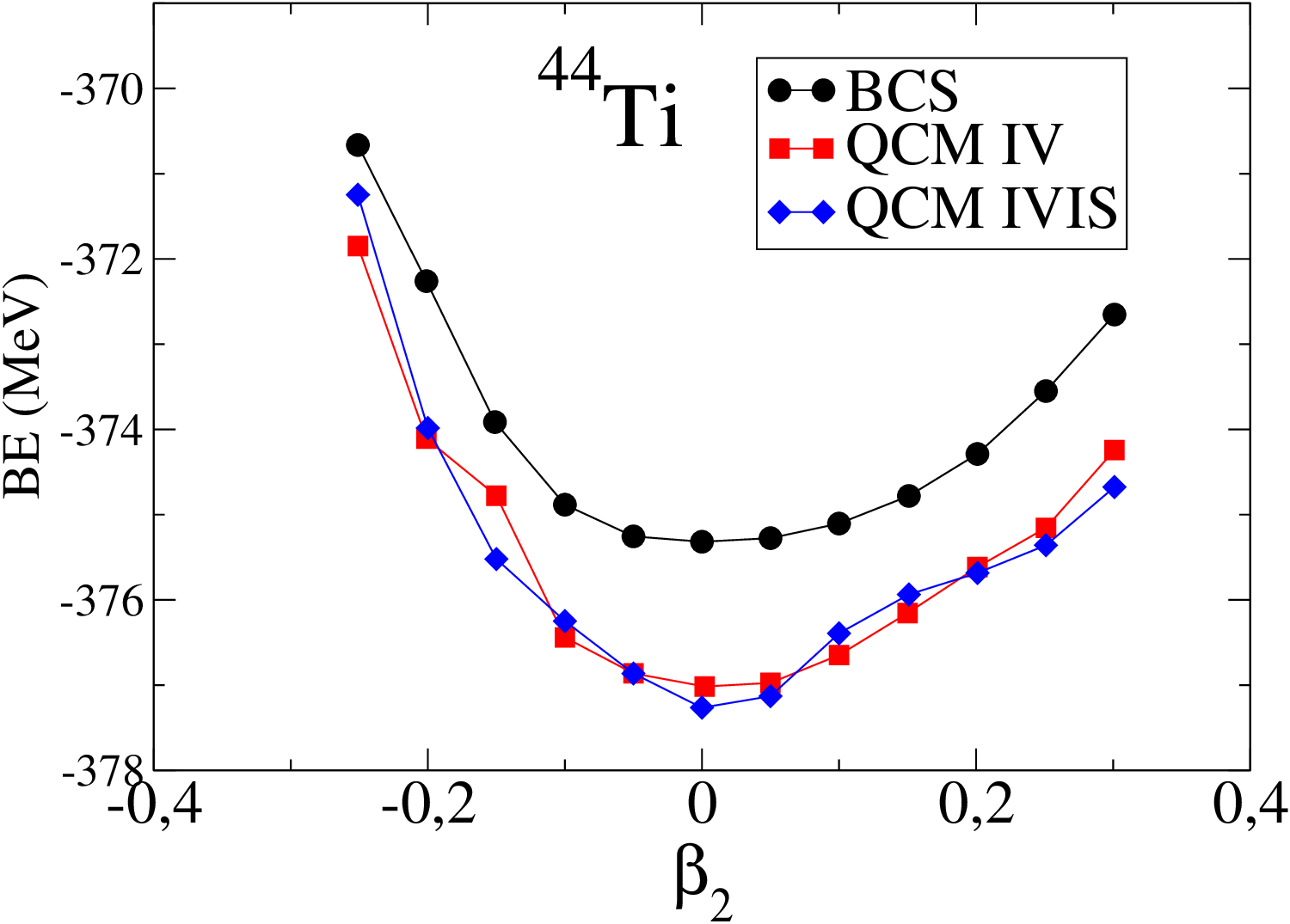}
    \end{subfigure}
    \begin{subfigure}{0.32\textwidth}
        \centering
        \includegraphics[width=\textwidth]{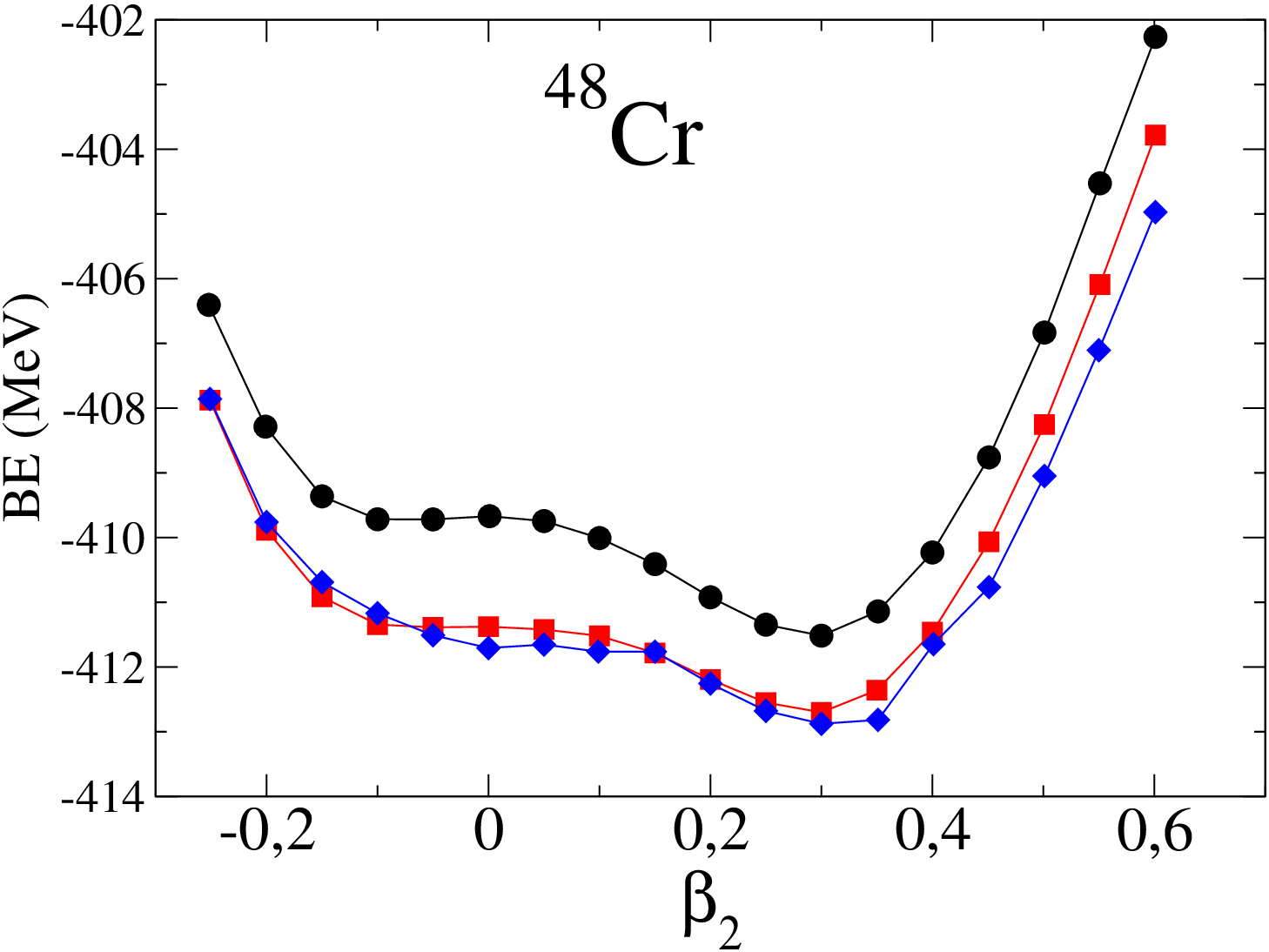}
    \end{subfigure}
    \begin{subfigure}{0.32\textwidth}
        \centering
        \includegraphics[width=\textwidth]{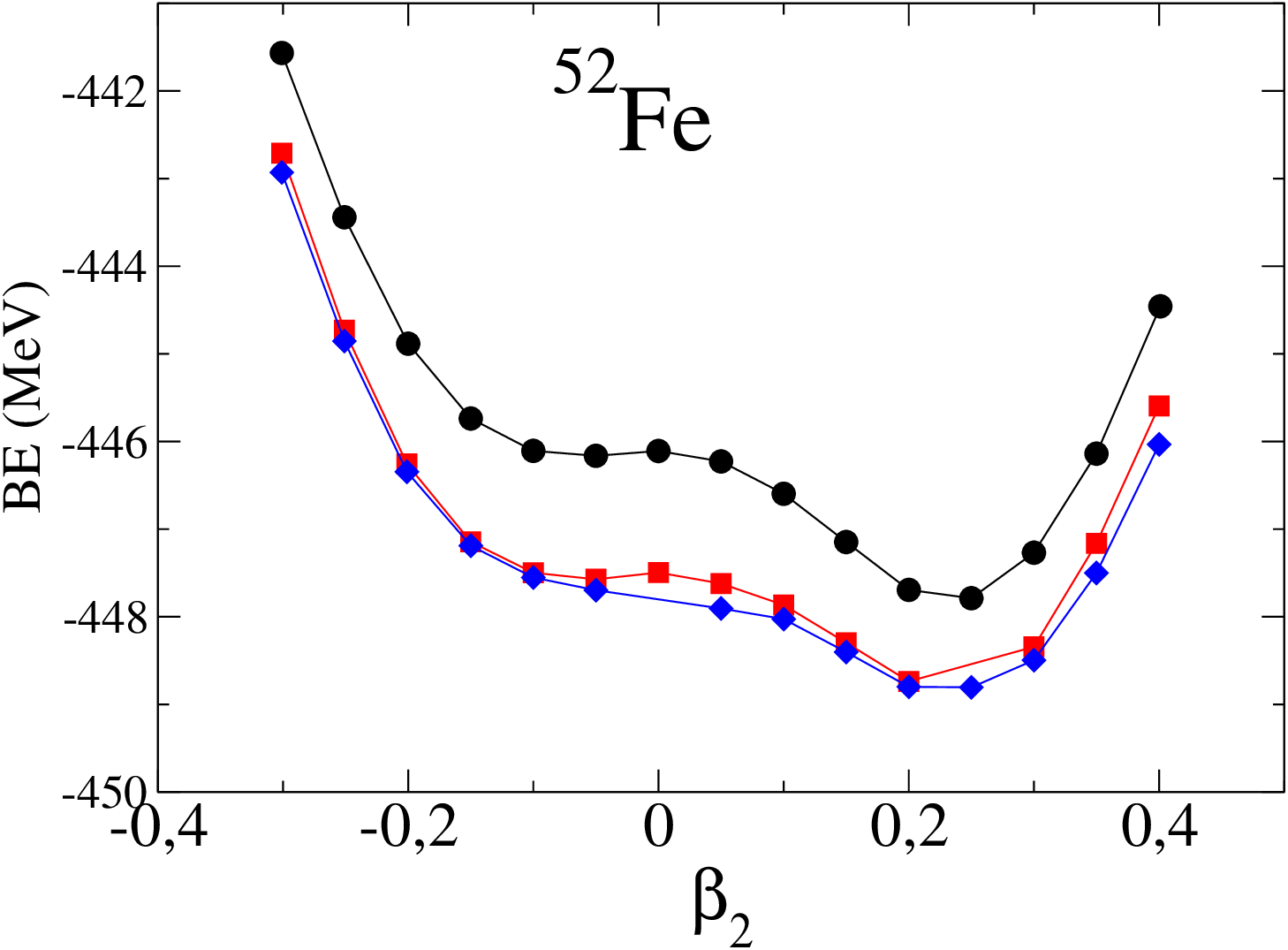}
    \end{subfigure}

    \begin{subfigure}{0.32\textwidth}
        \centering
        \includegraphics[width=\textwidth]{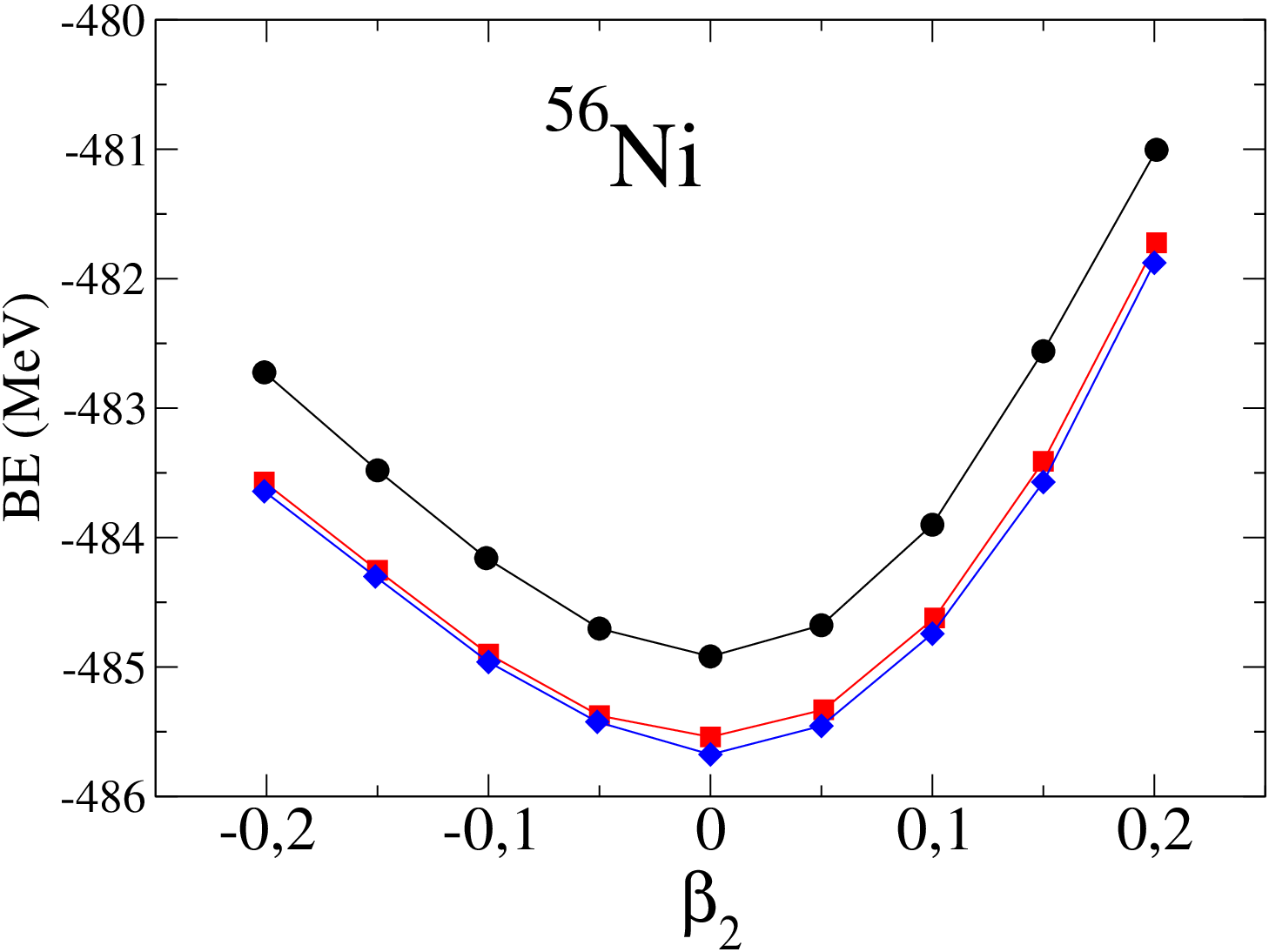}
    \end{subfigure}
    \begin{subfigure}{0.32\textwidth}
        \centering
        \includegraphics[width=\textwidth]{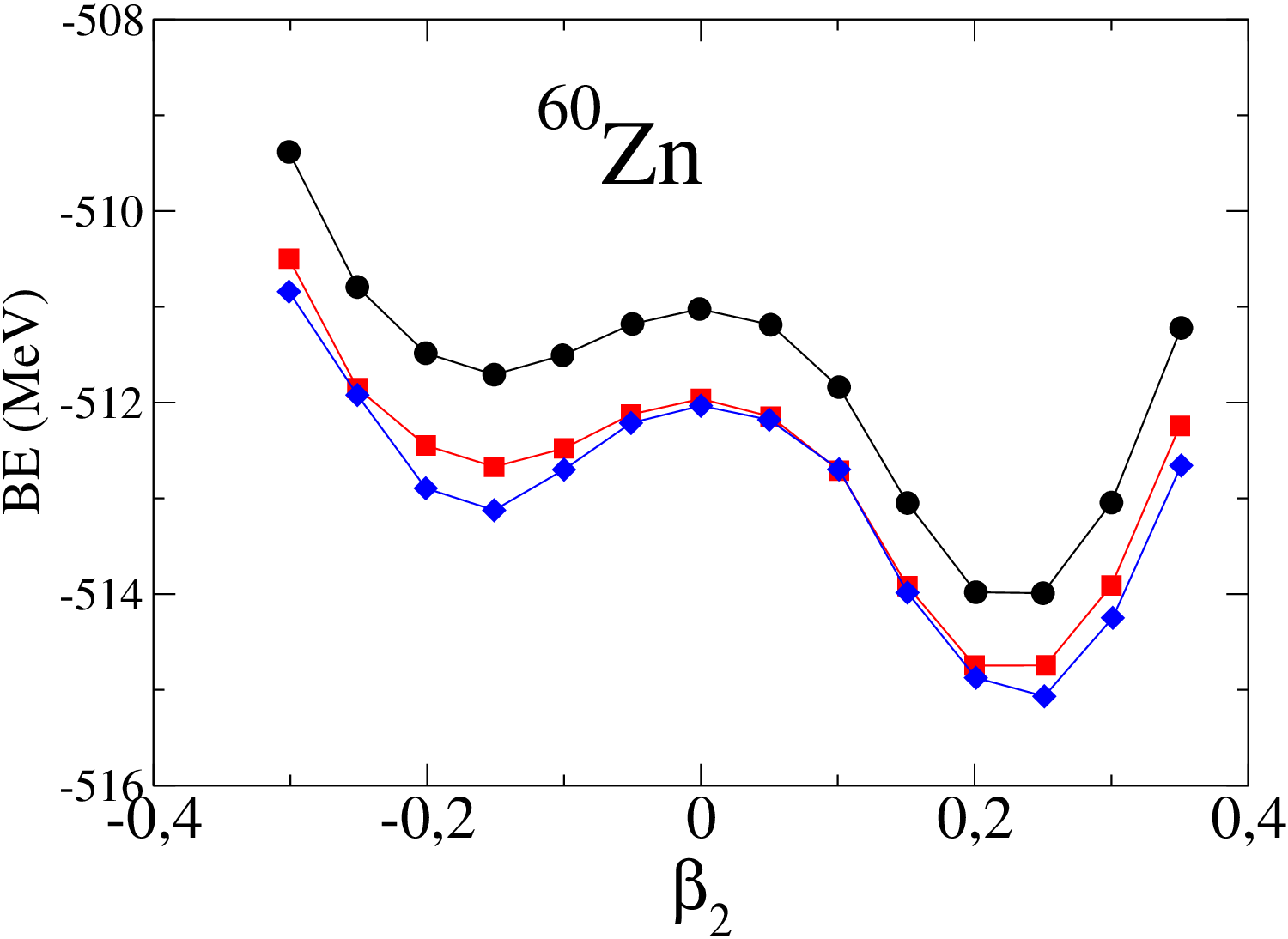}
    \end{subfigure}
    \begin{subfigure}{0.32\textwidth}
        \centering
        \includegraphics[width=\textwidth]{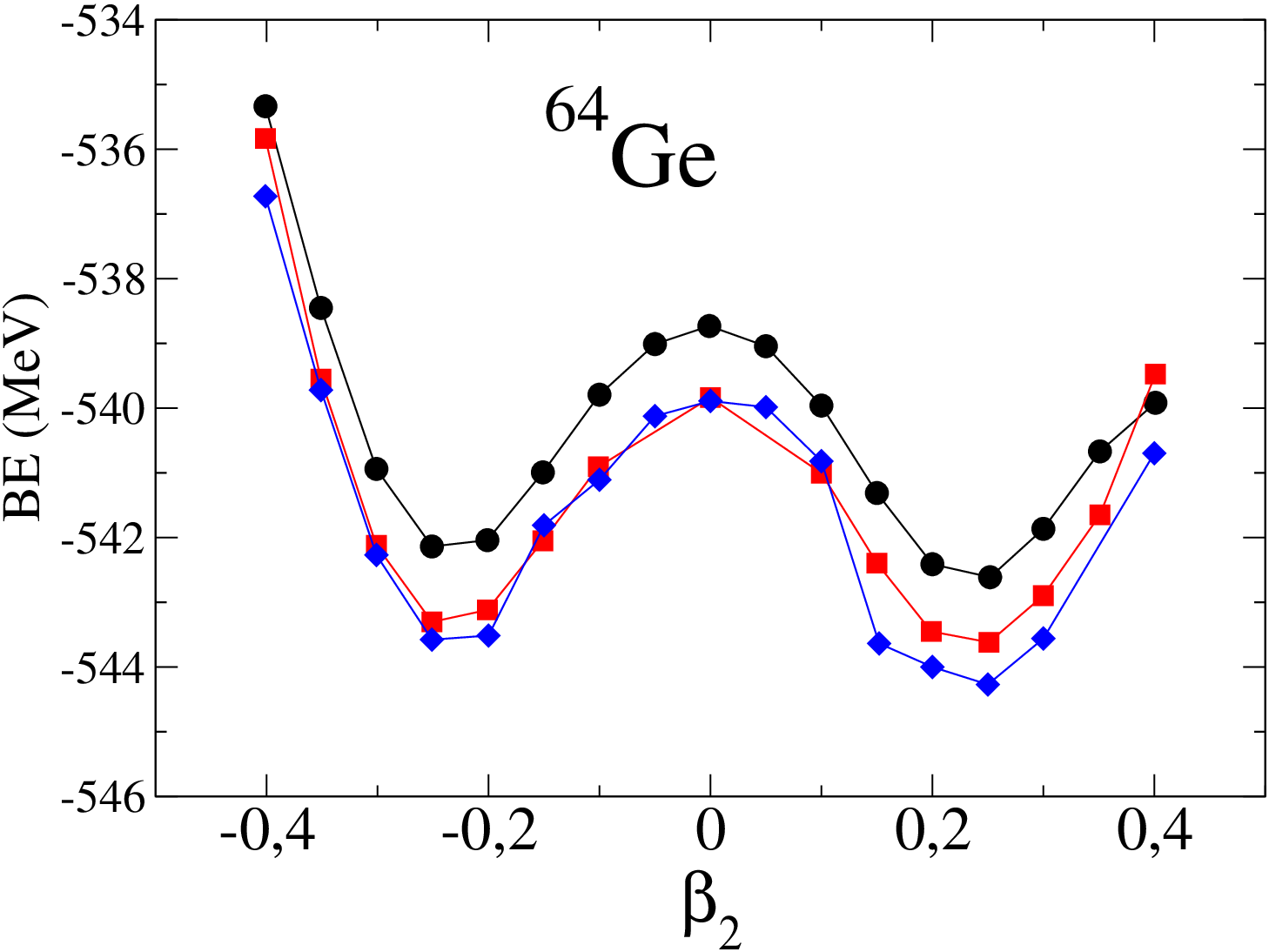}
    \end{subfigure}

    \begin{subfigure}{0.32\textwidth}
        \centering
        \includegraphics[width=\textwidth]{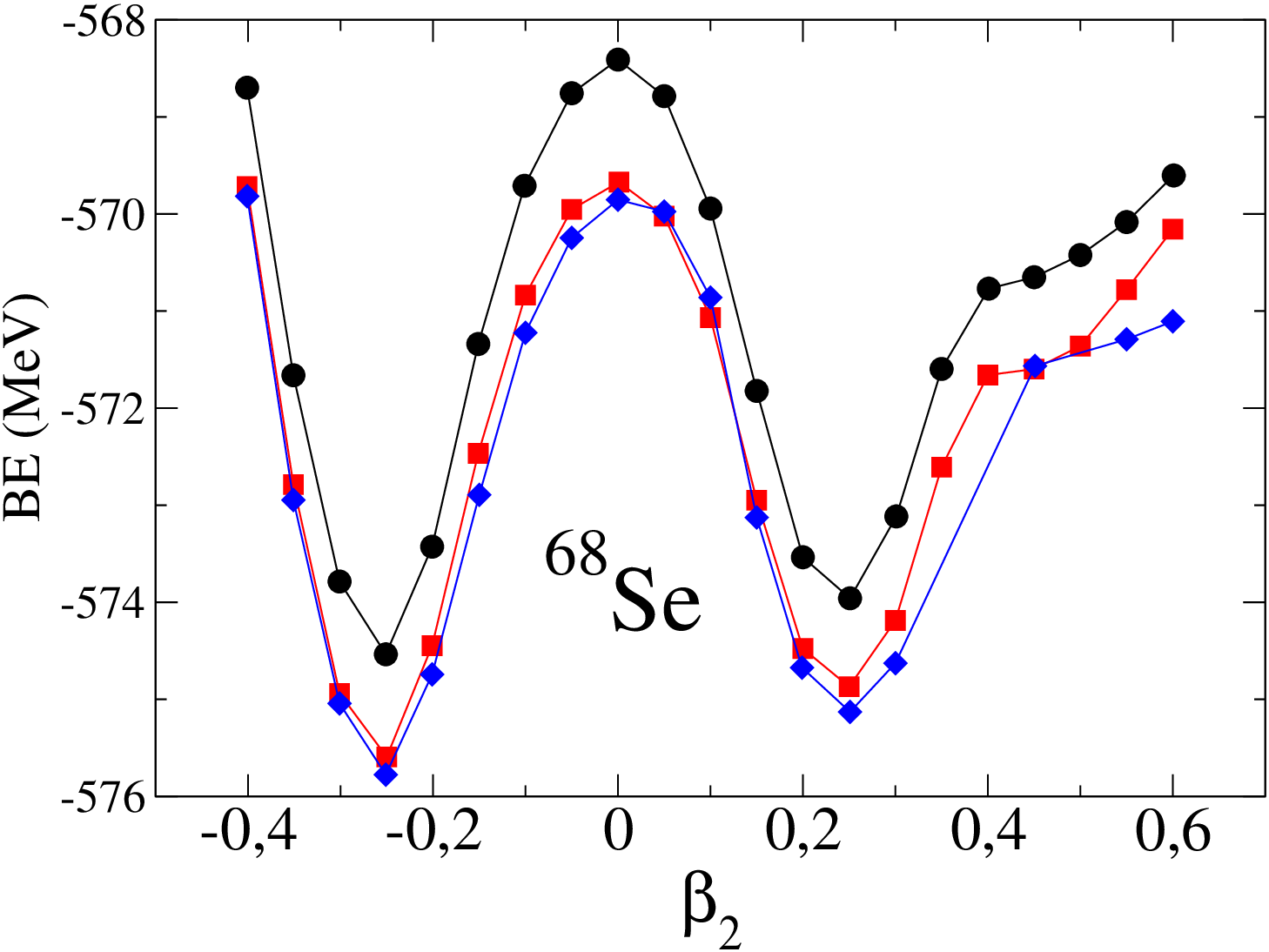}
    \end{subfigure}
    \begin{subfigure}{0.32\textwidth}
        \centering
        \includegraphics[width=\textwidth]{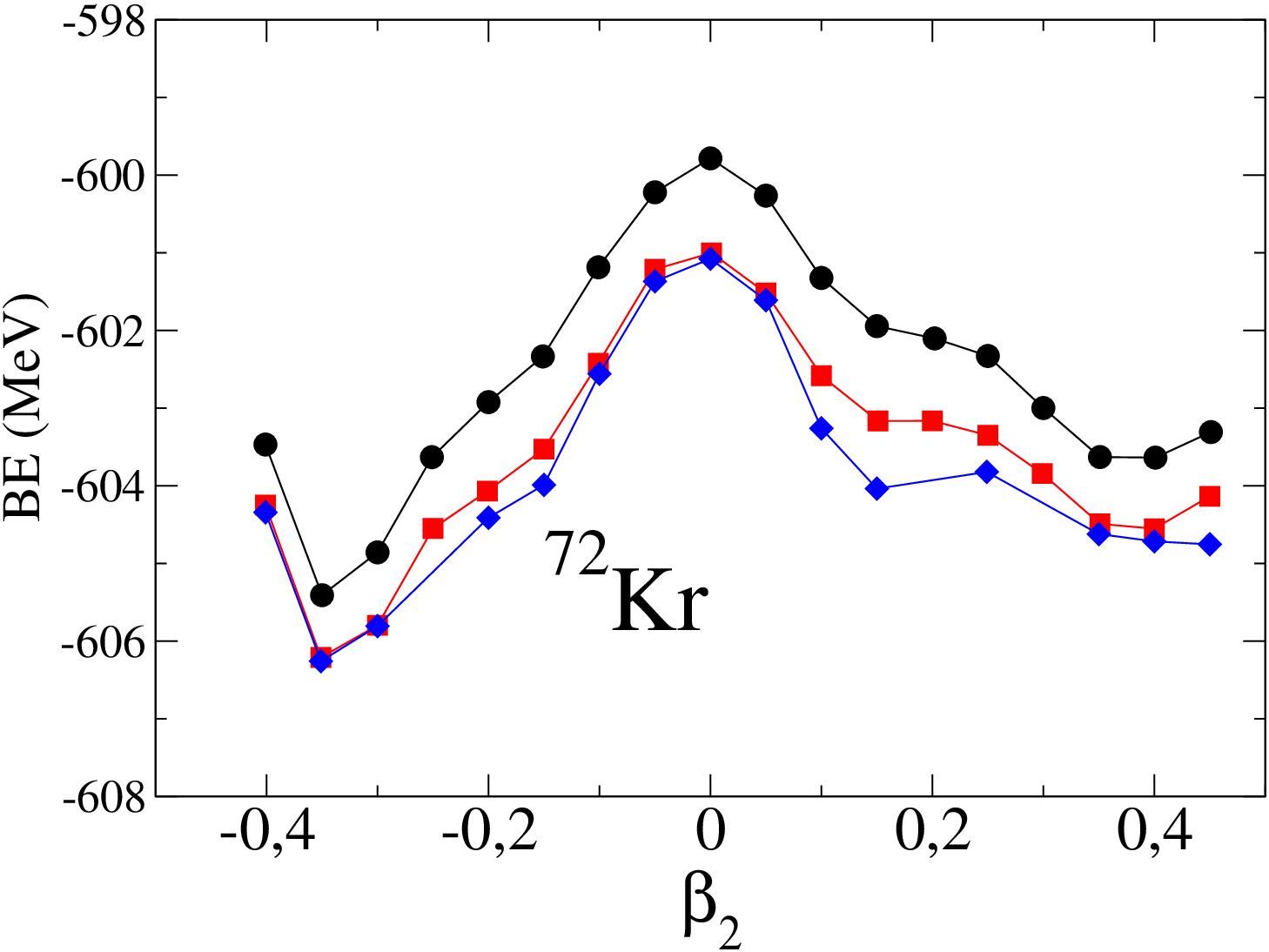}
    \end{subfigure}
    \begin{subfigure}{0.32\textwidth}
        \centering
        \includegraphics[width=\textwidth]{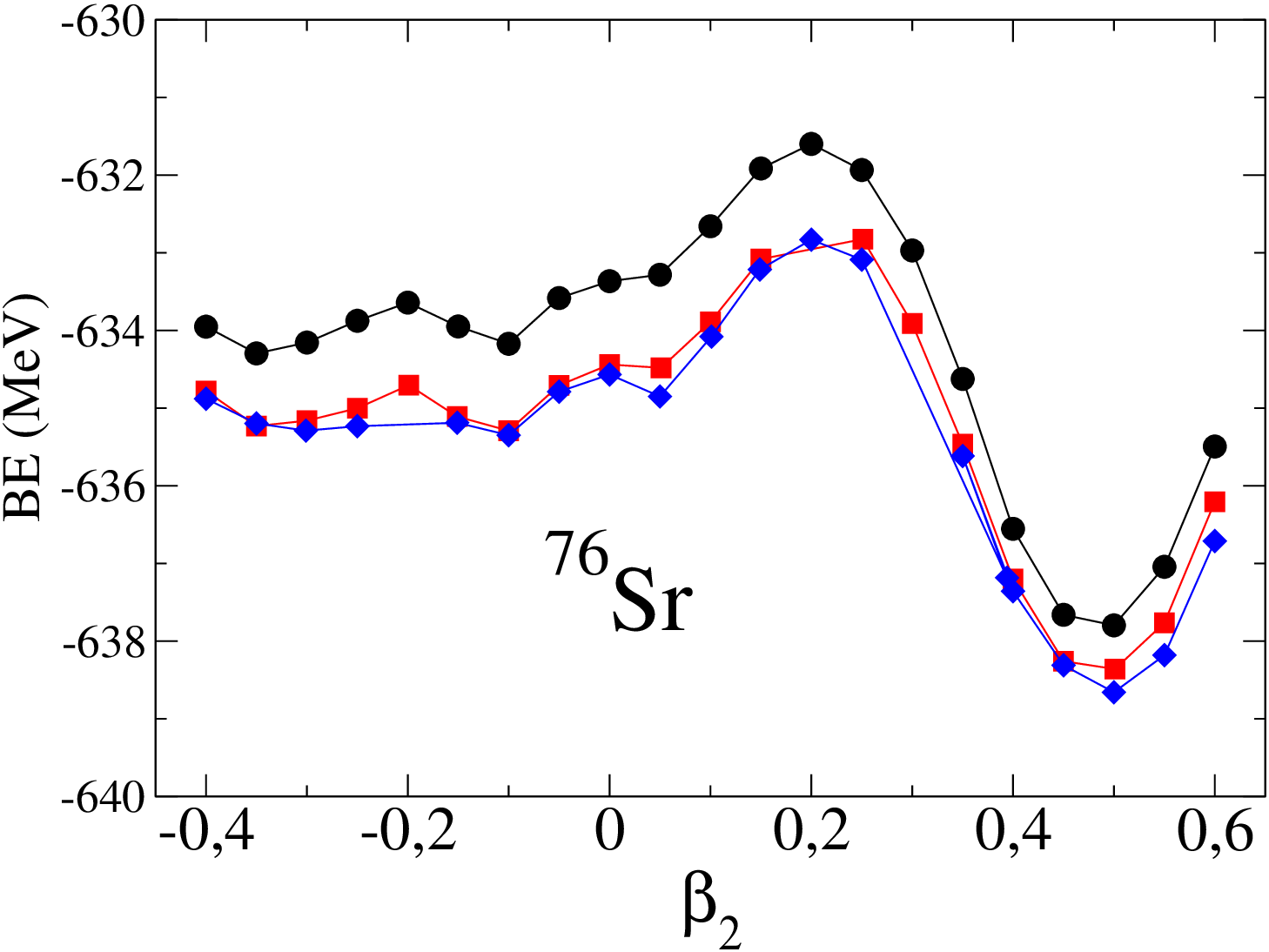}
    \end{subfigure}

    \begin{subfigure}{0.32\textwidth}
        \centering
        \includegraphics[width=\textwidth]{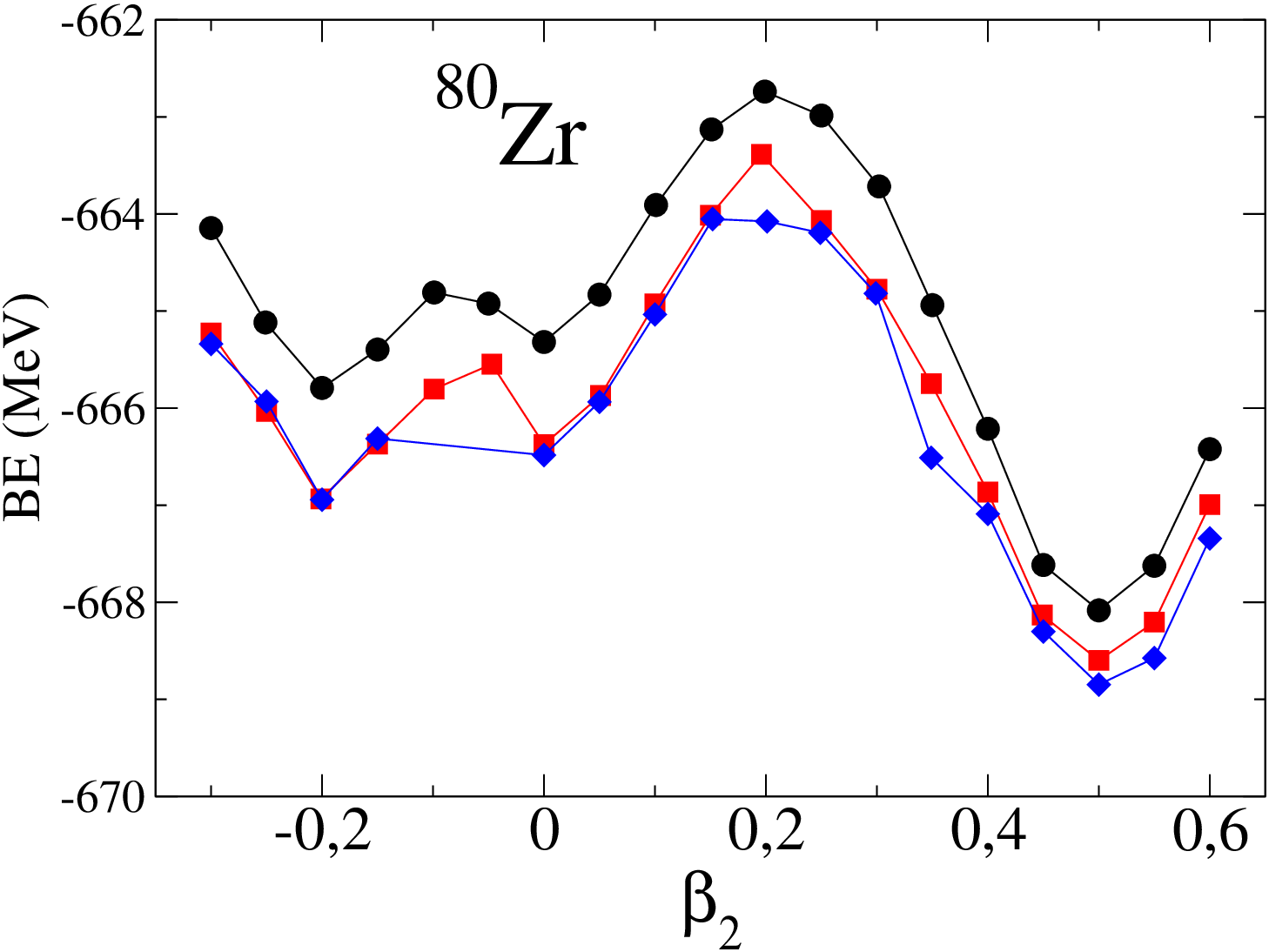}
    \end{subfigure}
    \begin{subfigure}{0.32\textwidth}
        \centering
        \includegraphics[width=\textwidth]{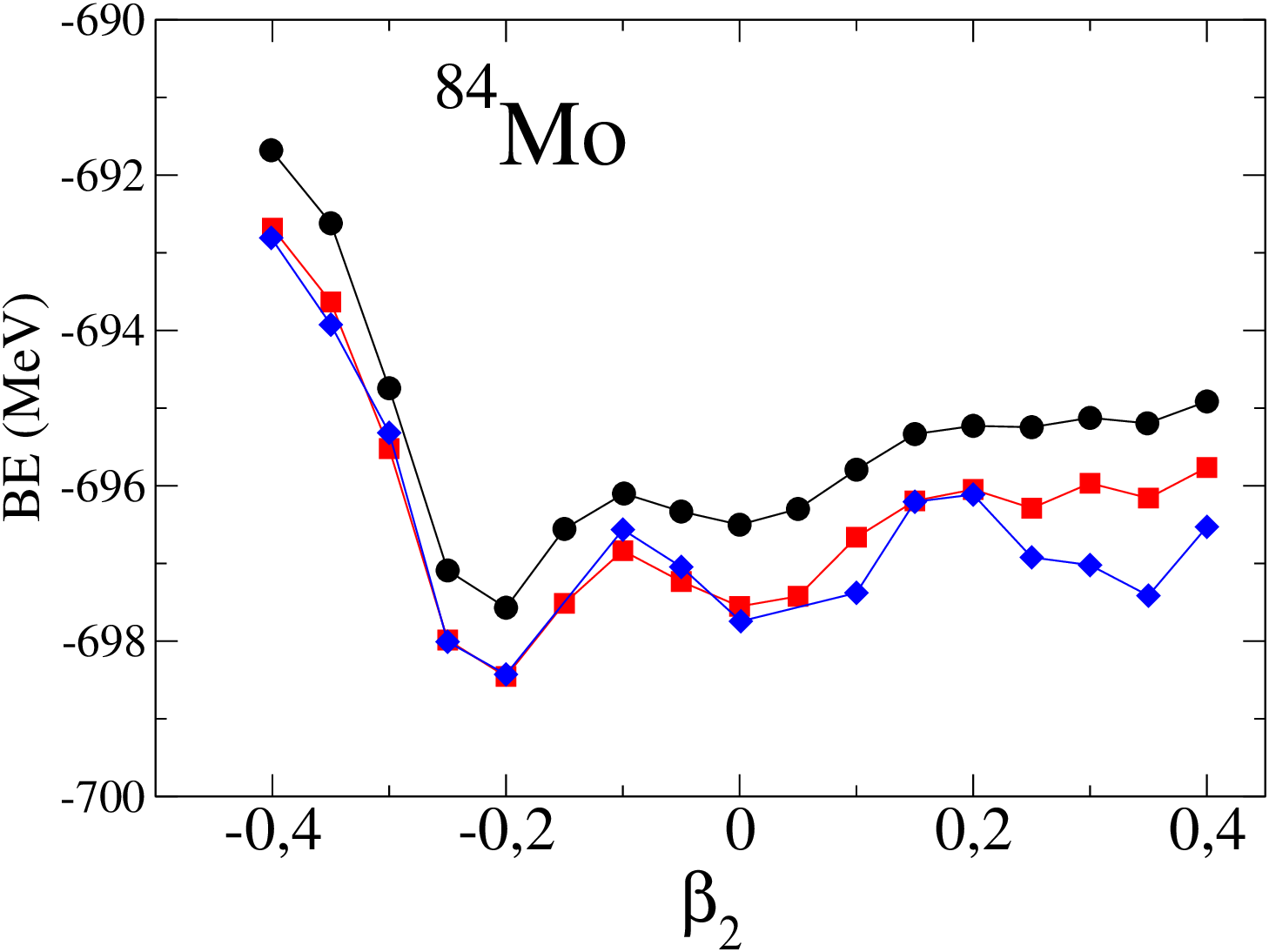}
    \end{subfigure}
    \begin{subfigure}{0.32\textwidth}
        \centering
        \includegraphics[width=\textwidth]{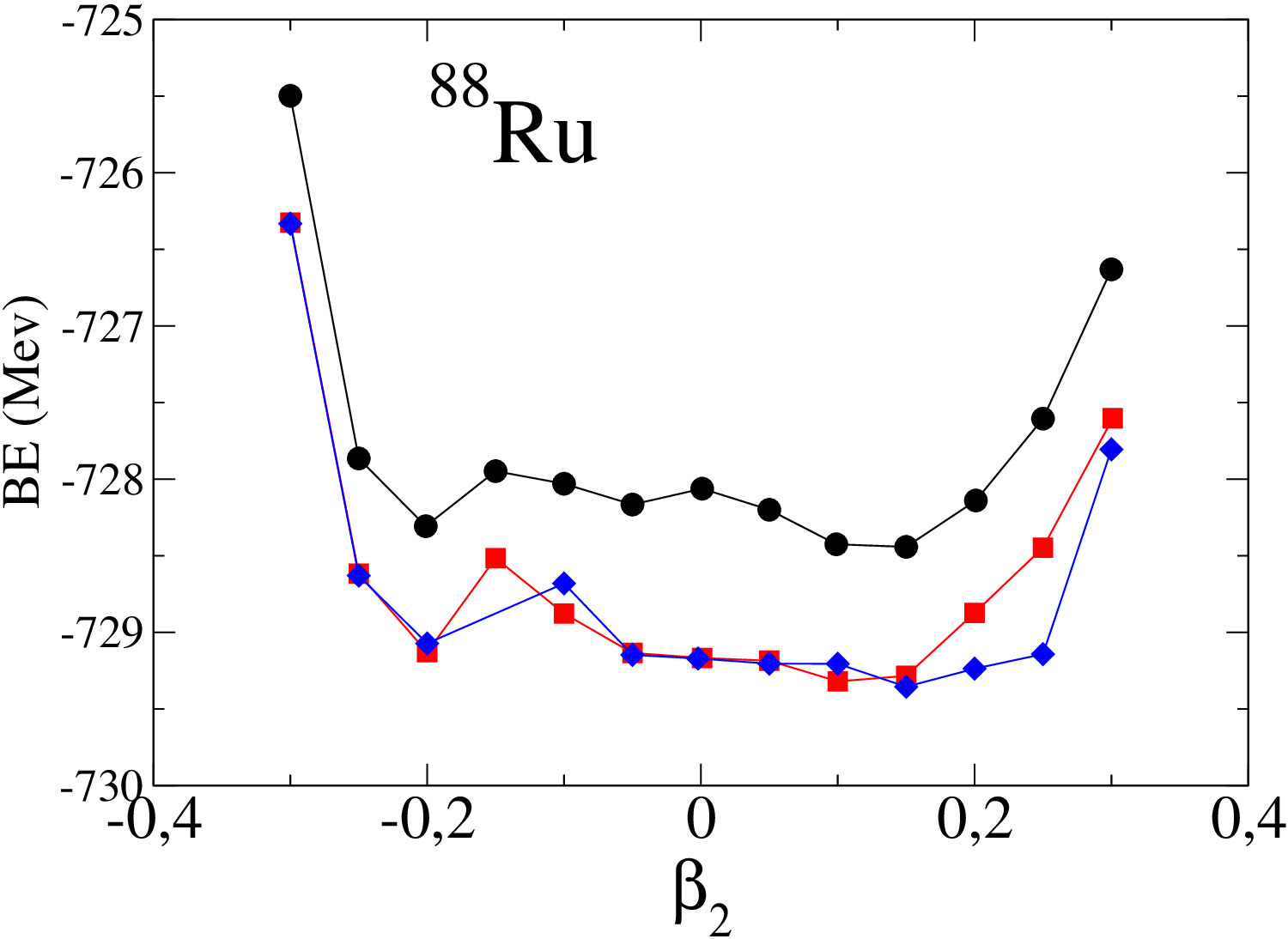}
    \end{subfigure}

    \begin{subfigure}{0.32\textwidth}
        \centering
        \includegraphics[width=\textwidth]{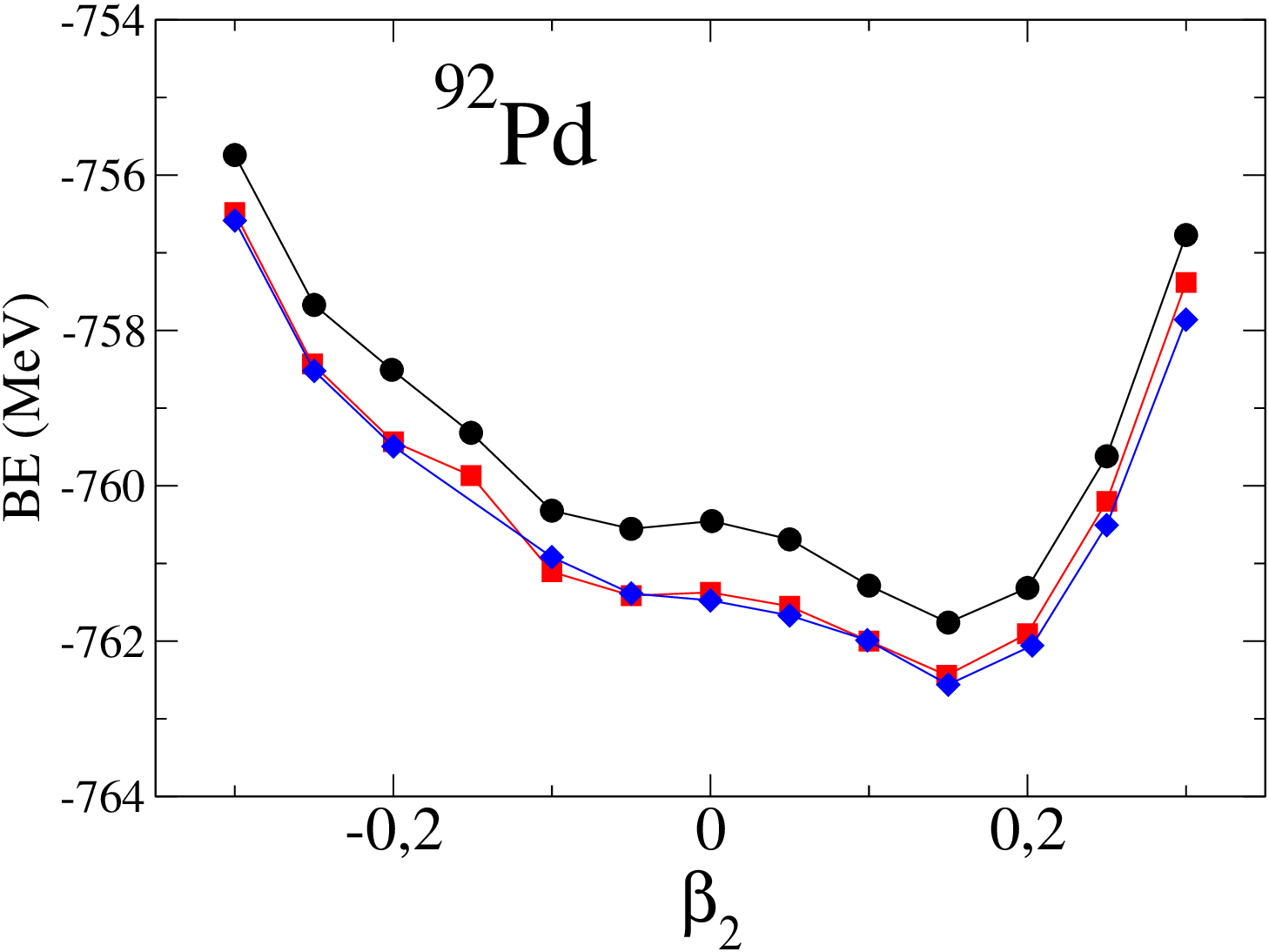}
    \end{subfigure}
    \begin{subfigure}{0.32\textwidth}
        \centering
        \includegraphics[width=\textwidth]{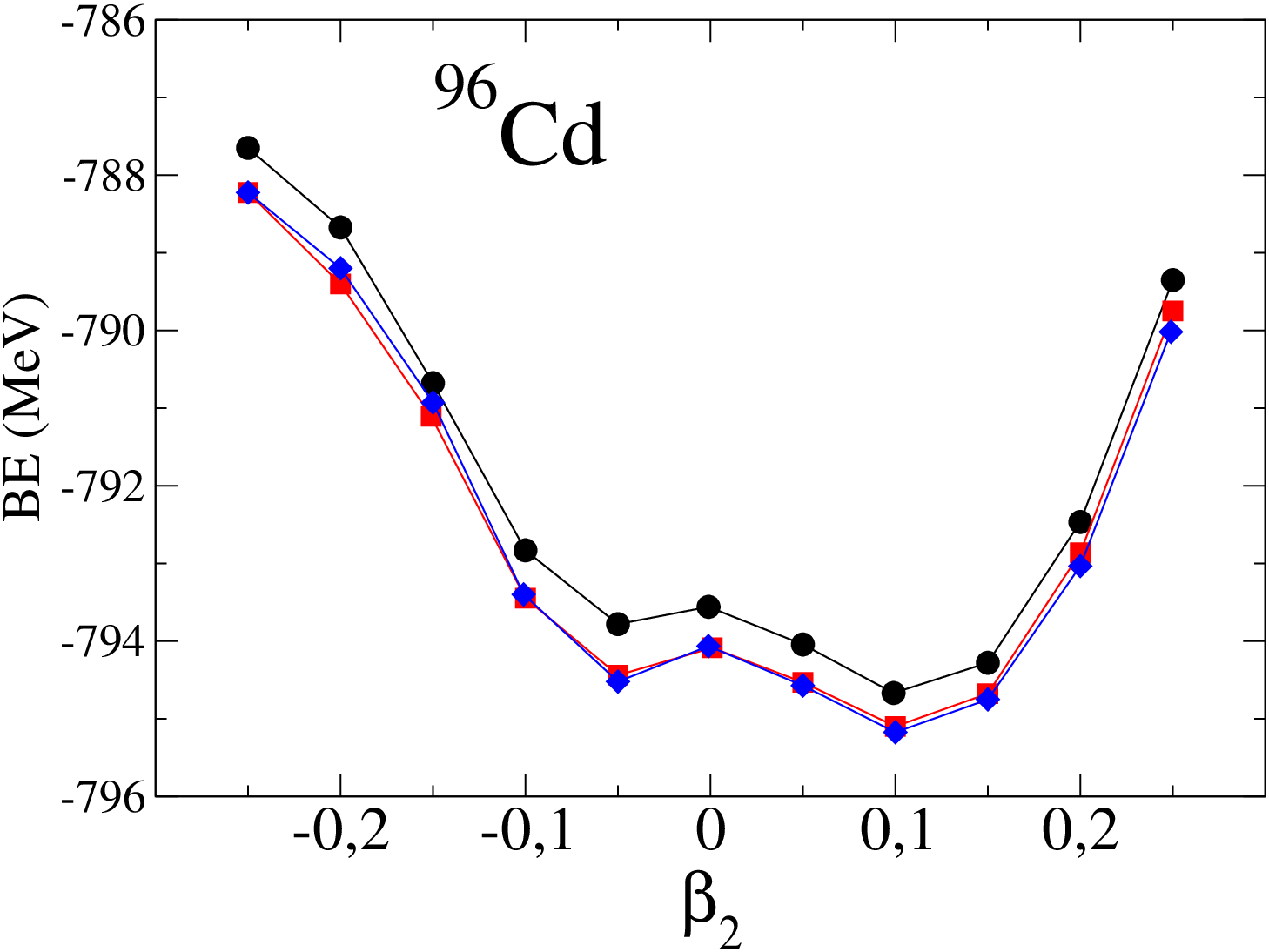}
    \end{subfigure}
    \begin{subfigure}{0.32\textwidth}
        \centering
        \includegraphics[width=\textwidth]{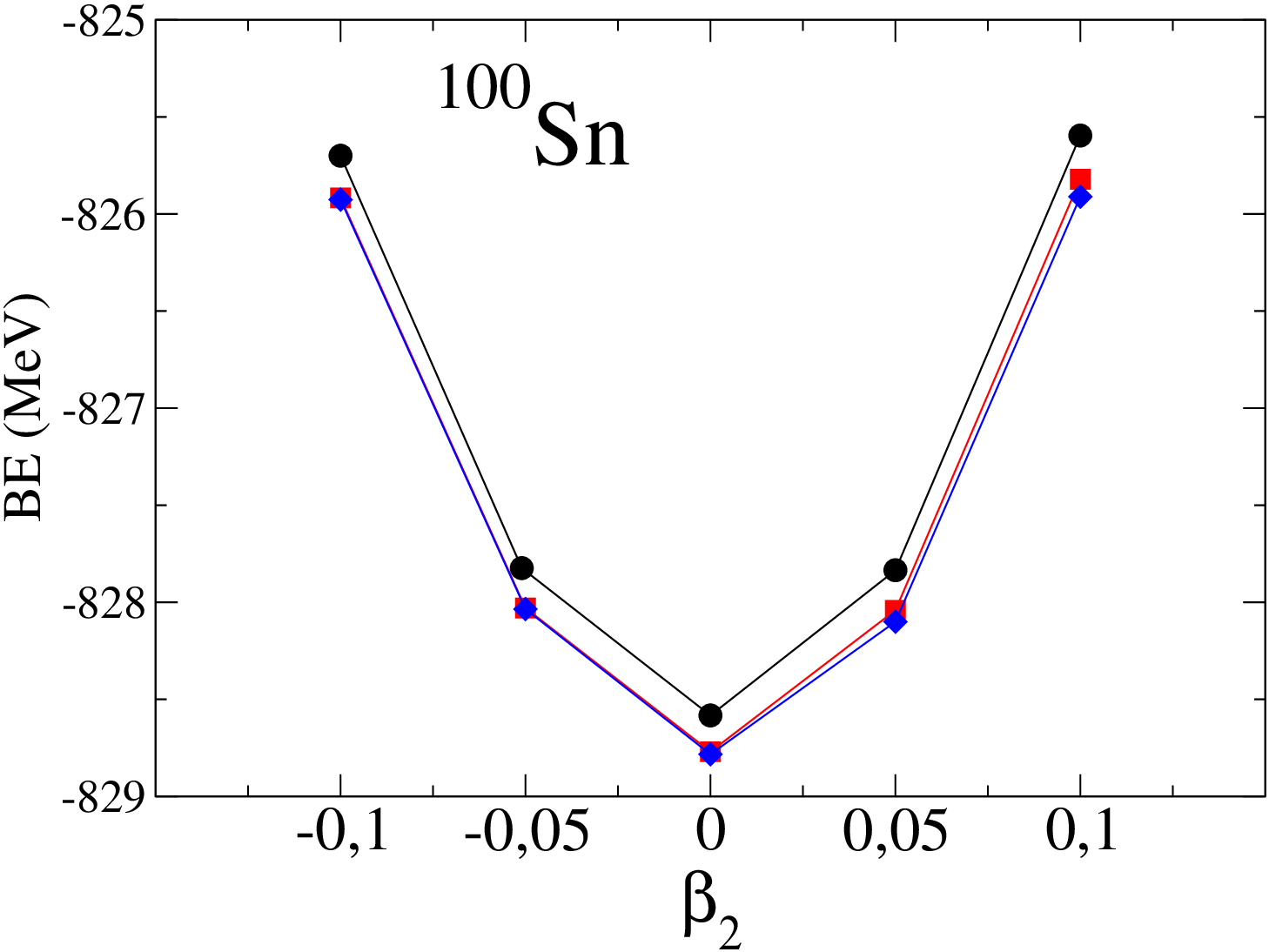}
    \end{subfigure}
    \caption{Binding energies as a function of quadrupole deformation
 for $pfg$-shell nuclei.}
    \label{fig:BE_deformation_pf}
\end{figure}

Compared with the QMC+BCS results, the binding energies of the majority of nuclei increase significantly when pairing correlations are treated within the QMC+QCM approach. The 
contribution of the isoscalar pairing channel is generally small, except in the case of 
$^{64}$Ge, where, as shown in Fig.~8, the isoscalar pairing energy is the largest. 

For nuclei between $^{56}$Ni and  $^{92}$Pd, the binding energies are improved by about 1–2 MeV relative to the BCS results. In contrast, for the $f$-shell nuclei $^{44}$Ti, $^{48}$Cr,
$^{52}$Fe and $^{56}$Ni, the binding energies are overestimated by approximately 1–2 MeV. As discussed above, this behavior is most likely due to an overestimation of the binding energies of these nuclei already at the QMC mean-field level.

Figure~\ref{fig:BE_deformation_pf} shows the dependence of the binding energy on quadrupole deformation for the $pfg$-shell nuclei. As in the case of the $sd$-shell nuclei, the inclusion of isovector pairing shifts the binding-energy curves downward by a comparable amount for all deformations relative to the BCS results. In general, the isovector pairing has a modest effect and does not significantly alter the overall dependence of the binding energy on deformation.

For some nuclei, two minima with very similar energies are observed, indicating possible shape coexistence. This is the case for
$^{64}$Ge and $^{68}$Se. For $^{88}$Ru the binding-energy curve is rather flat in the deformation range 
$-0.2 \leq \beta_2 \leq 0.2$ making it difficult to draw reliable conclusions about the ground-state shape and possible shape coexistence.

The deformations at the energy minima obtained from unconstrained QCM calculations 
are listed in Table~3 together with the experimental values and the predictions of FRDM. 
With the exception of  $^{44}$Ti and $^{56}$Ni, the QCM deformation are in reasonable agreement
with the experimental values.

\begin{table}[hbtp]
    \centering
    \renewcommand{\arraystretch}{1.2} 
    \setlength{\tabcolsep}{10pt}
    \begin{tabular}{|cccc|cccc|}
        \hline
        \textbf{A} & \textbf{QCM} & \textbf{FRDM} &  \textbf{Exp} & \textbf{A} & \textbf{QCM} & \textbf{FRDM} &  \textbf{Exp} \\
        \hline
        44  & 0.000  & 0.00 & 0.288 & 72  & -0.336 &-0.37 & 0.332  \\
        48  & 0.291  & 0.23 & 0.351 & 76  & 0.485  & 0.40 & - \\
        52  & 0.230  & 0.12 & 0.229 & 80  & 0.499  & 0.43 & - \\
        56  & 0.004  & 0.00 & 0.127 & 84  &-0.218  &-0.23 & - \\
        60  & 0.228  & 0.16 & -     & 88  &-0.217  &-0.24 & - \\
        64  & 0.238  & 0.21 & 0.258 & 92  & 0.156  & 0.00 & - \\
        68  & -0.259 & 0.23 & 0.242 & 96  & 0.108  &-0.02 & - \\
        \hline
    \end{tabular}

    \caption{Quadrupole deformation from QCM calculations, compared
to FRDM \cite{frdm} and experimental values \cite{nudat}.}
    \label{tab:beta2_comparison_pf}
\end{table}

\subsection{Nuclei with $ 100 < A \leq 120 $ }

The HFB calculations of Ref.~\cite{gezerlis_bertsch}, based on a spherically symmetric Woods–Saxon potential and a zero-range pairing force, indicate that some heavy nuclei with mass number 
$A>100$ exhibit an isoscalar pairing phase in the ground state. Subsequent HFB calculations \cite{gezerlis_def}, performed with a deformed Woods–Saxon potential, showed that these results depend strongly on the deformation, which in that study is treated as a parameter.
As an illustrative example, the case of $^{108}$Xe is discussed: this nucleus is found to be in a T=1 pairing phase from zero deformation  up to $\beta_2 \approx 0.11$, after which it undergoes a rather sudden transition to a triplet 
T=0 pairing phase at $\beta_2 \approx 0.12$. For deformations greater than $\beta_2 =0.1$
it is predicted that all other heavier $N=Z$ nuclei up to $A=150$ are in a triplet $T=0$
pairing phase.

The HFB calculations discussed above are performed with a fixed Woods–Saxon mean field and therefore do not account for the interplay between deformation and pairing correlations. In this section, we examine these predictions within the self-consistent QMC+QCM framework, in which the deformed ground-state mean field is determined consistently together with the pairing field. Moreover, in contrast to the HFB approach, both particle number and isospin are exactly conserved.

Nuclei above $^{100}$Sn are unstable and lie beyond the proton drip line; apart from 
$^{104}$Te and $^{108}$Xe, their binding energies are not experimentally known. In the present
calculations, we focus on these nuclei as well as on heavier
$N=Z$ nuclei with mass numbers below $A=120$, which are closer to the proton drip line.

\begin{figure}[h]
    \centering
    \begin{minipage}{0.45\textwidth}
        \centering
        \includegraphics[width=\textwidth]{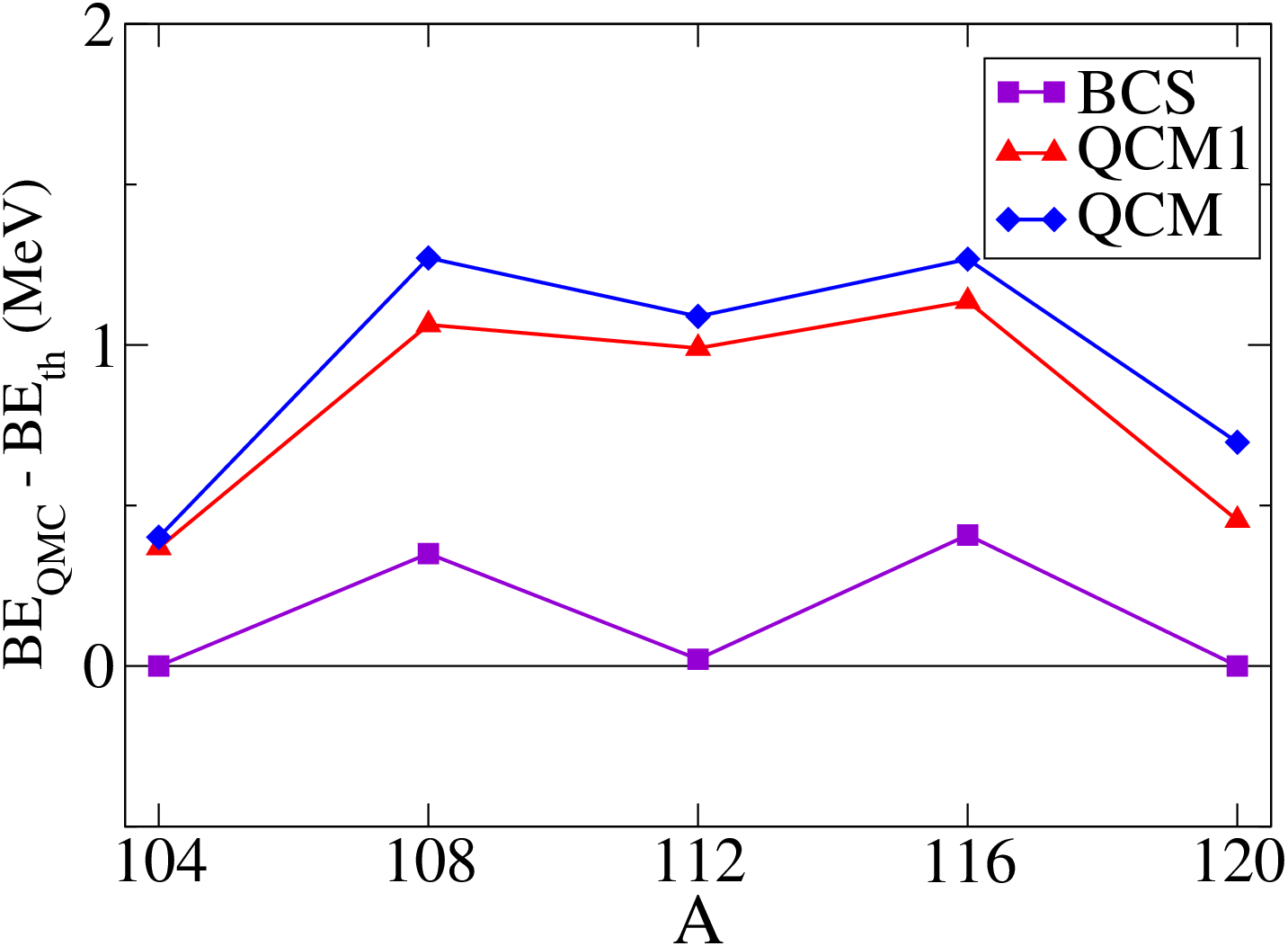}
        \caption{Binding energy residuals relative to the QMC results.}
        \label{fig:BE_5082}
    \end{minipage}
    \hfill
    \begin{minipage}{0.45\textwidth}
        \centering
        \includegraphics[width=\textwidth]{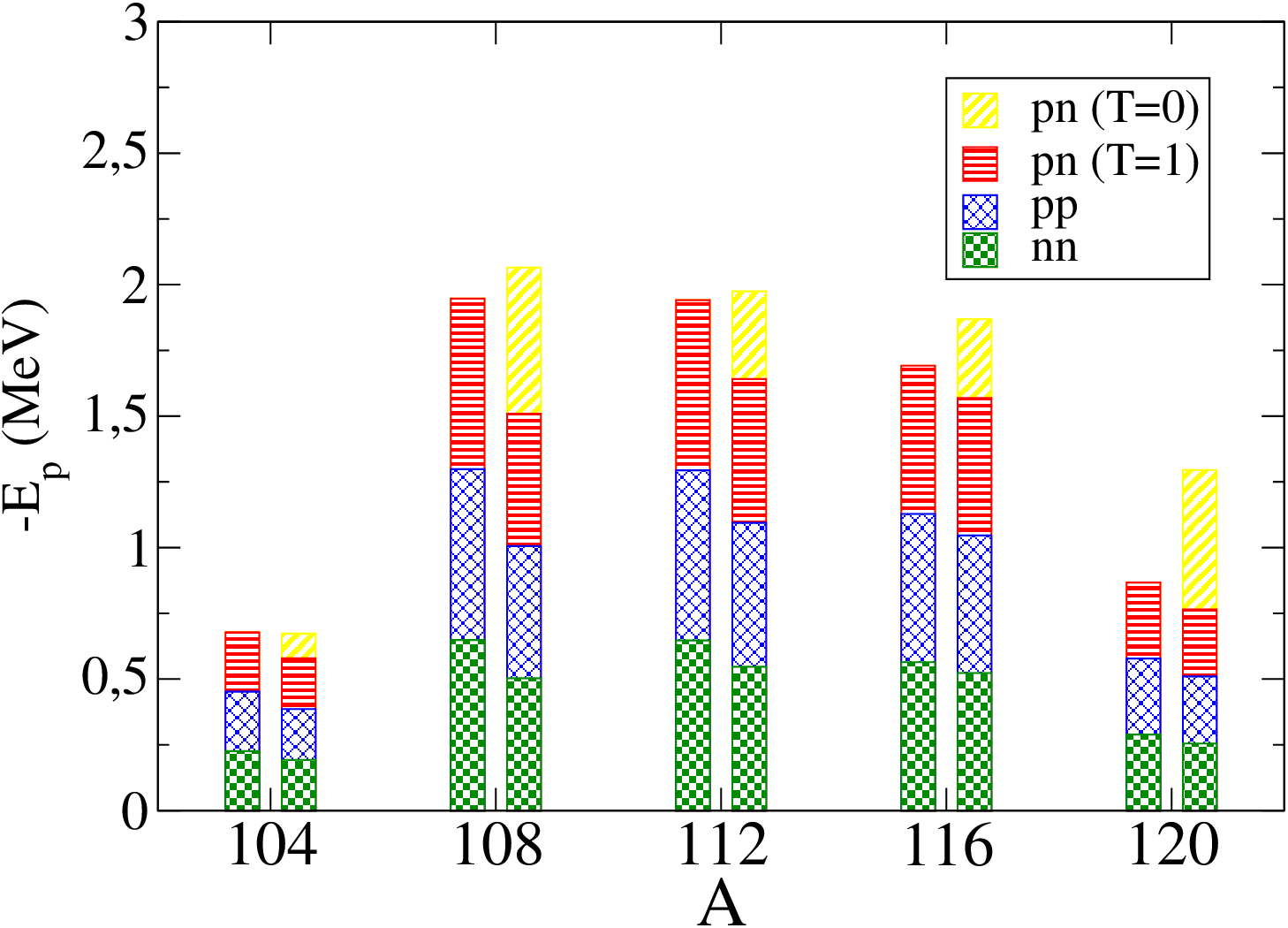}

        \caption{Pairing energies for nuclei with $A > 100$. For each nucleus, results are shown for QCM1 (left) and QCM (right).}

        \label{fig:PE_5082}
    \end{minipage}
\end{figure}
\begin{figure}[htbp]
    \centering
    \begin{subfigure}{0.48\textwidth}
        \centering
        \includegraphics[width=\textwidth]{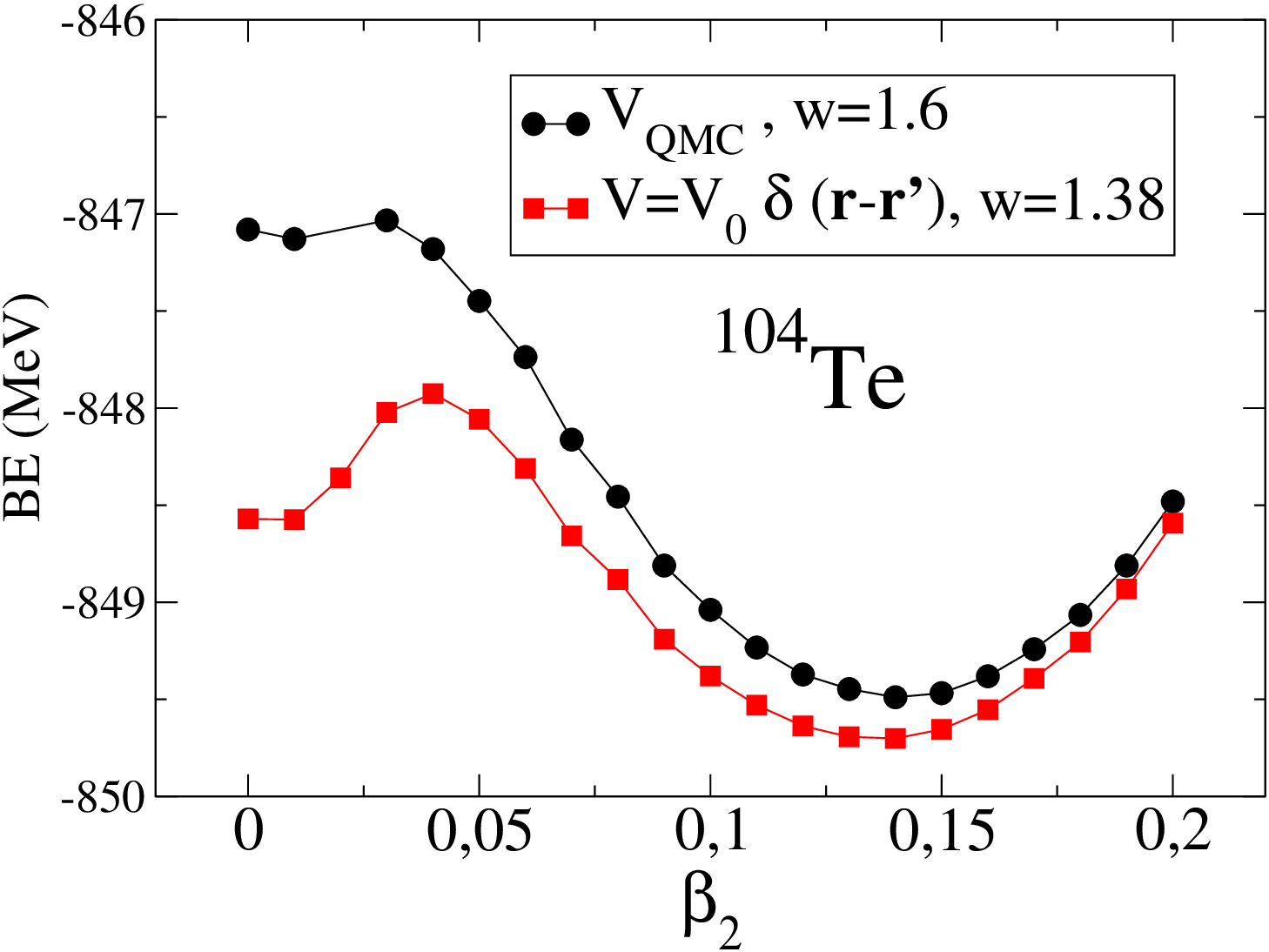}
    \end{subfigure}
    \begin{subfigure}{0.48\textwidth}
        \centering
        \includegraphics[width=\textwidth]{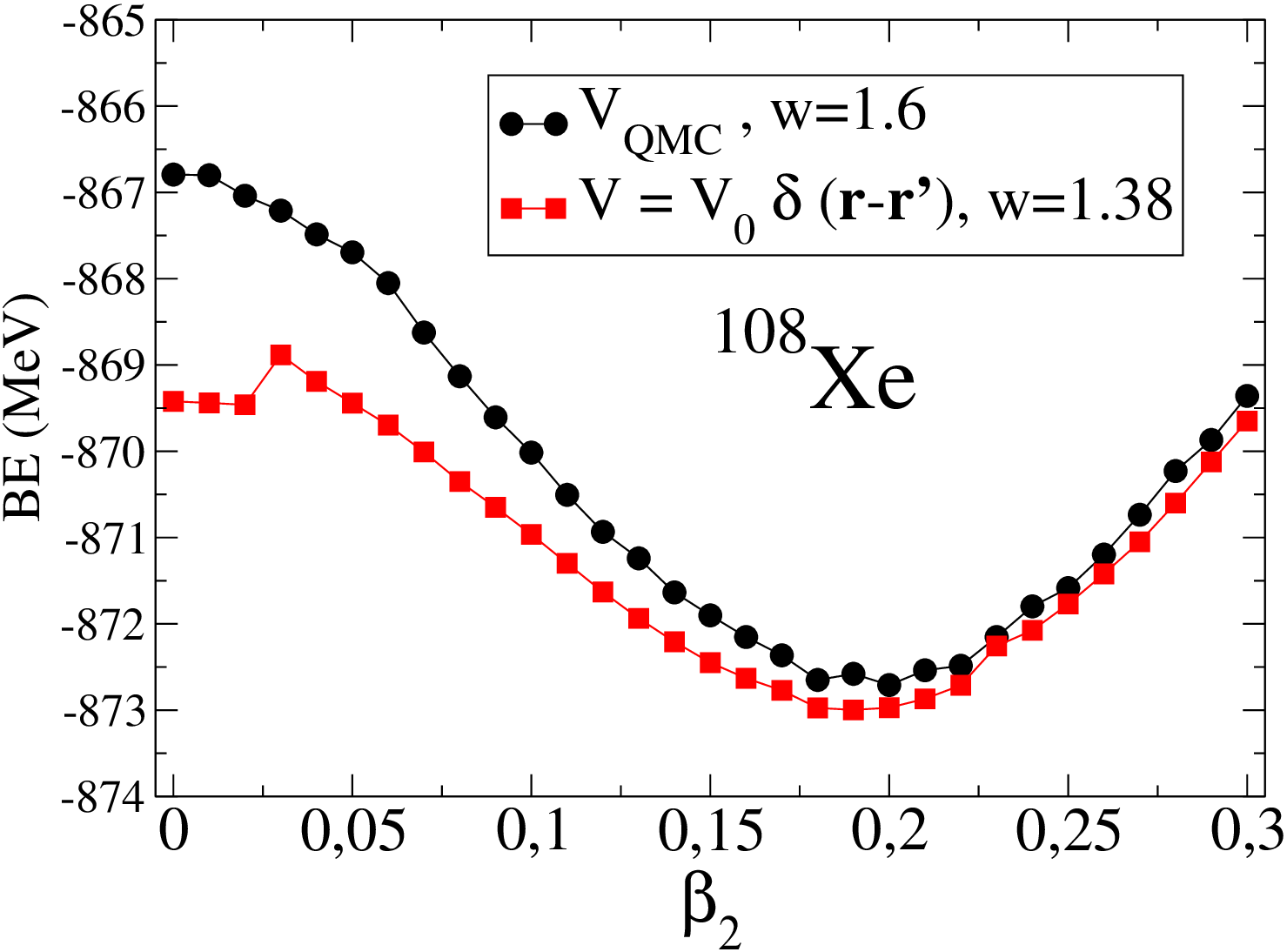}
    \end{subfigure}
   \caption{Binding energy as a function of quadrupole deformation for $^{104}$Te and $^{108}$Xe. For comparison, the QMC+QCM results calculated with
the zero-range pairing force employed in Ref. \cite{gezerlis_def} are also shown.}
    \label{fig:BE_deformation_104_108}
\end{figure}

To assess the effect of pairing on the binding energies—and given that
experimental binding energies are available only for $^{104}$Te and $^{108}$Xe—we present
in Fig.~10 the binding-energy residuals relative to the QMC mean-field results. 
It can be seen that the inclusion of $pn$ pairing increases the binding energies by 0.5-1 MeV.

The pairing energies displayed in Fig.~11 indicate that the isoscalar $pn$ pairing contribution
is large for $^{108}$Xe and $^{120}$Nd . Although it exceeds the isovector 
$pn$ pairing component, it remains smaller than the total isovector pairing energy.

Figure~12 shows, for $^{104}$Te and $^{108}$Xe, the dependence of the binding energies on 
quadrupole deformation. 
At the equilibrium deformation, the QCM calculations overestimate the binding energies of
both nuclei, by $\approx$ 1.15 MeV for $^{104}$Te and 0.5 MeV for $^{108}$Xe.

The pairing energies for $^{108}$Xe as a function of deformation are shown in Fig.~13. Far from the minimum, 
at $\beta_2 \approx  0.3$, the isoscalar pairing energy becomes comparable to the total isovector pairing energy; however, it does not surpass the total isovector pairing sufficiently to signal a transition to a dominant isoscalar phase. In fact, the variation of the pairing energies with deformation is smooth and does not exhibit any sudden transitions, in contrast to the behavior observed in HFB calculations. This latter feature is an artifact arising from the lack of exact particle-number conservation in the HFB approach.

\begin{figure}[htbp]
    \centering
    \begin{subfigure}{0.48\textwidth}
        \centering
        \includegraphics[width=\textwidth]{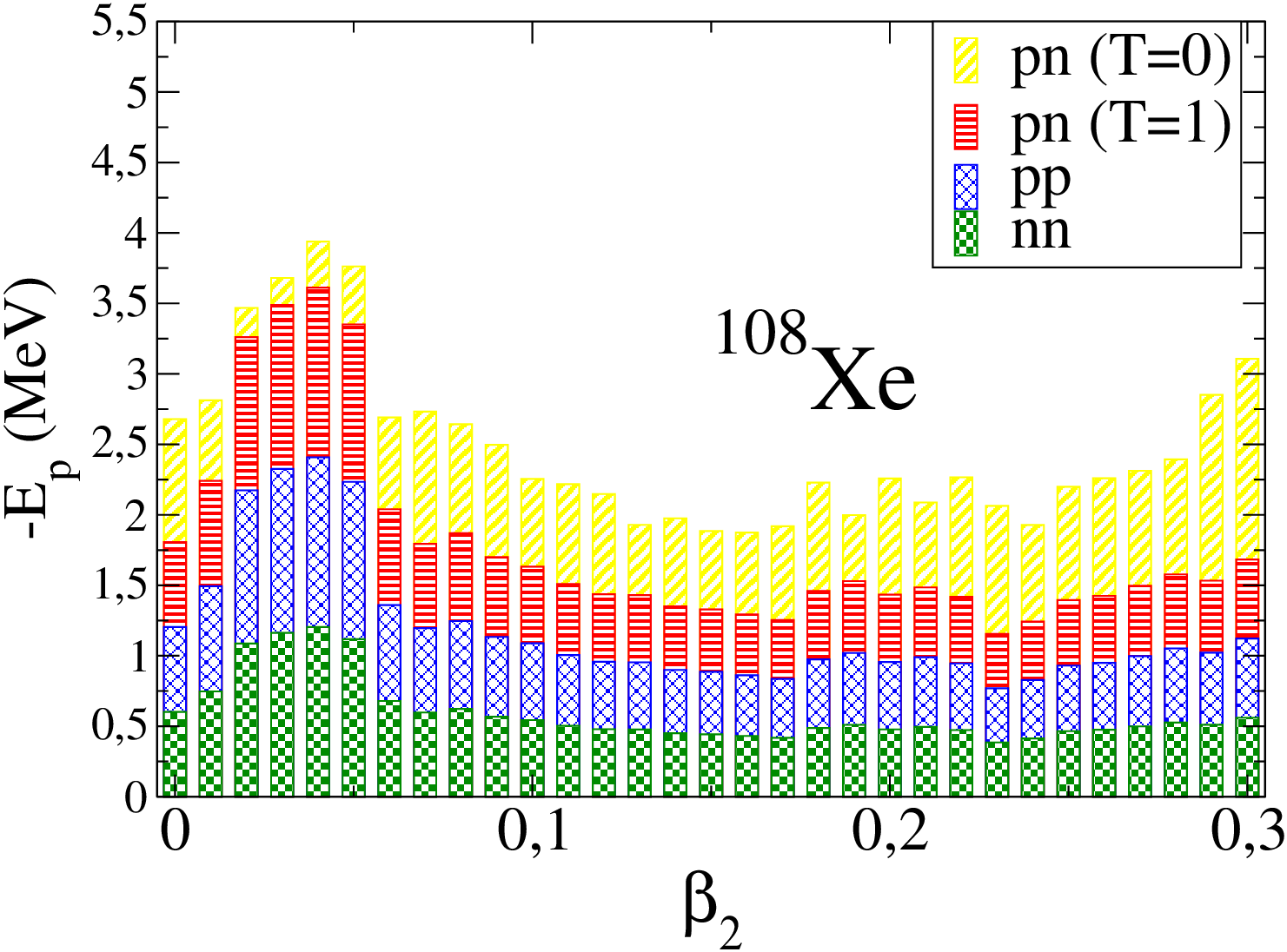}
    \end{subfigure}
    \begin{subfigure}{0.48\textwidth}
        \centering
         \includegraphics[width=\textwidth]{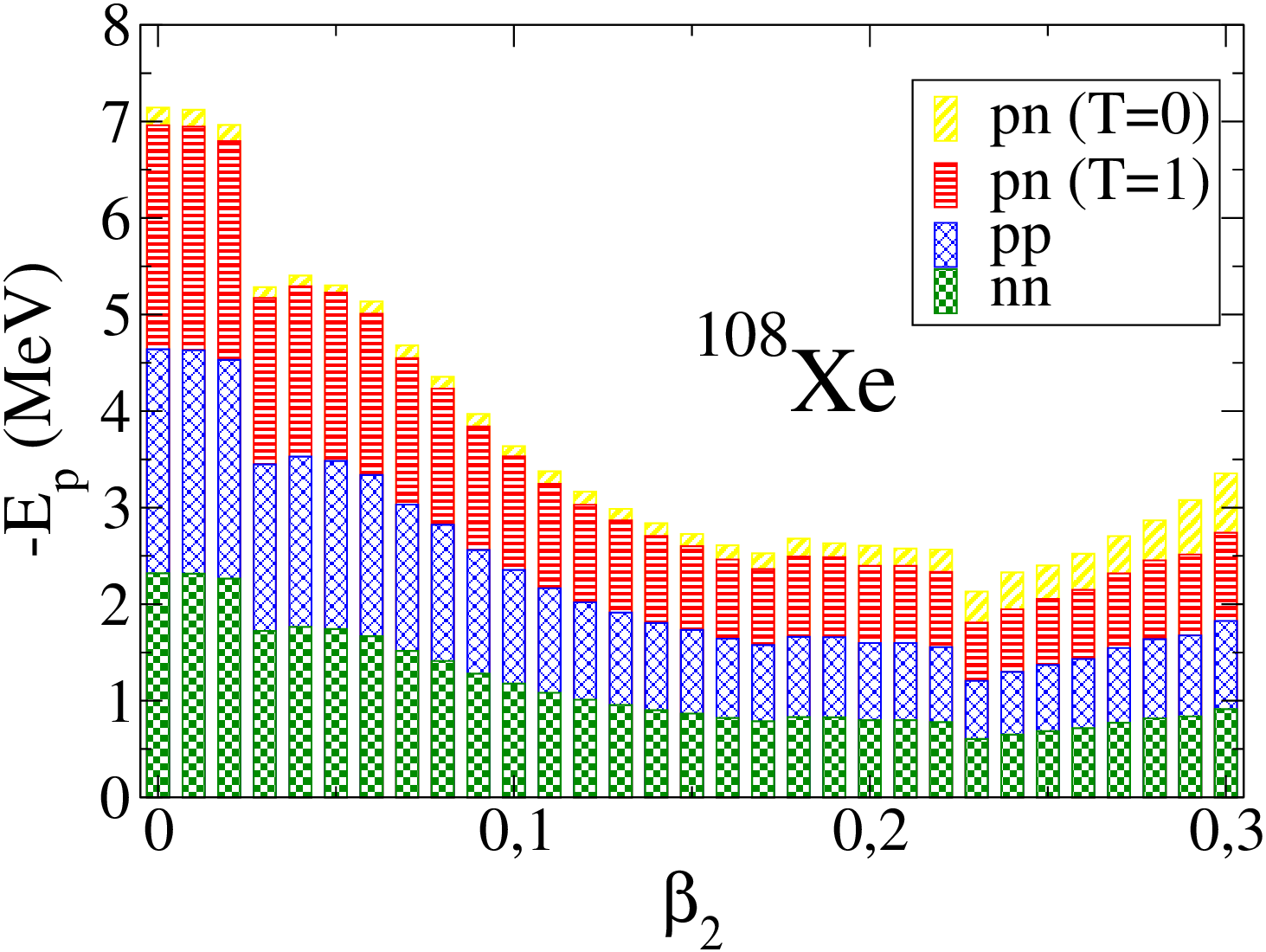}
    \end{subfigure}
\caption{Pairing energy as a function of quadruple deformation for $^{108}$Xe. For comparison, the right panel displays the QMC+QCM results obtained with the 
zero-range pairing force employed in Ref. \cite{gezerlis_def}. }
    \label{fig:PE_deformation}
\end{figure}

Figures~12 and 13 also present the QMC+QCM results obtained using the pairing interaction 
employed in the HFB calculations of Ref.~\cite{gezerlis_def}, namely a density-independent 
zero-range force with strength $V_0$=300 and an isoscalar pairing component scaled by a 
factor  $w$=1.38. The binding energies calculated with this interaction are  similar to
those obtained with the density-dependent zero-range pairing force of Eqs.~(12) and (14). 
In contrast, the isoscalar pairing energy is significantly smaller when the interaction of Ref.~\cite{gezerlis_def} is used. This demonstrates how sensitive the predictions for
proton–neutron pairing are to the treatment of pairing correlations and to the use of 
self-consistent mean-field calculations beyond the standard HFB approximation.

\subsection{ Double-magic nuclei}

In double-magic nuclei, pairing correlations vanish at the BCS and HFB levels due to the
closed-shell structure. However, this is no longer the case in approaches that treat pairing
correlations beyond the BCS approximation, such as particle-number–projected HFB (proj-HFB) 
and the QCM.

For instance, proj-HFB calculations of Ref. \cite{projHFB_stoicev}, based on a Skyrme EDF
and a density-dependent zero-range pairing interaction, predict for $^{40}$Ca 
a neutron pairing energy of about 2.2 MeV and a proton pairing energy of approximately 2–3 MeV. 
A sizable neutron pairing energy, of the order of 3 MeV, is also obtained for $^{132}$Sn. 
It should be recalled, however, that particle-number projection within the EDF framework
is affected by conceptual ambiguities and numerical difficulties, in particular those
related to the possible vanishing of overlaps between projected states generated by the
particle-number projection operator (e.g., see \cite{projHFB_robledo} and the references cited
therein).

In this section, we analyze how the binding energies of the  N=Z double-magic nuclei 
$^{16}$O, $^{40}$Ca, $^{56}$Ni and $^{100}$Sn, are modified by pairing correlations
when these are treated within the QCM approach, as in the case of the open-shell nuclei discussed above,
and with the particle-number-projected BCS (PBCS) approximation, which conserve particle
number but does not restore isospin symmetry.

The PBCS calculations are performed using the wave function
\begin{equation}
| \mathrm{PBCS} \rangle =
(\Gamma_n^\dagger)^{(N-N_0)/2}
(\Gamma_p^\dagger)^{(Z-Z_0)/2}
| - \rangle ,
\end{equation}
where $N_0$ and $Z_0$ denote the numbers of neutrons and protons in the inert 
core $| - \rangle$ . The operators $\Gamma_t^\dagger$ ( t=n,p), are collective pair operators 
that scatter like-particle pairs in time-reversed states, defined as
\begin{equation}
\Gamma_t^\dagger = \sum_i x_i^{(t)},
a_{ti}^\dagger a_{t\bar{i}}^\dagger .
\end{equation}
The mixing amplitudes $x_i^{(t)}$ are determined variationally by minimizing the expectation value of
the pairing Hamiltonian (1), without the $pn$ interaction, with respect to the PBCS wave function
under the normalization condition $\langle \mathrm{PBCS} \mid \mathrm{PBCS} \rangle$ =1.

\begin{figure}[h]
    \centering
    \begin{minipage}{0.45\textwidth}
        \centering
        \includegraphics[width=\textwidth]{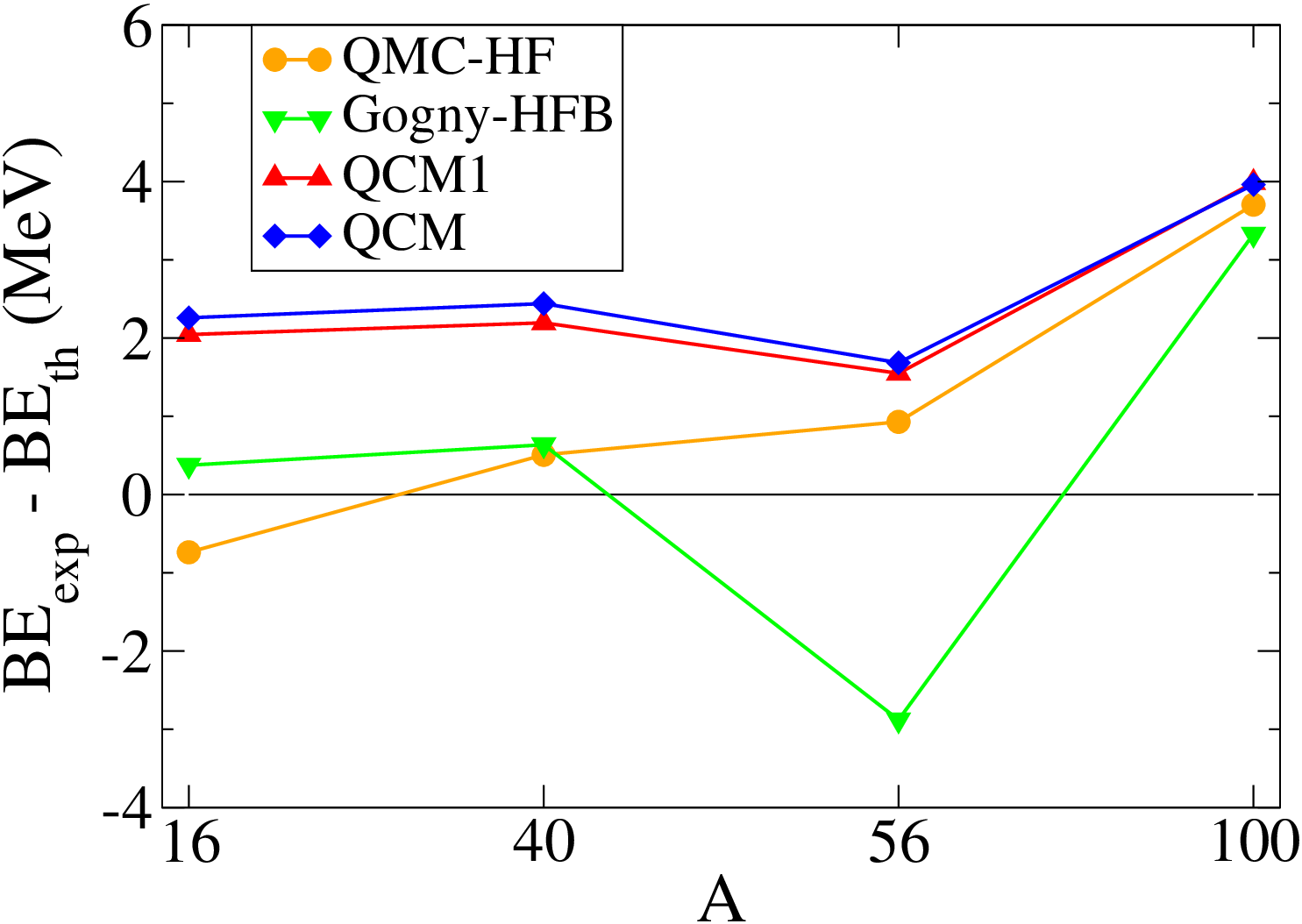}
        \caption{Binding-energy residuals for double-magic nuclei in the approximations indicated
in the figure.}
        \label{fig:BE_sd}
    \end{minipage}
    \hfill
    \begin{minipage}{0.45\textwidth}
        \centering
        \includegraphics[width=\textwidth]{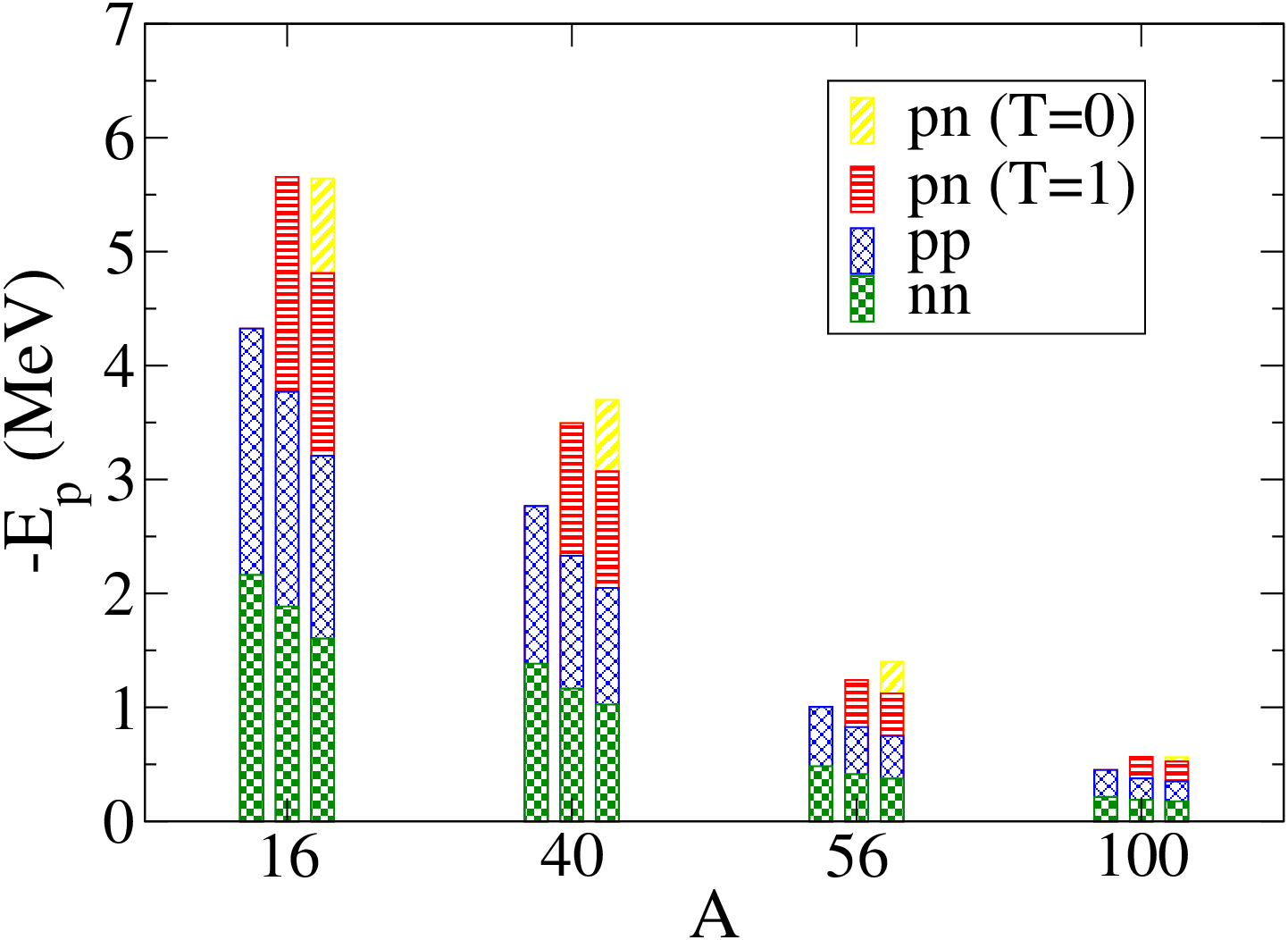}
        \caption{Pairing energies for double-magic nuclei. For each nucleus, results are shown 
        (from right to left) for QCM, QCM1 and PBCS.}
       \label{fig:PE_sd}
    \end{minipage}
\end{figure}

The QCM and PBCS calculations are performed using the isovector pairing interaction (12) with the following scaling factors: s=1.3 for $^{16}$O, s=1.5 for $^{40}$Ca and $^{56}$Ni, and s=1.8 for $^{100}$Sn.
For the isoscalar pairing channel, in the QCM calculations we employ the scaling factor 
$w$=1.6 as in the case of the open-shell N=Z nuclei discussed above. 

In both the QCM and PBCS wave functions, an inert core with 
$(N-6, Z-6)$ nucleons is assumed for all nuclei. The pairing 
correlations are treated within a model space comprising ten
single-particle levels above this core. In addition, in pairing 
calculations the proton single-particle energies are assumed
to be identical to those of the neutrons. 

In the PBCS and QCM calculations, the ground state of all nuclei 
have spherical symmetry. As an illustration, Figs.~6 and 9 display 
the dependence of binding energy on deformation for $^{40}$Ca, $^{56}$Ni and
$^{100}$Sn.

The residual binding energies are presented in Figs.~14.
At the QMC mean-field level, $^{16}$O is underbound by about 0.75 MeV, whereas the 
other three nuclei are overbound by approximately 0.5, 0.9, and 3.7 MeV, respectively. 
In contrast, the HFB calculations overbind $^{16}$O by about 0.4 MeV and underbind 
$^{56}$Ni by roughly 3 MeV. 

The QMC+QCM results indicate that the inclusion of pairing correlations increases the binding 
energies of all nuclei relative to the QMC results, by about 2.5, 2.0, 0.8, and 0.2 MeV,
respectively.

As shown in Fig.~15, the like-particle pairing energies predicted by PBCS are significant for all nuclei, highlighting the important role of particle-number restoration. A similar behavior is found in the 
proj-HFB calculations. In particular, as mentioned above, the proj-HFB results of 
Ref.~\cite{projHFB_stoicev} 
yield neutron and proton pairing energies for $^{40}$Ca that are even larger than those obtained in the
present PBCS calculations.

When isospin symmetry is further restored within the isovector QCM framework, an additional gain in pairing energy, relative to PBCS, arises from the isovector $pn$ contribution. This extra energy, however, does not coincide with the 
$pn$ pairing energy itself. As seen in the figure, the like-particle pairing energy decreases in QCM due to its competition with the $pn$ pairing correlations. A similar interplay occurs when isoscalar 
$pn$ pairing is included: the isovector pairing correlations are correspondingly reduced.

The finding that pairing correlations contribute significantly to the binding energies of 
double-magic nuclei when treated beyond the BCS/HFB approximation—such as in proj-HFB,
PBCS or QCM—has important implications for calculations based on Skyrme, Gogny, or QMC energy density
functionals (EDFs). The parameters of these mean-field models are typically adjusted to reproduce
the binding energies of double-magic nuclei, under the assumption that such systems are 
unaffected by pairing correlations at the BCS/HFB level.

Consequently, when these EDFs are combined with pairing approaches that go beyond BCS/HFB 
and generate nonvanishing pairing energies in double-magic nuclei, a refit of both the 
mean-field and pairing parameters is required for consistency. Otherwise, pairing effects may
already be effectively absorbed into the fitted EDF parameters, leading to double counting and, 
ultimately, to an overbinding of nuclei when beyond-BCS correlations are included, as observed
in the results presented above.

\section{Summary and Conclusions}

In this work, we have extended mean-field calculations based on the QMC EDF to include proton–neutron ($pn$) pairing correlations in both the isovector (T=1) and isoscalar (T=0) channels. Pairing correlations are treated within the QCM approach, which conserves both particle number and isospin exactly. Axially deformed, self-consistent QMC+QCM calculations have been performed for the 
ground states of N=Z nuclei in the mass range $16 \leq A \leq 120$.

In the QCM calculations, the isovector pairing interaction is taken as a zero-range force, 
with a density-dependent term derived consistently within the QMC framework. Its strength is fitted to reproduce odd–even mass differences in neutron-rich semi-magic proton nuclei, 
with the neutrons occupying the same major shell as in the corresponding 
N=Z systems. The isoscalar pairing interaction is assumed to be proportional to the 
isovector one, with a proportionality factor of 1.6.

The main results of this study can be summarized as follows:

(a) For the majority of nuclei, the inclusion of isovector pairing significantly improves the calculated binding energies, by about 2 MeV in $sd$-shell nuclei and by 1–2 MeV in 
$pfg$-shell nuclei, bringing them closer to the experimental values.

(b) For most nuclei, the total binding energies are only weakly affected by switching on the isoscalar pairing interaction. The largest additional contributions, of the order 
of 300–600 keV, are found in $^{20}$Ne, $^{24}$Mg, $^{64}$Ge and $^{108}$Xe.

(c) Isovector and isoscalar pairing correlations coexist in all studied nuclei. This feature, which is not always obtained in HFB calculations, is a general 
property of the QCM approach and is directly related to the exact conservation of particle number and isospin.

(d) In some nuclei, such as $^{20}$Ne, $^{24}$Mg, $^{64}$Ge and $^{76}$Sr, the $pn$ isoscalar pairing energy exceeds the corresponding isovector $pn$ pairing energy, but not the total isovector pairing
energy. Since this behavior is observed across different mass regions, and neighboring nuclei often exhibit significantly different isoscalar pairing energies, it is unlikely to be associated 
with global properties such as the spin–orbit splitting. Instead, it most likely reflects
the values of the off-diagonal matrix elements of the T=0 interaction between 
single-particle states near the Fermi level, which present strong oscillations
between positive and negative values.

(e) The deformation dependence of the binding energy suggests the possible occurrence of shape coexistence in nuclei such as $^{20}$Ne, $^{32}$S, $^{36}$Ar, $^{64}$Ge and $^{68}$Se. With the exception of $^{20}$Ne, the 
binding-energy curves and the degree of shape coexistence are only weakly influenced by the inclusion of isoscalar pairing.

(f) For nuclei above $^{100}$Sn, in particular for $^{108}$Xe, we do not find evidence for a 
clearly dominant isoscalar pairing phase, in contrast to the HFB prediction \cite{gezerlis_def}.

(g) In doubly magic $N=Z$ nuclei, pairing correlations treated within the  QCM
framework provide a non-negligible contribution, of about several MeV, to the total binding energy. 
This is in contrast with the BCS and HFB in which pairing has no contribution.

In the present study, we have focused exclusively on N=Z nuclei, for which $pn$
pairing correlations have the largest impact. As a perspective, we plan to
extend the QMC+QCM approach to investigate $pn$ correlations in nuclei with a
small neutron excess beyond the  $N=Z$ line, where the effects of proton–neutron pairing are
reduced but remain non-negligible. For this purpose, we will employ the
generalized QCM formalism described in Ref. \cite{qcm_t0t1_ngz}, which has recently been 
applied in Skyrme+QCM calculations \cite{skyrme_qcm}.

\newpage
\vskip 0.2 cm
\noindent
{\bf Acknowledgments}
\vskip 0.2cm
\noindent
We thank A. W. Thomas  for providing the QMC code - developed in collaboration with 
his co-workers - and for insightful discussions related to this work.
 N.S. acknowledges the hospitality of Institute of Modern Physics, 
Cantabria University, Spain, where this paper was primarily conceived and written. 
This work was supported by the Romanian Ministry of Education and Research through 
the project PN-23-21-01-01/2023. T. P. was also supported by a Ph.D. fellowship from 
the University of Bucharest.

\appendix
\section{Appendix: Matrix Element of Pairing Interaction }
\subsection{Single-particle states: definitions}
We consider axially deformed single-particle states $\psi_K$, obtained from the mean-field calculations.
Their expressions in term of spin projections are given by : 
\begin{equation}
\psi_K({\mathbf{r}})= \left[ \varphi_{K \uparrow }(r_\perp,z)e^{i\Lambda_K^-\phi}|\uparrow\rangle+\varphi_{K \downarrow}(r_\perp,z)e^{i\Lambda_K^+\phi}|\downarrow\rangle\right]
\label{Eq:hbcs2}
\end{equation}
where $K = (a,\Lambda_K^\pm)$, and $\Lambda_K^\pm=\Theta_K \pm 1/2$ represents the projection of the total
angular momentum onto the symmetry axes, while $(r_\perp,z,\phi)$ denote the cylindrical coordinates.

The corresponding time-reversed  states are obtained  by applying the
operator $\hat{T}=i\sigma_y \hat{K}$, where $\hat{K}$ denotes  the complex conjugate operator. This gives:

\begin{equation} \label{time}
\begin{split}
\psi_{\bar{K}}({\mathbf{r}})&=\hat{T}\psi_K({\mathbf{r}})=i\sigma_y \hat{K}\psi_K({\mathbf{r}})=i\sigma_y  \left[ \varphi_{K \uparrow }(r_\perp,z)e^{-i\Lambda_K^-\phi}|\uparrow\rangle+\varphi_{K \downarrow}(r_\perp,z)e^{-i\Lambda_K^+\phi}|\downarrow\rangle\right] \\
&=  \left[- \varphi_{K \uparrow }(r_\perp,z)e^{-i\Lambda_K^-\phi}|\downarrow\rangle+\varphi_{K \downarrow}(r_\perp,z)e^{-i\Lambda_K^+\phi}|\uparrow\rangle\right]\\
&=  \left[\varphi_{{\bar{K}} \downarrow }(r_\perp,z)e^{-i\Lambda_K^-\phi}|\downarrow\rangle+\varphi_{{\bar{K}} \uparrow}(r_\perp,z)e^{-i\Lambda_K^+\phi}|\uparrow\rangle\right]\\.
 \end{split}
\end{equation}

In summary, the single-particle states and their corresponding  time-reversed states, required for pairing
calculations, are given by :

\begin{eqnarray*}
\psi_K({\mathbf{r}})&=& \left[ \varphi_{K \uparrow }(r_\perp,z)e^{i\Lambda_K^-\phi}|\uparrow\rangle+\varphi_{K \downarrow}(r_\perp,z)e^{i\Lambda_K^+\phi}|\downarrow\rangle\right]\\
\psi^{\dagger}_K({\mathbf{r}})&=& \left[ \varphi_{K \uparrow }(r_\perp,z)e^{-i\Lambda_K^-\phi}|\uparrow\rangle^*+\varphi_{K \downarrow}(r_\perp,z)e^{-i\Lambda_K^+\phi}|\downarrow\rangle^*\right]\\
\psi_{\bar{K}}({\mathbf{r}})&=& \left[- \varphi_{K \uparrow }(r_\perp,z)e^{-i\Lambda_K^-\phi}|\downarrow\rangle+\varphi_{K \downarrow}(r_\perp,z)e^{-i\Lambda_K^+\phi}|\uparrow\rangle\right]\\
&=&\left[ \varphi_{ \bar{K} \downarrow}(r_\perp,z)e^{-i\Lambda_K^-\phi}|\downarrow\rangle+\varphi_{ \bar{K} \uparrow}(r_\perp,z)e^{-i\Lambda_K^+\phi}|\uparrow\rangle\right]\\
\psi^{\dagger}_{\bar{K}}({\mathbf{r}})&=& \left[- \varphi_{K \uparrow }(r_\perp,z)e^{i\Lambda_K^-\phi}|\downarrow\rangle^*+\varphi_{K \downarrow}(r_\perp,z)e^{i\Lambda_K^+\phi}|\uparrow\rangle^*\right]\\
&=& \left[ \varphi_{\bar{K} \downarrow  }(r_\perp,z)e^{i\Lambda_K^-\phi}|\downarrow\rangle^*+\varphi_{\bar{K} \uparrow }(r_\perp,z)e^{i\Lambda_K^+\phi}|\uparrow\rangle^*\right]
\end{eqnarray*}

\subsection{Two-body matrix element of the pairing interaction}

For the case of a mean field with axial symmetry, the pairing force $V_P$ is considered to scatter  two nucleons in time-conjugate states described by a the two-body wave function $|K \bar{K}>$. The corresponding matrix elements are of the form:
\begin{equation*}
 \bra{K \bar{K}} V_P (1-P_r P_{\sigma} P_{\tau}) \ket{L \bar{L}}
\end{equation*}
where the exchange term is expressed by the exchange operators for the radial coordinates, spin and isospin.
For a zero range pairing force of the type $V_P=V\delta (r_1 -r_2)$, considered here, $P_r =1$ and
the matrix elements become:
\begin{equation*}
\bra{K \bar{K}} V_P (1-P_{\sigma} P_{\tau}) \ket{L \bar{L}}
\end{equation*}
The exchange operators for the spin and the isospin have the standard expressions:
\begin{equation*}
P_\sigma = \frac{1}{2} (1+ \sigma_1 \sigma_2) \hskip 0.5cm P_\tau =\frac{1}{2} (1+ \tau_1 \tau_2)
\end{equation*}
The expression $(1-P_\sigma P_\tau)$ can be written as:
\begin{eqnarray*}
 (1-P_\sigma P_\tau)= 
 2P_{S=1}P_{T=0}+2P_{S=0}P_{T=1}
 \nonumber
\end{eqnarray*} 
where $P_S$ and $P_T$ are the projection operators on the total spin S=0,1 and total isospin $T=0,1$:
\begin{equation*}
P_S = \frac{1}{2} (1 - (-1^S P_\sigma) \hskip 0.5cm  P_T =\frac{1}{2} (1 - (-1)^T P_\tau)
\end{equation*}
Therefore the matrix elements of the pairing interaction can be split in two terms, representing
the spin-isospin channels $(S=0, T=1)$ and $(S=1, T=0)$. 

In conclusion, for a zero range force of the form $V_P=V \delta(r_1-r2)$ the matrix elements we need to evaluate are the following:

\begin{equation}\label{Integral}
 V^{TS}_{KL}=2\langle K  \bar K| V \delta P_S P_T| L \bar L\rangle=
 2\int d\mathbf{r} \psi_K^\dagger({\mathbf{r}}) \psi_{\bar{K}}^\dagger({\mathbf{r}}) V(\mathbf{r})P_S P_T \psi_L({\mathbf{r}}) \psi_{\bar L}({\mathbf{r}})=
\end{equation}

\begin{equation}
 =2\int r dr dz d \phi
 \left[ \varphi_{K \uparrow }(r_\perp,z)e^{-i\Lambda_K^-\phi}|\uparrow\rangle^*+\varphi_{K \downarrow}(r_\perp,z)e^{-i\Lambda_K^+\phi}|\downarrow\rangle^*\right] \nonumber
\end{equation}
\begin{displaymath}
 \left[- \varphi_{K \uparrow }(r_\perp,z)e^{i\Lambda_K^-\phi}|\downarrow\rangle^*+\varphi_{K \downarrow}(r_\perp,z)e^{i\Lambda_K^+\phi}|\uparrow\rangle^*\right] V(r) P_S\left[ \varphi_{L \uparrow }(r_\perp,z)e^{i\Lambda_L^-\phi}|\uparrow\rangle+\varphi_{L \downarrow}(r_\perp,z)e^{i\Lambda_L^+\phi}|\downarrow\rangle\right]
\end{displaymath}
\begin{displaymath}
 \left[- \varphi_{L \uparrow }(r_\perp,z)e^{-i\Lambda_L^-\phi}|\downarrow\rangle+\varphi_{L \downarrow}(r_\perp,z)e^{-i\Lambda_L^+\phi}|\uparrow\rangle\right]
 \langle \tau_K \tau_{\bar{K}} | P_T|\tau_L \tau_{\bar{L}}\rangle
\end{displaymath}
When the two-body wave functions have the proper isospin symmetry, in the two channels (S=0,T=1) and
(S=1,T=0) the matrix elements $\langle \tau_K \tau_{\bar{K}} | P_T|\tau_L \tau_{\bar{L}}\rangle=1$.

To obtain the matrix elements (24), we  first evaluate the left and the right side of the operator $P_S$. 
For the left side term we get: 
\begin{displaymath}
  \left[ \varphi_{K \uparrow }(r_\perp,z)e^{-i\Lambda_K^-\phi}|\uparrow\rangle^*+\varphi_{K \downarrow}(r_\perp,z)e^{-i\Lambda_K^+\phi}|\downarrow\rangle^*\right] \nonumber
 \left[- \varphi_{K \uparrow }(r_\perp,z)e^{i\Lambda_K^-\phi}|\downarrow\rangle^*+\varphi_{K \downarrow}(r_\perp,z)e^{i\Lambda_K^+\phi}|\uparrow\rangle^*\right]
\end{displaymath}
\begin{displaymath}
 =
 \left[- \varphi_{K \uparrow }(r_\perp,z)e^{-i\Lambda_K^-\phi}\varphi_{K \uparrow }(r_\perp,z)e^{i\Lambda_K^-\phi}|\uparrow\rangle^* |\downarrow\rangle^*\right]+   \left[ \varphi_{K \uparrow }(r_\perp,z)e^{-i\Lambda_K^-\phi}\varphi_{K \downarrow}(r_\perp,z)e^{i\Lambda_K^+\phi}|\uparrow\rangle^*|\uparrow\rangle^*\right]+
\end{displaymath}
\begin{displaymath}
 \left[- \varphi_{K \downarrow}(r_\perp,z)e^{-i\Lambda_K^+\phi}\varphi_{K \uparrow }(r_\perp,z)e^{i\Lambda_K^-\phi}|\downarrow\rangle^*|\downarrow\rangle^*\right]+   \left[\varphi_{K \downarrow}(r_\perp,z)e^{-i\Lambda_K^+\phi}\varphi_{K \downarrow}(r_\perp,z)e^{i\Lambda_K^+\phi}|\downarrow\rangle^*|\uparrow\rangle^*\right]
\end{displaymath}
\begin{displaymath}
 = \left[- \varphi_{K \uparrow }(r_\perp,z)\varphi_{K \uparrow }(r_\perp,z)|\uparrow\rangle^* |\downarrow\rangle^*\right]+
  \left[ \varphi_{K \uparrow }(r_\perp,z)\varphi_{K \downarrow}(r_\perp,z)e^{i(\Lambda_K^+-\Lambda_K^-)\phi}|\uparrow\rangle^*|\uparrow\rangle^*\right]+
\end{displaymath}
\begin{displaymath}
 \left[- \varphi_{K \downarrow}(r_\perp,z)\varphi_{K \uparrow }(r_\perp,z)e^{i(\Lambda_K^--\Lambda_K^+)\phi}|\downarrow\rangle^*|\downarrow\rangle^*\right]+
  \left[\varphi_{K \downarrow}(r_\perp,z)\varphi_{K \downarrow}(r_\perp,z)|\downarrow\rangle^*|\uparrow\rangle^*\right] \equiv A_1
\end{displaymath}

The right side term can be written as:
\begin{displaymath}
 \left[ \varphi_{L \uparrow }(r_\perp,z)e^{i\Lambda_L^-\phi}|\uparrow\rangle+\varphi_{L \downarrow}(r_\perp,z)e^{i\Lambda_L^+\phi}|\downarrow\rangle\right]
 \left[- \varphi_{L \uparrow }(r_\perp,z)e^{-i\Lambda_L^-\phi}|\downarrow\rangle+\varphi_{L \downarrow}(r_\perp,z)e^{-i\Lambda_L^+\phi}|\uparrow\rangle\right]=
\end{displaymath}
\begin{displaymath}
  \left[- \varphi_{L \uparrow }(r_\perp,z)\varphi_{L \uparrow }(r_\perp,z)|\uparrow\rangle |\downarrow\rangle\right]+
  \left[ \varphi_{L \uparrow }(r_\perp,z)\varphi_{L \downarrow}(r_\perp,z)e^{-i(\Lambda_L^+-\Lambda_L^-)\phi}|\uparrow\rangle|\uparrow\rangle\right]+
\end{displaymath}
\begin{displaymath}
 \left[- \varphi_{L \downarrow}(r_\perp,z)\varphi_{L \uparrow }(r_\perp,z)e^{-i(\Lambda_L^--\Lambda_L^+)\phi}|\downarrow\rangle|\downarrow\rangle\right]+
  \left[\varphi_{L \downarrow}(r_\perp,z)\varphi_{L \downarrow}(r_\perp,z)|\downarrow\rangle|\uparrow\rangle\right
]\end{displaymath}

\subsubsection{Matrix elements for the isovector  pairing interaction}

To obtain the matrix elements of the isovector pairing interaction, we apply the operator 
$P_{S=0}$ to the right side term provided above. We observe that, under the action of $P_S$ the components in which both spins are
aligned (i.e., both up or both down) vanish and therefore do not contribute. One thus obtains:
\begin{displaymath}
 \frac{1}{2}(1-P_\sigma) \Big\{\left[- \varphi_{L \uparrow }(r_\perp,z)\varphi_{L \uparrow }(r_\perp,z)|\uparrow\rangle |\downarrow\rangle\right]+
  \left[ \varphi_{L \uparrow }(r_\perp,z)\varphi_{L \downarrow}(r_\perp,z)e^{-i(\Lambda_L^+-\Lambda_L^-)\phi}|\uparrow\rangle|\uparrow\rangle\right]+
\end{displaymath}
\begin{displaymath}
 \left[- \varphi_{L \downarrow}(r_\perp,z)\varphi_{L \uparrow }(r_\perp,z)e^{-i(\Lambda_L^--\Lambda_L^+)\phi}|\downarrow\rangle|\downarrow\rangle\right]+
  \left[\varphi_{L \downarrow}(r_\perp,z)\varphi_{L \downarrow}(r_\perp,z)|\downarrow\rangle|\uparrow\rangle\right]\Big\}=
  \end{displaymath}
\begin{align*}
 \frac{1}{2}(1-P_\sigma) \Big\{\left[- \varphi_{L \uparrow }(r_\perp,z)\varphi_{L \uparrow }(r_\perp,z)|\uparrow\rangle |\downarrow\rangle\right]
+
  \left[\varphi_{L \downarrow}(r_\perp,z)\varphi_{L \downarrow}(r_\perp,z)|\downarrow\rangle|\uparrow\rangle\right]\Big\}=
  \end{align*}
  \begin{align*}
 \frac{1}{2} \Big\{-\left[ \varphi_{L \uparrow }(r_\perp,z)\varphi_{L \uparrow }(r_\perp,z)+ \varphi_{L \downarrow}(r_\perp,z)\varphi_{L \downarrow}(r_\perp,z)\right] |\uparrow\rangle |\downarrow\rangle
+ \\
  \left[\varphi_{L \downarrow}(r_\perp,z)\varphi_{L \downarrow}(r_\perp,z)
  + \varphi_{L \uparrow }(r_\perp,z)\varphi_{L \uparrow }(r_\perp,z) \right]|\downarrow\rangle |\uparrow\rangle
  \Big\}
  \end{align*}
  Taking the inner product with the left side expression $A_1$, we observe that only the first and fourth term survive, e.g.
  \begin{displaymath}
  \frac{1}{2}\Big\{
  \left[- \varphi_{K \uparrow }(r_\perp,z)\varphi_{K \uparrow }(r_\perp,z)|\uparrow\rangle^* |\downarrow\rangle^*\right]+
  \left[\varphi_{K \downarrow}(r_\perp,z)\varphi_{K \downarrow}(r_\perp,z)|\downarrow\rangle^*|\uparrow\rangle^*\right]\Big\}\times
  \end{displaymath}
  \begin{align*}
  \Big\{-\left[ \varphi_{L \uparrow }(r_\perp,z)\varphi_{L \uparrow }(r_\perp,z)+ \varphi_{L \downarrow}(r_\perp,z)\varphi_{L \downarrow}(r_\perp,z)\right] |\uparrow\rangle |\downarrow\rangle
+ \\
  \left[\varphi_{L \downarrow}(r_\perp,z)\varphi_{L \downarrow}(r_\perp,z)
  + \varphi_{L \uparrow }(r_\perp,z)\varphi_{L \uparrow }(r_\perp,z) \right]|\downarrow\rangle |\uparrow\rangle
  \Big\}
  \end{align*}
  
  \begin{displaymath}
   = \frac{1}{2}\Big\{
   \varphi_{K \uparrow }(r_\perp,z)\varphi_{K \uparrow }(r_\perp,z)  \left[\varphi_{L \uparrow }(r_\perp,z)\varphi_{L \uparrow }(r_\perp,z)+ \varphi_{L \downarrow}(r_\perp,z)\varphi_{L \downarrow}(r_\perp,z) \right]+
  \end{displaymath}
\begin{displaymath}
  \varphi_{K \downarrow}(r_\perp,z)\varphi_{K \downarrow}(r_\perp,z)\left[ \varphi_{L \uparrow }(r_\perp,z)\varphi_{L \uparrow }(r_\perp,z)+ \varphi_{L \downarrow}(r_\perp,z)\varphi_{L \downarrow}(r_\perp,z) \right]\Big\}
\end{displaymath}
\begin{displaymath}
 = \frac{1}{2}\left[
   \varphi_{K \uparrow }(r_\perp,z)\varphi_{K \uparrow }(r_\perp,z)+\varphi_{K \downarrow}(r_\perp,z)\varphi_{K \downarrow}(r_\perp,z)
   \right]
   \left[ \varphi_{L \uparrow }(r_\perp,z)\varphi_{L \uparrow }(r_\perp,z)+ \varphi_{L \downarrow}(r_\perp,z)\varphi_{L \downarrow}(r_\perp,z) \right]
\end{displaymath}

In conclusion, the matrix elements of the isovector pairing interaction are given by (note that the factor 1/2 from the expressions above is canceled by the factor 2 in front of the integral \ref{Integral}):
\begin{align*}
 V^{T=1}_{KL} =\frac{1}{2}\int r dr dz d \phi V(r,z)
  \left[
   \varphi_{K \uparrow }(r_\perp,z)\varphi_{K \uparrow }(r_\perp,z)+\varphi_{K \downarrow}(r_\perp,z)\varphi_{K \downarrow}(r_\perp,z)
   \right] \times \\
   \left[ \varphi_{L \uparrow }(r_\perp,z)\varphi_{L \uparrow }(r_\perp,z)+ \varphi_{L \downarrow}(r_\perp,z)\varphi_{L \downarrow}(r_\perp,z) \right]
\end{align*}

\subsubsection{Matrix elements for the isoscalar pairing interaction}

In this case one must evaluate the action of $P_{S=1}$ on the right side term. One thus obtains:

\begin{equation*}
P_{S=1} \Big\{ \left[ \varphi_{L \uparrow }(r_\perp,z)e^{i\Lambda_L^-\phi}|\uparrow\rangle+\varphi_{L \downarrow}(r_\perp,z)e^{i\Lambda_L^+\phi}|\downarrow\rangle\right] \left[- \varphi_{L \uparrow }(r_\perp,z)e^{-i\Lambda_L^-\phi}|\downarrow\rangle+\varphi_{L \downarrow}(r_\perp,z)e^{-i\Lambda_L^+\phi}|\uparrow\rangle\right] \Big\}
\end{equation*}
\begin{equation*}
=\frac{1}{2}(1+P_\sigma) \Big\{ \left[ \varphi_{L \uparrow }(r_\perp,z)e^{i\Lambda_L^-\phi}|\uparrow\rangle+\varphi_{L \downarrow}(r_\perp,z)e^{i\Lambda_L^+\phi}|\downarrow\rangle\right] \left[- \varphi_{L \uparrow }(r_\perp,z)e^{-i\Lambda_L^-\phi}|\downarrow\rangle+\varphi_{L \downarrow}(r_\perp,z)e^{-i\Lambda_L^+\phi}|\uparrow\rangle\right] \Big\}
\end{equation*}

\begin{align*}
=\frac{1}{2} \Big\{ -\left[ \varphi_{L \uparrow }(r_\perp,z) \varphi_{L \uparrow }(r_\perp,z)|\uparrow\rangle |\downarrow\rangle \right] -\left[\varphi_{L \downarrow}(r_\perp,z) \varphi_{L \uparrow }(r_\perp,z)e^{i(\Lambda_L^+ - \Lambda_L^-)\phi}|\downarrow\rangle |\downarrow\rangle \right] \\
 + \left[ \varphi_{L \uparrow }(r_\perp,z) \varphi_{L \downarrow}(r_\perp,z)e^{i(\Lambda_L^- - \Lambda_L^+)\phi} |\uparrow\rangle |\uparrow\rangle \right] + \left[ \varphi_{L \downarrow}(r_\perp,z) \varphi_{L \downarrow}(r_\perp,z) |\downarrow\rangle |\uparrow\rangle \right] \\
  - \left[ \varphi_{L \uparrow }(r_\perp,z) \varphi_{L \uparrow }(r_\perp,z)|\downarrow\rangle |\uparrow\rangle \right] + \left[ \varphi_{L \downarrow}(r_\perp,z) \varphi_{L \downarrow}(r_\perp,z) |\uparrow\rangle |\downarrow\rangle \right] \\
  - \left[\varphi_{L \downarrow}(r_\perp,z) \varphi_{L \uparrow }(r_\perp,z)e^{i(\Lambda_L^+ - \Lambda_L^-)\phi}|\downarrow\rangle |\downarrow\rangle \right] + \left[ \varphi_{L \uparrow }(r_\perp,z) \varphi_{L \downarrow}(r_\perp,z)e^{i(\Lambda_L^- - \Lambda_L^+)\phi} |\uparrow\rangle |\uparrow\rangle \right]  \Big\}
\end{align*}

\begin{multline*}
=\frac{1}{2}  \Big\{ \left[ \varphi_{L \downarrow }(r_\perp,z) \varphi_{L \downarrow }(r_\perp,z) - \varphi_{L \uparrow }(r_\perp,z) \varphi_{L \uparrow }(r_\perp,z) \right] |\uparrow\rangle |\downarrow\rangle + \\
\left[ \varphi_{L \downarrow }(r_\perp,z) \varphi_{L \downarrow }(r_\perp,z) - \varphi_{L \uparrow }(r_\perp,z) \varphi_{L \uparrow }(r_\perp,z) \right] |\downarrow\rangle |\uparrow\rangle \\
 + 2 \left[ \varphi_{L \uparrow }(r_\perp,z) \varphi_{L \downarrow}(r_\perp,z) e^{i(\Lambda_L^- - \Lambda_L^+)\phi} \right] |\uparrow\rangle |\uparrow\rangle - 2 \left[\varphi_{L \downarrow}(r_\perp,z) \varphi_{L \uparrow }(r_\perp,z)e^{i(\Lambda_L^+ - \Lambda_L^-)\phi} \right] |\downarrow\rangle |\downarrow\rangle \Big\}
\end{multline*}

We now take the product with the left side term  $A_1$ and use the fact that
$\Lambda_I^+ - \Lambda_I^- = 1$.  This yields the following expressions for the possible spin-spin configurations:

a) Spins up-up 
\begin{equation*}  
uu = \varphi_{K \uparrow }(r_\perp,z)\varphi_{K \downarrow}(r_\perp,z)\varphi_{L \uparrow }(r_\perp,z) \varphi_{L \downarrow}(r_\perp,z) 
\end{equation*}

b) Spins down-down 
\begin{equation*}
dd = \varphi_{K \uparrow }(r_\perp,z)\varphi_{K \downarrow}(r_\perp,z)\varphi_{L \downarrow}(r_\perp,z) \varphi_{L \uparrow }(r_\perp,z)
\end{equation*}

c) Spins up-down
\begin{equation*}
  ud = -\frac{1}{2}\left[\varphi_{K \uparrow }(r_\perp,z)\varphi_{K \uparrow }(r_\perp,z)\varphi_{L \downarrow }(r_\perp,z) \varphi_{L \downarrow }(r_\perp,z) -\varphi_{K \uparrow }(r_\perp,z)\varphi_{K \uparrow }(r_\perp,z)\varphi_{L \uparrow }(r_\perp,z) \varphi_{L \uparrow }(r_\perp,z) \right]
\end{equation*}

d) Spins down-up 
\begin{equation*}
  du = \frac{1}{2}\left[ \varphi_{K \downarrow}(r_\perp,z)\varphi_{K \downarrow}(r_\perp,z)\varphi_{L \downarrow }(r_\perp,z) \varphi_{L \downarrow }(r_\perp,z) - \varphi_{K \downarrow}(r_\perp,z)\varphi_{K \downarrow}(r_\perp,z)\varphi_{L \uparrow }(r_\perp,z) \varphi_{L \uparrow }(r_\perp,z)  \right]
\end{equation*}

Summing  all terms, we obtain: 
\begin{multline*}
 2\varphi_{K \uparrow }(r_\perp,z)\varphi_{K \downarrow}(r_\perp,z)\varphi_{L \uparrow }(r_\perp,z) \varphi_{L \downarrow}(r_\perp,z) \\
+ \frac{1}{2} \left[ \varphi_{K \downarrow}(r_\perp,z)\varphi_{K \downarrow}(r_\perp,z) - \varphi_{K \uparrow }(r_\perp,z)\varphi_{K \uparrow }(r_\perp,z) \right] \left[ \varphi_{L \downarrow }(r_\perp,z) \varphi_{L \downarrow }(r_\perp,z) - \varphi_{L \uparrow }(r_\perp,z) \varphi_{L \uparrow }(r_\perp,z) \right] 
\end{multline*}

In conclusion, the matrix elements of the isoscalar pairing interaction are given by (note again 
that the factor 1/2 from the expressions above is canceled by the factor 2 in front of the integral \ref{Integral}):
\begin{multline*}
V^{T=0}_{KL} =\int r dr dz d \phi V(r,z) (
 4\varphi_{K \uparrow }(r_\perp,z)\varphi_{K \downarrow}(r_\perp,z)\varphi_{L \uparrow }(r_\perp,z) \varphi_{L \downarrow}(r_\perp,z) \\
+ \left[ \varphi_{K \downarrow}(r_\perp,z)\varphi_{K \downarrow}(r_\perp,z) - \varphi_{K \uparrow }(r_\perp,z)\varphi_{K \uparrow }(r_\perp,z) \right] \left[ \varphi_{L \downarrow }(r_\perp,z) \varphi_{L \downarrow }(r_\perp,z) - \varphi_{L \uparrow }(r_\perp,z) \varphi_{L \uparrow }(r_\perp,z) \right] )
\end{multline*}

\end{document}